\documentclass[final,3p,times]{elsarticle}

\usepackage{amsmath,amssymb,amsthm, mathrsfs}
\usepackage{mathtools}
\usepackage{graphicx}
\usepackage{stmaryrd}
\usepackage{multirow}
\usepackage{showlabels}
\usepackage[ruled,vlined]{algorithm2e}
\usepackage{hyperref}
\hypersetup{
    colorlinks=true,       
}
\usepackage{wrapfig}

\newcommand{\mynabla}{\widetilde{\nabla}} 
\newcommand{\jump}[1]{[\![#1]\!]} 

\journal{}
\makeatletter
\def\ps@pprintTitle{%
 \let\@oddhead\@empty
 \let\@evenhead\@empty
 \def\@oddfoot{}%
 \let\@evenfoot\@oddfoot}
\makeatother

\begin{document}

\begin{frontmatter}


\title{Adaptive Hybridizable Discontinuous Galerkin discretization of the Grad-Shafranov equation by extension from polygonal subdomains}



\author[nyu]{Tonatiuh S\'anchez--Vizuet}
\ead{tonatiuh@cims.nyu.edu}
\address[nyu]{New York University, Courant Institute of Mathematical Sciences}
\fntext[fn1]{A.J.C. and T. S-V have been partially funded by the US Department of Energy. Grant No. DE-FG02-86ER53233.}
\author[ms,ms2]{Manuel E. Solano}
\ead{msolano@ing-mat.udec.cl}
\address[ms]{Universidad de Concepci\'on, Department of Mathematical Engineering}
\address[ms2]{Universidad de Concepci\'on, Center for Research in Mathematical Engineering CI$^2$MA}
\fntext[fn2]{M. S. has been partially funded by CONICYT--Chile through FONDECYT project No. 1160320 and by Project AFB170001 of the PIA Program: Concurso Apoyo a Centros Cientificos y Tecnologicos de Excelencia con Financiamiento
Basal.}
\author[nyu]{Antoine J. Cerfon}
\ead{cerfon@cims.nyu.edu}

\begin{abstract}
We propose a high-order adaptive numerical solver for the semilinear elliptic boundary value problem modelling magnetic plasma equilibrium in axisymmetric confinement devices. In the fixed boundary case, the equation is posed on curved domains with piecewise smooth curved boundaries that may present corners. The solution method we present is based on the hybridizable discontinuous Galerkin method and sidesteps the need for  geometry-conforming triangulations thanks to a transfer technique that allows to approximate the solution using only a polygonal subset as computational domain. Moreover, the solver features automatic mesh refinement driven by a residual-based a posteriori error estimator. As the mesh is locally refined, the computational domain is automatically updated in order to always maintain the distance between the actual boundary and the computational boundary of the order of the local mesh diameter. Numerical evidence is presented of the suitability of the estimator as an approximate error measure for physically relevant equilibria with pressure pedestals, internal transport barriers, and current holes on realistic geometries.    
\end{abstract}

\end{frontmatter}


\section{Introduction} \label{sec:intro}

In toroidally axisymmetric configurations and in the absence of flows, the steady-state equations of magnetohydrodynamics (MHD) yield the following partial differential equation for the poloidal flux function $\psi=\psi(r,z)$, known as the \textit{Grad-Shafranov equation}  \cite{GrRu:1958,Shafranov:1958,LuSc:1957}
\begin{equation}\label{eq:GradShafranov}
-\Delta^*\psi = \mu_{0}r^2\frac{dp}{d\psi}+\frac{1}{2}\frac{dg^2}{d\psi}=:F(r,\psi).
\end{equation}
In Equation \eqref{eq:GradShafranov}, $r$ is the radial coordinate in the $(r,\phi,z)$ coordinate system naturally associated with the toroidal geometry, $\phi$ being the ignorable coordinate for the axisymmetric configurations we consider here, $\psi=\psi_{p}/2\pi$ where $\psi_{p}$ is the poloidal magnetic flux, $p=p(\psi)$ is the plasma pressure, $2\pi g(\psi) = -I_{p}$ is the net poloidal current flowing in the
plasma and the toroidal field coils, and the elliptic toroidal operator is defined by 
\[
\Delta^*\psi := r^2 div\left(\frac{1}{r^2}grad\,\psi\right) = r\partial_r\left(\frac{1}{r}\partial_r\psi\right) + \partial_z^2\psi= r\partial_r\left(\frac{1}{r}\partial_r\psi\right) + r\partial_z\left(\frac{1}{r}\partial_z\psi\right)=r\widetilde\nabla\cdot\left(\frac{1}{r}\widetilde\nabla\psi\right).
\] 
Above, for simplicity in the manipulations we have defined the operator $\widetilde\nabla:=(\partial_r,\partial_z)$ that acts formally like a vector of partial derivatives independent of the coordinate system. 

Both $p$ and $g$ are free functions of $\psi$, which are determined from other physical processes or experimental data, and taken as input to the partial differential equation. In general, $p$ and $g$ are such that $F(r,\psi)$ is a nonlinear function of $\psi$, so that the Grad-Shafranov equation is a semi-linear partial differential equation. Together with the boundary conditions, $p$ and $g$ determine the nature of the MHD equilibrium. Once $\psi$ is computed, the equilibrium magnetic configuration is fully determined, through the relations
\begin{align}
\mathbf{B}=\frac{1}{r}\nabla\psi\times \mathbf{e}_{\phi}+\frac{g(\psi)}r \mathbf{e}_{\phi}\label{eq:magfield_psi}\\
\mu_{0}\mathbf{J}=\frac{1}{r}\frac{dF}{d\psi}\nabla\psi\times \mathbf{e}_{\phi}-\frac{1}{r}\Delta^*\psi\,\mathbf{e}_{\phi}\label{eq:current_psi}
\end{align}
where $\mathbf{B}$ is the magnetic field and $\mathbf{J}$ is the current density, and $\mathbf{e}_{\phi}$ is the unit vector in the toroidal direction.

In this article, we will focus on fixed boundary equilibria, for which the boundary $\Gamma$ of the computational domain $\Omega$ is known, and corresponds to the boundary of the confinement region of the plasma. Physically, it must be a level set of $\psi$, and without loss of generality, we can let $\Gamma:=\partial\Omega$ be the level set $\psi=0$. We are therefore interested in the following Dirichlet boundary value problem
\begin{subequations}\label{eq:GradShafranovBVP}
\begin{alignat}{6}
-r\widetilde{\nabla}\cdot\left(\frac{1}{r}\widetilde\nabla\,\psi\right) =\, & F(r,\psi) & \qquad & \text{ in }\; \Omega\subset \mathbb R^2, \\
\psi =\,& 0  & \qquad & \text{ on }\; \Gamma.
\end{alignat}
\end{subequations}
For plasma physics applications, the solution $\psi$ to \eqref{eq:GradShafranovBVP}, the corresponding magnetic field $\mathbf{B}$ and the current density $\mathbf{J}$ given by Equations \eqref{eq:magfield_psi} and \eqref{eq:current_psi}, are used as input for stability, transport, and radio-frequency (RF) wave propagation and heating solvers \cite{Brambilla1999,Fable2013,HoSo:2014,Kerner1998,Lapillonne2009,Lee2017,Lee2019}. This gives stringent performance requirements on numerical solvers for the Grad-Shafranov equation. The solver should be fast, so that the time to compute the equilibrium configuration is negligible compared to the run time of the stability, transport, or RF wave solvers. The solver should also be accurate, because some of the physical quantities of interest depend not only on $\psi$, but also on the first derivatives of $\psi$, as is obvious for $\mathbf{B}$ and $\mathbf{J}$, and on the second derivatives of $\psi$, as is the case for the magnetic curvature for example. Because of the central role of the Grad-Shafranov equation in magnetic confinement fusion, many numerical solvers for Eq.\eqref{eq:GradShafranovBVP} have been developed in the last decades, relying on a vast range of formulations and numerical schemes. A good summary of early efforts can be found in \cite{TaTo:1991}. More recently, approaches based on bi-cubic finite elements \cite{Kerner1998,LuBoRo:1992,Lutjens1996}, on spectral elements \cite{HoSo:2014,PaKoFe:2016}, on the hybridizable discontinuous Galerkin method\cite{SaSo:2018}, and on integral equation methods \cite{PaCeFrGrOn:2013,LeCe:2015} have led to fast, high order accurate, and flexible solvers. Even so, none of these solvers simultaneously satisfy the four criteria required for optimal performance in magnetic confinement fusion applications, which can be listed as follows: 1) the solver must be fast; 2) it must be able to handle arbitrary boundaries $\Gamma$, which may or may not have corners (corresponding to magnetic X-points); 3) it must compute derivatives with high accuracy; and 4) it must have automatic adaptive refinement capabilities, in order to resolve the strong gradients in internal transport barriers and in edge pedestals without having a fine grid throughout the computational domain, where the solution is typically very smooth and requires few grid points for good accuracy. In this article, we present the first numerical Grad-Shafranov solver which satisfies these four performance requirements. 

To achieve this goal, we added adaptive refinement capabilities to the Hybridizable Discontinuous Galerkin (HDG) Grad-Shafranov solver we originally presented in \cite{SaSo:2018}. The solver relies on reformulating the problem as a first order system \cite{CoGoLa:2009}, which takes the form
\begin{subequations}\label{eq:FirstOrderSystem}
\begin{alignat}{6}
\boldsymbol q - \frac{1}{r}\mynabla\psi = \,& \boldsymbol 0 &\qquad&  \text{in }\, \Omega\subset \mathbb R^2,  \label{eq:FirstOrderSystema} \\
-\mynabla\cdot\boldsymbol q =\,& \frac{F}{r} &\qquad& \text{in }\, \Omega\subset \mathbb R^2, \label{eq:FirstOrderSystemb}\\
\psi =\, & 0 &\qquad& \text{on }\, \partial\Omega. \label{eq:FirstOrderSystemc}
\end{alignat}
\end{subequations}
The auxiliary variable $\boldsymbol q$ will be referred to as \textit{the flux}. This mixed formulation has the practical advantage of discretizing $\boldsymbol q$ directly, thus providing additional accuracy for the the physically meaningful quantity. Our enhanced solver exploits the natural suitability of discontinuous Galerkin methods for parallel computation, sidesteps the geometrical complexities by carrying on the computations on a polygonal subdomain of $\Omega$ discretized by a simple non-fitting and shape regular triangulation, handling the curved boundaries with a high order transferring technique, approximates the partial derivatives of $\psi$ directly, and features automatic mesh refinement driven by a local error estimator. 

The structure of the paper is as follows. Starting with geometric considerations, Section \ref{sec:discreteproblem} describes the discretization of the mixed form \eqref{eq:FirstOrderSystem}, introducing the basic concepts of the hybridizable discontinuous Galerkin method and the components of the solution process including the treatment of curved boundaries, the accelerated iteration process to handle the non-linearity of the Grad-Shafranov equation, and a post-processing step yielding an approximation to $\psi_h$ that converges with an additional order of accuracy. Throughout Section \ref{sec:discreteproblem} the problem is posed on a fixed computational domain with a given mesh. Adaptivity is then addressed in Section \ref{sec:adaptive}, which starts by introducing a residual-based local error estimator and by briefly discussing choices for the element marking strategy. The remaining part of the section discusses the issue of refining the embedded triangulation while maintaining the distance between the computational domain and the boundary always on the order of the local mesh parameter. The resulting strategy generates a sequence of updates of the computational domain that approximate the physical domain by exhaustion as the refinement progresses. In Section \ref{sec:numerical} we present numerical experiments to demonstrate the efficiency and reliability of the error estimator, as well as the convergence properties of the numerical solution. The experiments are carried out in realistic geometries first for a Solov'ev equilibrium for which an exact solution is available, and then for physically relevant benchmark problems with sharp and localized features, specifically an equilibrium with a pressure pedestal, and equilibrium with an internal transport barrier, and an equilibrium with a deep current hole. Concluding remarks are given in the final Section \ref{sec:conclusion}.
%
\section{The discrete problem}\label{sec:discreteproblem}
%
Before describing the adaptive algorithm, we will discuss the problem on a fixed, uniform and embedded polygonal mesh. Most of the details have been described in \cite{SaSo:2018}. However, starting with this standard case will allow us to introduce the notation and the fundamental ideas underpinning our HDG approach, as well as our treatment of curved boundaries and the iterative method to treat the non-linearity of the equation. Moreover, the solution of the problem in this setting will constitute the starting point for the adaptive algorithm. It is therefore worth repeating the key elements of the numerical scheme for the standard situation here. 

The HDG formulation of the problem depends on the specific spatial discretization of the domain where the equation is posed. Therefore, we will first describe the choice of polygonal subdomain where the problem will be discretized. The use of a polygonal subset of $\Omega$ as the computational domain creates the need to communicate the Dirichlet boundary conditions from the ``true" boundary to that of the polygonal subdomain where the computations are carried out. Hence, we will then describe the high order transfer scheme that will be used to impose the boundary conditions on the computational domain. With all these ingredients in place, it will be then possible to pose the discrete problem and describe the numerical method in detail. A rigorous analysis of the method for general semilinear elliptic equations is the subject of ongoing work \cite{SaSaSo:2018}.   
\subsection{The computational domain}\label{sec:ComputationalDomain}
%
We will start by defining the polygonal subdomain where the discrete system will be posed, henceforward the \textit{computational domain}, and the grids that will be used for approximation. Following \cite{CoSo:2012}, the computational domain $\Omega^{h}$ will be chosen to be a polygonal subdomain of $\Omega$ obtained from a regular background triangulation as follows.

\begin{figure}\centering
\begin{tabular}{ccc}
Background mesh  \qquad & \qquad  Computational mesh \qquad & \qquad  Companion mesh \\
\includegraphics[height = 0.18\linewidth]{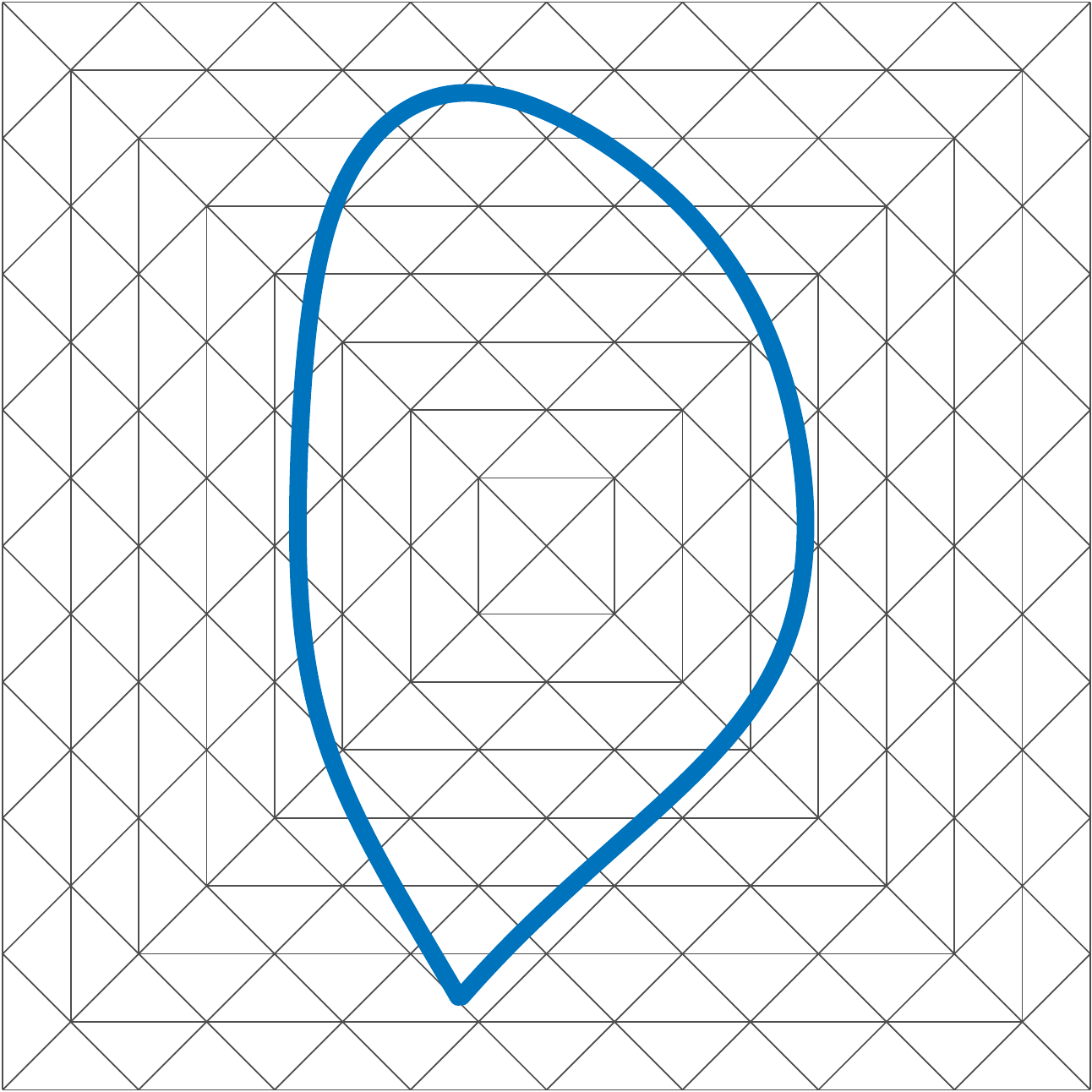}  \qquad & \qquad  \includegraphics[height = 0.18\linewidth]{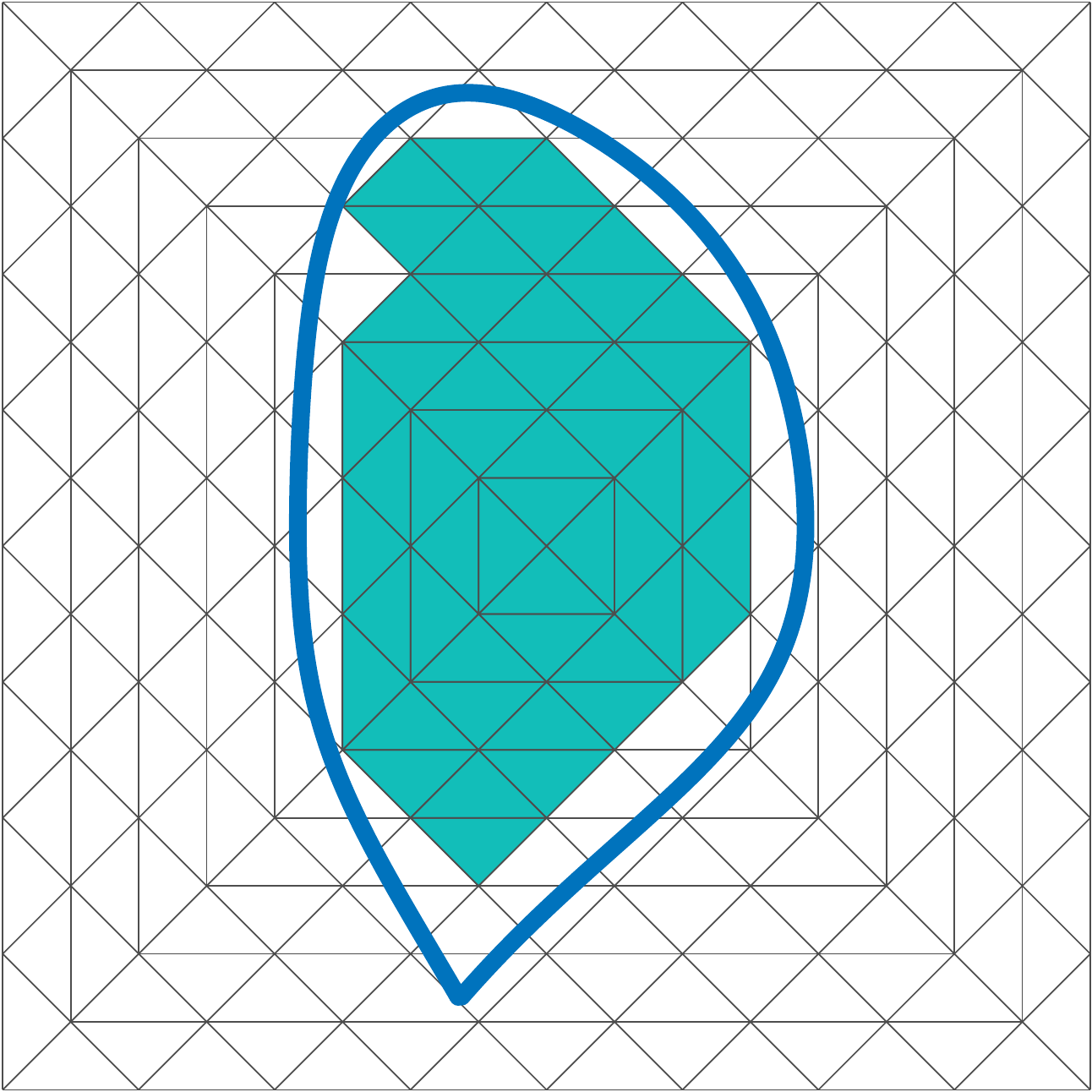}  \qquad & \qquad  \includegraphics[height = 0.18\linewidth]{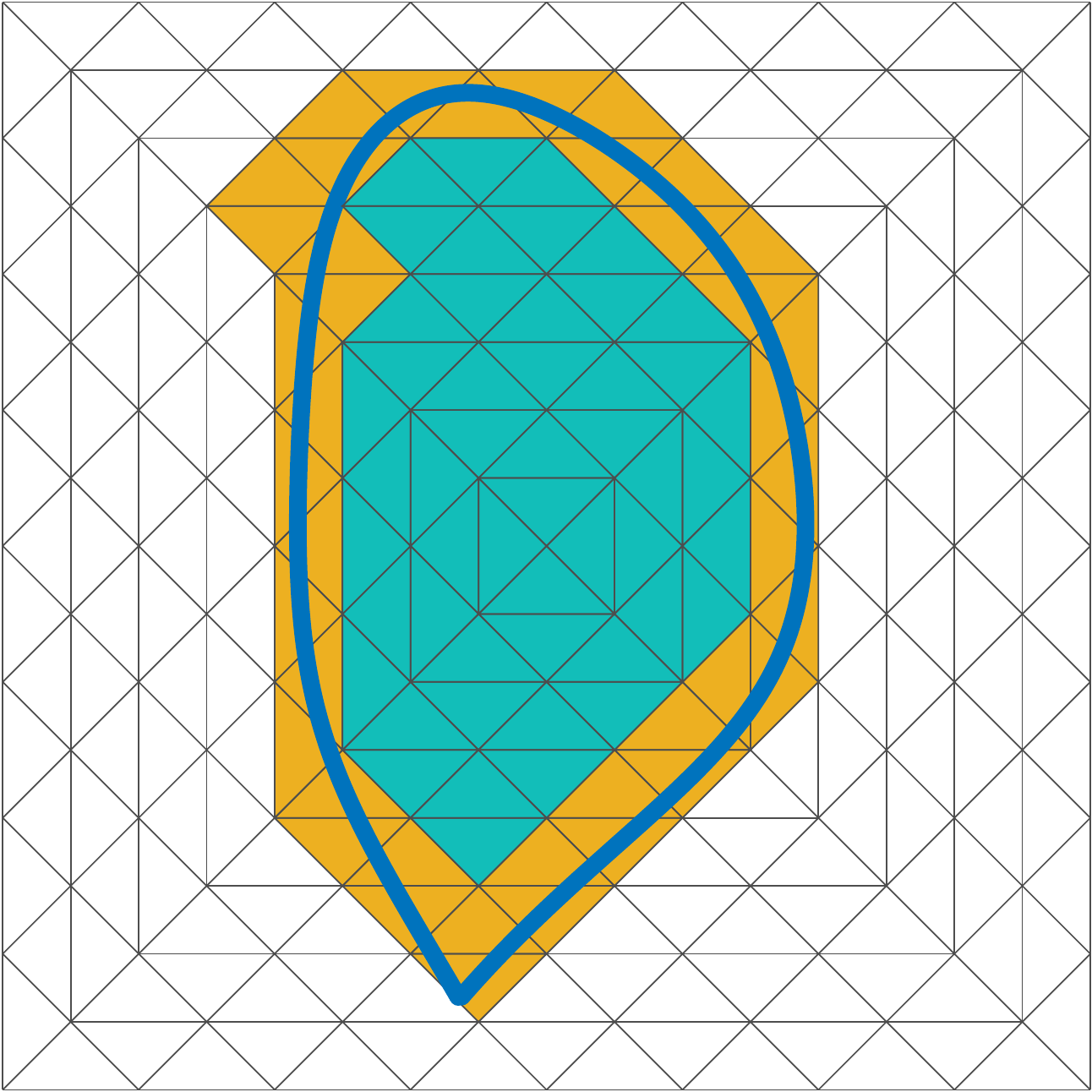} 
\end{tabular}
\caption{A uniform, shape-regular background mesh is used to define the initial computational and companion meshes (left). The background elements completely contained in $\Omega$ define the computational mesh $\mathsf T^{h}$ (center). The union of the background elements defining a minimal cover of $\overline{\Omega}$ will define the companion mesh $\mathsf T^{h}_c$ (right). (Colors online).}\label{fig:InternalAndExternalMesh}
\end{figure}

Consider $\mathcal T^{h}$ to be a triangulation of a polygonal domain containing $\overline{\Omega}$ and consisting of uniformly shape-regular triangles $K$ as in Figure \ref{fig:InternalAndExternalMesh} (left). The computational mesh $\mathsf T^{h}$ and the computational domain $\Omega^{h}$ (Figure \ref{fig:InternalAndExternalMesh} center) are, respectively, the set of triangles completely contained in $\Omega$ and its interior. More precisely, we define
\[
\mathsf T^{h} : = \left\{ K\in\mathcal T^{h} \,:\, \overline{K}\subset \Omega \right\}, \quad \partial\mathsf T^{h} : = \left\{ \partial K \, : \, K \in\mathsf T^{h} \right\}, \quad \text{and} \quad \Omega^{h}:= \left(\cup_{K\in \mathsf T^{h}}\overline{K}\right)^\circ.
\]
The boundary of the computational domain will be denoted by $\Gamma^{h}$ and set of all edges $e$ commonly referred to as the \textit{skeleton}  of the triangulation will be denoted by $\mathcal E^{h}$. We note that the  skeleton can be decomposed as $\mathcal E^{h} = \mathcal E^\partial \cup \mathcal E^\circ$ where
\[
\mathcal E^\partial := \left\{e\in \mathcal E^{h} \,:\, e\subset \Gamma^{h}\right\} \quad \text{ and } \quad \mathcal E^\circ_0 := \left\{e\in \mathcal E^{h} \,:\, e\nsubset \Gamma^{h}\right\}
\]
are the set of boundary edges and interior edges respectively. In addition, a companion grid consisting of those elements in $\mathcal T^{h}$ that constitute a minimal cover of $\overline{\Omega}$ will be defined
\[
\mathsf T_c^{h}  := \left\{ K\in\mathcal T^{h} \,|\, K\cap\overline{\Omega} \neq \varnothing\right\}, 
\]
the companion mesh consists of all the elements in $\mathsf T^{h}$ together with those background elements that intersect with the boundary of $\Omega$, as depicted on the right end of Figure \ref{fig:InternalAndExternalMesh}. The need for this additional companion mesh will become apparent in Section \ref{sec:meshrefine}, where the mesh refinement strategy will be discussed.

\subsection{Extension from subdomains}
%
The definition of the computational domain $\Omega^{h}$ as an unfitted and embedded subdomain seems to leave open the problem of defining the approximate solutions in the intermediate region $\Omega\setminus\overline{\Omega^{h}}$ corresponding to the area between the ``true" boundary and the computational boundary. In particular, one must deal with the problem of defining the boundary conditions on $\Gamma^{h}$. This in fact can be dealt with in a natural way through the following transfer procedure.

\begin{figure}\centering
\begin{tabular}{ccc}
 \qquad 
\includegraphics[height = 0.16\linewidth]{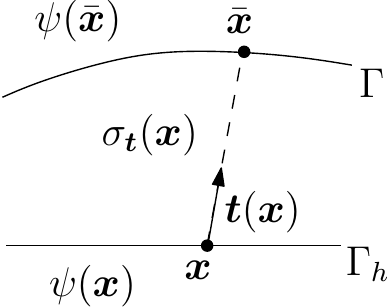} \qquad & \qquad 
\includegraphics[height = 0.18\linewidth]{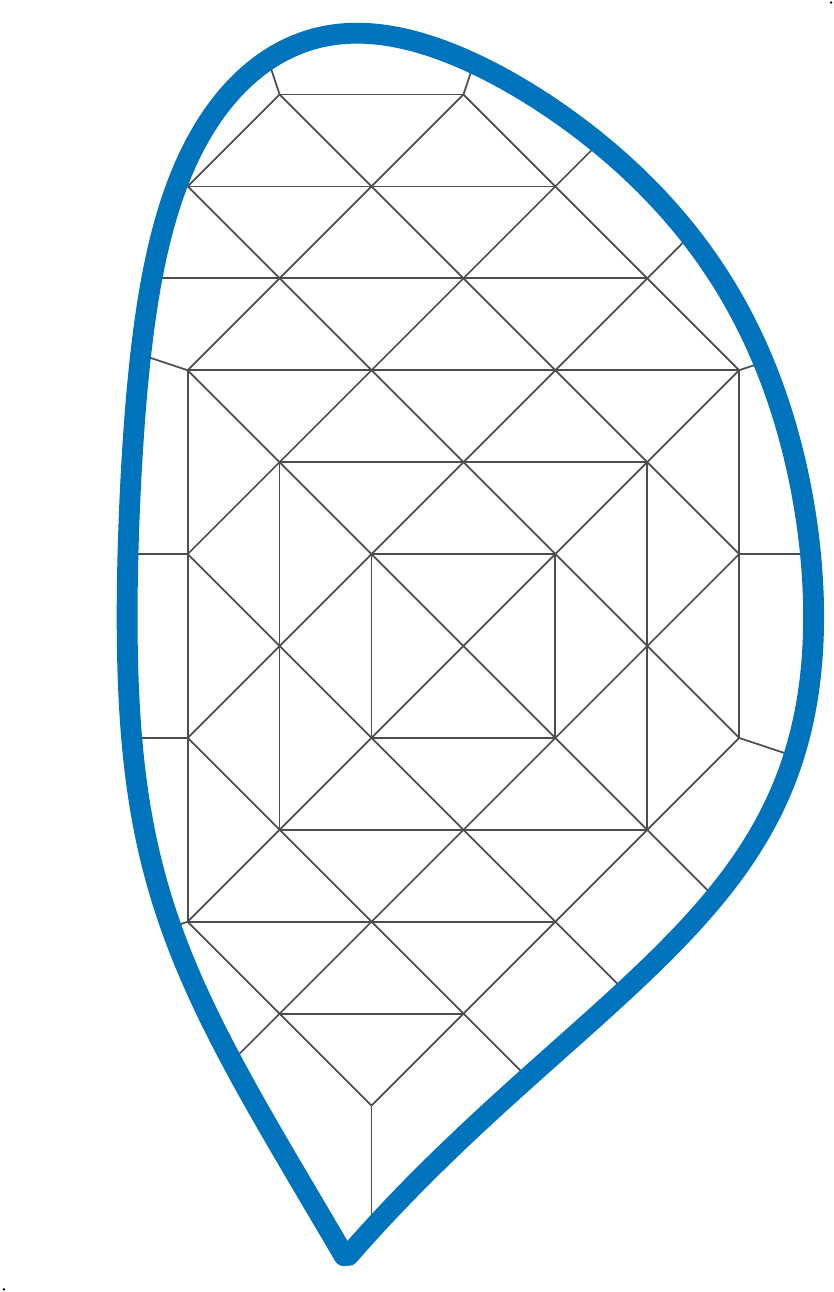} \qquad & \qquad \includegraphics[height = 0.16\linewidth]{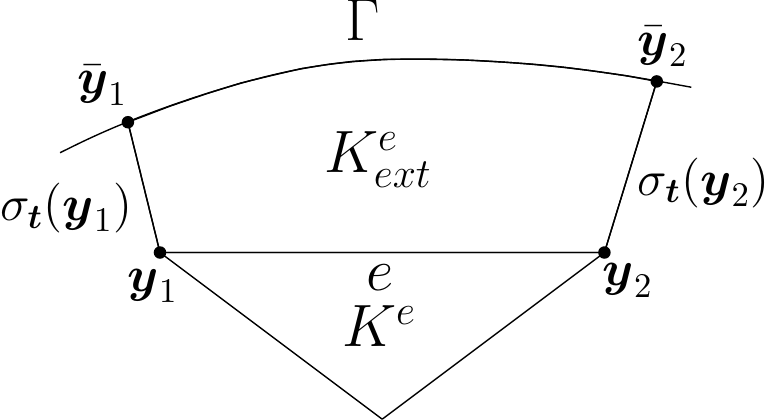}
\end{tabular}
\caption{Data on the boundary $\Gamma$ can be transferred to the boundary of the computational domain using the transfer paths $\boldsymbol \sigma_t$ (left). The region $\Omega\setminus\overline{\Omega^{h}}$ is divided into patches $K^{e}_{ext}$ where the flux will be extrapolated from the neighboring element $K^e$ (right) giving rise to a full tesselation of $\Omega$ (center). }\label{fig:extendedmesh}
\end{figure}

Consider a point $\overline{\boldsymbol x}=(\overline r,\overline z)$ on the true boundary $\Gamma$ and a point $\boldsymbol x=(r,z)$ on the computational boundary $\Gamma^{h}$. We will denote by $\boldsymbol t$ the normalized vector anchored at $\boldsymbol x$ pointing towards $\overline{\boldsymbol x}$ and   by $\sigma_{\boldsymbol t}(\boldsymbol x)$ the line segment connecting them, henceforward the \textit{transfer path} -- as in Figure \ref{fig:extendedmesh} (right). Then, integrating equation \eqref{eq:FirstOrderSystema} along a transfer path with direction vector given by $\boldsymbol{t}(\boldsymbol{x})$ it follows that 
\[
\psi(\boldsymbol{x})= \psi(\bar{\boldsymbol{x}}) - \int_{0}^ {d(\boldsymbol{x},\overline{\boldsymbol{x}})} \! r(\boldsymbol{x}+\boldsymbol{t}(\boldsymbol{x})s)\, \boldsymbol{q} (\boldsymbol{x}+\boldsymbol{t}(\boldsymbol{x})s)\cdot\boldsymbol{t}(\boldsymbol{x})\, ds,
\]
where $d(\boldsymbol{x},\overline{\boldsymbol{x}})$ is the Euclidean distance between $\boldsymbol{x}$ and $\overline{\boldsymbol{x}}$. Therefore, as long as $\boldsymbol q$ is known along the transfer path, it is possible to represent \textit{exactly} the value of $\psi$ at any point $\boldsymbol x$ of the computational boundary in terms of its value at one point $\overline{\boldsymbol x}$ of the physical boundary and the values of the flux. However, the value of $\boldsymbol q$ will be determined only within the computational domain $\Omega^{h}$ and thus we will resort to an approximation of $\boldsymbol q$ by extrapolation. In order to detail the extrapolation procedure we must first introduce the following extension of the computational domain.

Consider the set $\mathcal Y$ consisting of all endpoints of the edges $e\in \mathcal E^{\partial}_0$ and denote by $h_{loc}(\boldsymbol y)$ the minimum diameter $h$ over all triangles containing $\boldsymbol y$. To every $\boldsymbol y \in \mathcal Y$ we will assign a unique point $\overline{\boldsymbol y} \in \Gamma$ and will denote by $\sigma_{t}(\boldsymbol y)$ the straight line segment connecting them. The assignment must be done such that the Euclidean distance between them $d(\boldsymbol y,\overline{\boldsymbol y}) = \mathcal O(h_{loc}(\boldsymbol y))$, and that no two pair of such paths intersects.  This can be done in different ways, for instance, following the procedure described in \cite{CoSo:2012}. 

For a boundary edge $e\in \mathcal E^{\partial}_0$ we will denote by $K^{e}_{ext}$ the region enclosed by $\Gamma$, the paths corresponding to each endpoint of $e$, and the edge itself---as depicted on the right end of Figure \ref{fig:extendedmesh}. The union of all these patches covers the ``un-meshed" gap $\Omega\setminus\overline{\Omega^{h}}$, as can be seen in the center of Figure \ref{fig:extendedmesh}. Each of the patches $K^{e}_{ext}$ can be unambiguously identified with the unique element $K^e$ with which they share the edge $e$. This allows us to define the \textit{extension} $E(p): K^e \cup K^e_{ext} \rightarrow \mathbb R$ of a polynomial function $p: K^e \rightarrow \mathbb R$ as

\begin{alignat}{6}\label{eq:extp}
E(p): K^e \cup K^e_{ext} & \longrightarrow && \; \mathbb R \\
\nonumber
\boldsymbol x & \longmapsto && \; p(\boldsymbol x).
\end{alignat}

In other words, the extension $E(p)$ is a polynomial function with the same coefficients as $p$, but defined on the larger domain $K^e\cup K^e_{ext}$.  We can finally address the issue of transferring the boundary conditions to the computational boundary. Let $\boldsymbol q_h$ be a polynomial approximation to $\boldsymbol q$. Then,  for $\boldsymbol x \in \Gamma^{h} \cap e$ and $\overline{\boldsymbol x} \in \Gamma \cap K^e_{ext}$, the quantity
\begin{equation}\label{eq:transferBC}
\varphi_h(\boldsymbol x) := \psi(\bar{\boldsymbol{x}}) - \int_{0}^ {d(\boldsymbol{x},\overline{\boldsymbol{x}})} \! r(\boldsymbol{x}+\boldsymbol{t}(\boldsymbol{x})s)\, E(\boldsymbol{q}_h) (\boldsymbol{x}+\boldsymbol{t}(\boldsymbol{x})s)\cdot\boldsymbol{t}(\boldsymbol{x})\, ds 
\end{equation}
will be used as an approximation to the boundary value $\psi(\boldsymbol x)$. Note that the same formula can be used to define the approximation of the pointwise value of $\psi$ for \textit{any} point $x\in K^e_{ext}$.

\subsection{The discrete system}\label{sec:discretesystem}
We can now present the form of \eqref{eq:FirstOrderSystem} that will be discretized. Given a computational domain $\Omega^{h}$ with boundary $\Gamma^{h}$ and a regular embedded triangulation $\mathsf T^{h}$ as defined in Section \ref{sec:ComputationalDomain}, we look for functions $\psi$ and $\boldsymbol q$ defined on $\mathsf T^{h}$ and $\widehat\psi$ defined on $\mathcal E^{h}$ satisfying the system
\begin{subequations}\label{eq:HDGcontinuousSystem}
\begin{alignat}{6}
\label{eq:HDGcontinuousSystemA}
\boldsymbol q - \frac{1}{r}\mynabla\psi = \,& \boldsymbol 0 &\qquad& \text{ in } K \;\forall\, K\in\mathsf T^{h},  \\
\label{eq:HDGcontinuousSystemB}
-\mynabla\cdot\boldsymbol q =\,& \frac{F}{r} &\qquad& \text{ in } K \; \forall\, K\in\mathsf T^{h}, \\
\label{eq:HDGcontinuousSystemC}
\psi =\, & \widehat{\psi} &\qquad& \text{ on }\, \partial K \; \forall\, K\in\mathsf T^{h},  \\
\label{eq:HDGcontinuousSystemD}
 \jump{\boldsymbol q} =\,& 0 &\qquad& \text{ on } e\; \forall e\in \mathcal E^\circ_{h},  \\
 \label{eq:HDGcontinuousSystemE}
  \psi =\, & \varphi_h &\qquad& \text{ on }\, \Gamma^{h}. 
\end{alignat}
\end{subequations}
where $\varphi_h$ is given by \eqref{eq:transferBC} and the \textit{jump} of the flux across two elements $K^{+}, K^{-}$ with exterior normal vectors $\boldsymbol n^+, \boldsymbol n^-$ along a shared edge $e$ is defined in standard fashion as
\[\jump{\boldsymbol q}:=\boldsymbol q^+\cdot \boldsymbol n^+ + \boldsymbol q^-\cdot \boldsymbol n^-
\]. Note that for the remainder of this article, we will also use this notation for the jump of any scalar quantity $a$ across two elements, namely $\jump{a}:=a^+ - a^-$. In \eqref{eq:HDGcontinuousSystemC} and in what follows, quantities with the superscript $\;\widehat{\phantom{u}}\;$ must be understood as defined only on the skeleton of the mesh. The system \eqref{eq:HDGcontinuousSystem} is the restatement of \eqref{eq:FirstOrderSystem} as a collection of the local problems \eqref{eq:HDGcontinuousSystemA} and \eqref{eq:HDGcontinuousSystemB} satisfying the boundary conditions \eqref{eq:HDGcontinuousSystemC} or \eqref{eq:HDGcontinuousSystemE} on each element and ``glued" together by the continuity condition on the flux \eqref{eq:HDGcontinuousSystemD}. 

The introduction of the \textit{hybrid} unknown $\widehat\psi$ as a global quantity encoding the boundary conditions allows to fully decouple the local problems if an appropriate numerical flux is chosen (more on this below). Once the hybrid variable has been determined, the local problems for $\psi$ and $\boldsymbol q$ can be solved independently. This process is akin to the well known \textit{static condensation} technique used to decouple degrees of freedom on the edges/faces of an element from those on the interior which was first devised for Finite Elements \cite{Guyan:1965} and mixed formulations through hybridization \cite{Fraeijs:1965}. The connections between HDG and static condensation have been thoroughly discussed in \cite{Cockburn:2016}.  

When equation \eqref{eq:HDGcontinuousSystemB} is posed weakly and discretized, one is faced with the choice of a numerical approximation for the normal flux across the element edges. Due to the fact that it allows to express the weak forms of \eqref{eq:HDGcontinuousSystemA} and \eqref{eq:HDGcontinuousSystemB} entirely in terms of local quantities and also that it allows great freedom of choice for the approximation spaces, the choice
\begin{equation}\label{eq:NumericalFlux}
\widehat{\boldsymbol q}\cdot\boldsymbol n := \boldsymbol q\cdot\boldsymbol n- \tau(\psi-\widehat\psi).
\end{equation}
has become standard \cite{Cockburn:2016} and we will follow it in our discretization. Moreover, it is known that if the stabilization parameter $\tau>0$ remains of order $\mathcal O(1)$, the method achieves optimal convergence order. Therefore, for all the computations we will set it to be $\tau=1$. 
\subsection{The HDG discretization}
%
The HDG method \cite{Cockburn:2010} yields piecewise polynomial approximations $(\boldsymbol{q}_h,\psi_h,\widehat{\psi}_h)$ to the solutions  of the weak formulation obtained by testing \eqref{eq:HDGcontinuousSystem} with functions in the finite dimensional spaces
%
\begin{eqnarray}
\boldsymbol{V}_h & = & \{ \boldsymbol{v}\in\boldsymbol{L}^2(\mathsf{T}^h): \ \ \boldsymbol{v}|_K\in\mathbf{P}_k(K) \ \ \forall K\in\mathsf{T}^h \}, \\
W_h & = & \{ w \in L^2(\mathsf{T}^h): \ \ \ w|_K\in \mathbb{P}_k(K) \ \ \forall K\in\mathsf{T}^h \}, \\
M_h & = & \{ \mu\in L^2(\mathcal{E}_h): \ \ \mu|_e\in \mathbb{P}_k(e) \ \ \forall e\in\mathcal{E}_h \},
\end{eqnarray}
%
where the space of polynomials of degree $k$ defined on the triangle $K$ is denoted by $\mathbb{P}_k(K)$, the product space of two copies of itself is given by $\mathbf{P}(K):=[\mathbb{P}_k(K)]^2$ and $\mathbb{P}_k(e)$ is the space of polynomials of degree $k$ defined on a given edge $e$. The $L^2$ inner products in these spaces are given by  
\[
(\cdot,\cdot)_{\mathsf T^{h}}:= \sum_{K\in \mathsf T^{h}} (\cdot,\cdot)_{K}\, \qquad \qquad \langle\cdot,\cdot\rangle_{\partial \mathsf T^{h}} :=  \sum_{K\in \mathsf T^{h}} \langle\cdot,\cdot\rangle_{\partial K}, 
\]
where, as is customary,  $(\cdot,\cdot)_K$ and $\langle\cdot, \cdot\rangle_{\partial K}$ are the $L^2$ inner products on a single element $K$ and on its boundary $\partial K$ respectively.

Once the choice of trace of the numerical flux given by \eqref{eq:NumericalFlux} has been introduced and the system has been tested with functions $(\boldsymbol v, w,\mu) \in \boldsymbol V_h\times W_h\times M_h$,   the weak form of the first order system \eqref{eq:HDGcontinuousSystem} can be understood as consisting of two parts. The local equations
\begin{subequations}\label{eq:HDGsystem}
\begin{eqnarray}
(r\boldsymbol{q}_h,\boldsymbol{v})_{\mathsf{T}^h} + (\psi_h,\nabla\cdot\boldsymbol{v})_{\mathsf{T}^h} & = &  \langle\widehat{\psi}_h,\boldsymbol{v}\cdot\boldsymbol{n} \rangle_{\partial\mathsf{T}^h}, \\
-(\nabla\cdot\boldsymbol{q}_h,w)_{\mathsf{T}^h}   - \langle\tau\,\psi_h ,w\rangle_{\partial\mathsf{T}^h} & = & -\langle\tau\,\widehat{\psi}_h,w\rangle_{\partial\mathsf{T}^h} + (F/r,w)_{\mathsf{T}^h},
\end{eqnarray}
that are satisfied by $\boldsymbol q_h$ and $\psi_h$ independently on every element of the triangulation, and the global equations 
\begin{eqnarray}
\label{eq:intedge}
\langle\boldsymbol{q}_h \cdot\boldsymbol{n},\mu\rangle_{\partial\text{T}^h\setminus\Gamma^{h}} + \langle\tau\,\psi_h,\mu\rangle_{\partial\text{T}^h\setminus\Gamma^{h}}& = & \langle\tau\,\widehat\psi_h,\mu\rangle_{\partial\text{T}^h\setminus\Gamma^{h}}, \\ 
\label{eq:bdedge}
\langle \varphi_h,\mu\rangle_{\Gamma^{h}} & = & \langle\widehat{\psi}_h,\mu\rangle_{\Gamma^{h}},
\end{eqnarray}
\end{subequations}
that are satisfied by the hybrid unknown $\widehat\psi_h$ at the interior edges of the triangulation $\mathcal E^{\circ}_h$ \eqref{eq:intedge} and at the set of edges belonging to the boundary of the computational domain, $\mathcal E^\partial_h$ \eqref{eq:bdedge}. For each extended element $K_{ext}$ we can let $\overline{\boldsymbol x}\in\partial K_{ext}\cap\Gamma$ be the starting point of a transfer path and use the equation \eqref{eq:transferBC} together with  the fact that $\psi$ satisfies homogeneous Dirichlet boundary conditions to express the transferred boundary value $\varphi_h$ appearing on \eqref{eq:bdedge} as
\begin{equation}
\label{eq:transferapprox}
\varphi_h(\boldsymbol{x})=- \int_{0}^ {d(\boldsymbol{x},\overline{\boldsymbol{x}})} r(\boldsymbol{x}+\boldsymbol{t}(\boldsymbol{x})s)\, E(\boldsymbol{q}_h) (\boldsymbol{x}+\boldsymbol{t}(\boldsymbol{x})s)\cdot\boldsymbol{t}(\boldsymbol{x})\, ds.
\end{equation}
In the previous expression,  $E(\boldsymbol{q}_h)$ is the extrapolation of the polynomial $\boldsymbol{q}_h$ defined on $K^e$ to the neighboring exterior element $K_{ext}^e$ obtained by extending the domain of definition of $p$ from $K^e$ to $K^e_{ext}$ while keeping the same polynomial form.
\subsection{The solution method}\label{sec:solutionmethod}
%
In this section we will first consider the source term $F$ to be independent of $\psi$; once the solution method has been described for this simple, linear case, we will come back to address the nonlinear case, in which $F=F(r,\psi)$. We will denote the basis functions of the approximation spaces $\boldsymbol V_h$, $W_h$, and $M_h$ respectively by $\boldsymbol\phi_i, \phi_i$, and $\mu_i$, and define the Finite Element-style mass and convection matrices
\begin{alignat*}{8}
[r]_{ij} &:=\,&& (r\phi_j,\phi_i)_{\mathsf T^{h}} &&\qquad& [\nabla\cdot]_{ij} &:=\,&& (\nabla\cdot\boldsymbol\phi_j,\phi_i)_{\mathsf T^{h}} \\
[\tau]_{ij} &:=\,&& \langle\tau \phi_j,\mu_i\rangle_{\partial\mathsf T^{h}} &&\qquad& [\boldsymbol n]_{ij} &:=\,&& \langle\boldsymbol\phi_j\cdot\mathbf n,\mu_i\rangle_{\partial\mathsf T^{h}}.
\end{alignat*}
Moreover, if $\boldsymbol q_h, \psi_h, \widehat \psi_h, \varphi_h$, and $F/r$ denote respectively the coefficient vectors of the system unknowns, the transferred boundary conditions, and the source term, we can write the two parts of the HDG system succinctly in matrix form as
\begin{subequations}\label{eq:LinearSistem}
\begin{alignat}{6}
\label{eq:localLinearSystem}
\left[\begin{array}{cc}
r & \nabla\cdot^\top \\ -\,\nabla\cdot & -\tau \end{array}\right]\,\left[\begin{array}{c} \boldsymbol q_h \\ \psi_h \end{array}\right]   & =\,&&   \left[\begin{array}{c} \boldsymbol n ^\top \\ -\tau \end{array}\right]\, \left[\begin{array}{c} \widehat \psi_h \end{array}\right]  +  \left[\begin{array}{c} 0 \\ F/r \end{array}\right] \qquad & \text{(Local equations)},\\
\nonumber  & &&  & \\
\label{eq:globalLinearSystem}
\left[\begin{array}{c} 0 \\ \varphi_h \end{array}\right] + \left[\begin{array}{cc} \boldsymbol n\,\mathbf I^\circ_e & -\tau\,\mathbf I^\circ_e \\ 0 & 0 \end{array}\right]\,\left[\begin{array}{c} \boldsymbol q_h \\ \psi_h \end{array}\right] & =\,&&   \left[\begin{array}{c} -\tau\,\mathbf I^\circ_e \\ \mathbf I^\Gamma_e \end{array}\right]\, \left[\begin{array}{c} \widehat \psi_h \end{array}\right]  \qquad & \text{(Global equations)},
\end{alignat}
\end{subequations}
where the operators $\mathbf I_e^\circ$ and $\mathbf I_e^\Gamma$ above are the discrete counterparts of the restriction to the interior edges $\widehat\psi|_{\partial\text{T}^h\setminus\Gamma^{h}}$ and boundary edges $\widehat\psi|_{\Gamma^{h}}$, respectively.  Now, following equation \eqref{eq:transferapprox} the first term on the left hand side of \eqref{eq:globalLinearSystem} comes from integrating the extrapolation $r\boldsymbol E(q_h)$ along the transfer paths; it can therefore be expressed in the form 
\[
\left[\begin{array}{c} 0 \\ \varphi_h \end{array}\right] = \left[\begin{array}{cc} 0 & 0\\ -\mathbf Q & 0 \end{array}\right]\left[\begin{array}{c} \boldsymbol q_h \\ \psi_h \end{array}\right]
\]
where $\mathbf Q = \widetilde{\mathbf Q}\; \mathbf I_{K}^\Gamma$ is the composition of two discrete operators: 1) a restriction to the elements with at least one edge on the computational boundary, denoted as $ \mathbf I_{K}^\Gamma$ , and 2) the combination of a line integral along the transfer paths and the inner product with the basis of $M_h$ defined on the skeleton of the mesh, which can be represented as
\[
[\widetilde{\mathbf Q}]_{ij} := \langle\textstyle\int_{0}^ {d(\boldsymbol{x},\overline{\boldsymbol{x}})} r(\boldsymbol{x}+\boldsymbol{t}(\boldsymbol{x})s)\, E(\boldsymbol{\phi}_i) (\boldsymbol{x}+\boldsymbol{t}(\boldsymbol{x})s)\cdot\boldsymbol{t}(\boldsymbol{x})\, ds,\, \mu_i(\boldsymbol x)\rangle_{\partial K}.
\] 
At the implementation level, both integrals (the line integral and the inner product) involved in $\widetilde{\mathbf Q}$ are approximated by quadrature rules with matching orders of accuracy. Therefore the global system can be written in matrix form as
\begin{equation}
\label{eq:globalLinearSystemMat}
 \left[\begin{array}{cc} \boldsymbol n\,\mathbf I^\circ_e & -\tau\,\mathbf I^\circ_e \\ -\mathbf Q & 0 \end{array}\right]\,\left[\begin{array}{c} \boldsymbol q_h \\ \psi_h \end{array}\right]  =   \left[\begin{array}{c} -\tau\,\mathbf I^\circ_e \\ \mathbf I^\Gamma_e \end{array}\right]\, \left[\begin{array}{c} \widehat \psi_h \end{array}\right],
\end{equation}
where the top row is satisfied by the degrees of freedom of $\widehat \psi_h$ lying on the internal edges of the skeleton and the bottom row is satisfied by those on the edges corresponding to the boundary of the computational domain.  Solving formally the linear system \eqref{eq:localLinearSystem} for $(\boldsymbol q_h,\psi_h)$ we obtain
\begin{equation}
\label{eq:LocalSolve}
\left[\begin{array}{c} \boldsymbol q_h \\ \psi_h \end{array}\right]   =  \left[ \begin{array}{cc} r & \nabla\cdot^\top \\ -\nabla\cdot & -\tau \end{array}\right]^{-1}\left[\begin{array}{c} \boldsymbol n ^\top \\ -\tau \end{array}\right]\left[\begin{array}{c} \widehat \psi_h \end{array}\right] + \left[ \begin{array}{cc} r & \nabla\cdot^\top \\ -\nabla\cdot & -\tau \end{array}\right]^{-1}\left[\begin{array}{c} 0 \\ F/r \end{array}\right].
\end{equation}
This expression for $(\boldsymbol q_h,\psi_h)$ can then be substituted into \eqref{eq:globalLinearSystemMat} and, from the resulting system, it follows that 
\begin{equation}
\label{eq:GlobalSolve}
\Bigg(\left[\begin{array}{cc} \boldsymbol n\,\mathbf I^\circ_e & -\tau\,\mathbf I^\circ_e \\ -\mathbf Q & 0 \end{array}\right]\left[\begin{array}{cc} r & \nabla\cdot^\top \\ -\nabla\cdot & -\tau \end{array}\right]^{-1}\left[\begin{array}{c} \boldsymbol n ^\top \\ -\tau \end{array}\right] - \left[\begin{array}{c} -\tau\,\mathbf I^\circ_e \\  \mathbf I^\Gamma_e \end{array}\right]\Bigg)\left[\begin{array}{c} \widehat \psi_h \end{array}\right] = - \left[\begin{array}{cc} \boldsymbol n\,\mathbf I^\circ_e & -\tau\,\mathbf I^\circ_e \\ -\mathbf Q & 0 \end{array}\right]\left[\begin{array}{cc} r & \nabla\cdot^\top \\ -\nabla\cdot & -\tau \end{array}\right]^{-1}\left[\begin{array}{c} 0 \\ F/r \end{array}\right].
\end{equation}
From this equation one can obtain $\widehat\psi_h$ and back-substitute in \eqref{eq:LocalSolve} to obtain $(\boldsymbol q_h,\psi_h)$. The relevance of the last two equations stems from the following observations. First, the matrices and the corresponding linear solves appearing on the right hand side of equation \eqref{eq:LocalSolve} are entirely in terms of local quantities, and can therefore be processed fully in parallel. Second, despite the fact that it involves a global unknown, the system \eqref{eq:GlobalSolve} is sparse, for it includes only degrees of freedom associated to either the skeleton of the mesh or elements with at least one edge on the computational boundary. Moreover, the linear solves appearing on each side of equation \eqref{eq:GlobalSolve} are the same as those in \eqref{eq:LocalSolve} and therefore have to be computed only once. The solution process can then be split into three steps: 
\begin{enumerate}
\item Locally (i.e. in parallel) solve the systems
\[
\left[ \begin{array}{cc} r & \nabla\cdot^\top \\ -\nabla\cdot & -\tau \end{array}\right]^{-1}\left[\begin{array}{c} \boldsymbol n ^\top \\ -\tau \end{array}\right] \quad \text{ and } \quad \left[ \begin{array}{cc} r & \nabla\cdot^\top \\ -\nabla\cdot & -\tau \end{array}\right]^{-1}\left[\begin{array}{c} 0 \\ F/r \end{array}\right]
\]
appearing in equation \eqref{eq:LocalSolve} and store them.
\item Using the local vectors obtained in the first step, assemble the matrices on both sides of equation \eqref{eq:GlobalSolve} and solve the resulting global system, thus recovering the hybrid unknown $\widehat\psi_h$.

\item Distribute the relevant parts of $\widehat\psi_h$ over local elements and use the local  solvers obtained on the first step to recover $(\boldsymbol q_h,\psi_h)$ fully in parallel.
\end{enumerate}
\paragraph{Accelerated Picard iterations}\label{sec:nonlinear}

In order to deal with the non-linear nature of the Grad-Shafranov equation, we will resort to a simple, yet effective, iterative strategy consisting of accelerated Picard iterations. The standard Picard or fixed point iteration goes as follows. Given a guess for the solution $\psi^n$, the source term can be evaluated yielding a source $F:=F(r,\psi^n)$ that is independent of the solution. The resulting system \eqref{eq:LinearSistem} is a linear problem that can be solved as described above yielding an update $\psi^{n+1}$.  The source term is then updated by evaluation at the newly computed solution and the process is repeated iteratively until the relative difference between successive updates falls below a certain predetermined tolerance.

This strategy is simple to implement but may require a large number of iterations to converge for small values of the tolerance. However, the convergence rate for Picard iterations can be improved by means of a device known as \textit{Anderson acceleration} \cite{Anderson:1965}. Anderson's idea is to improve convergence through the use of information from more than one previous iterate. This is achieved by defining the update $\psi_h^{n+1}$ to be an optimized convex linear combination of the solutions to \eqref{eq:LinearSistem} obtained on a predetermined number of previous iterations. The coefficients of the convex linear combination are chosen so that the difference between the solutions and the updates is minimized. This requires the storage of $m$ previous updates and $m$ previous solutions and the solution of a small $(m+1)\times(m+1)$ system at every iteration in order to determine the optimal coefficients.

Below we describe algorithmically the simplest form of the acceleration---which is the version implemented in our solver---but we refer the reader to the works by Kelly and Toth \cite{ToKe:2015}, and Walker and Ni \cite{WaNi:2011}, where the method is studied in detail. If we denote by $\epsilon$ a prescribed tolerance, by $\psi_0$ the initial input, by $\left(-\Delta^*\right)^{-1}$ the solution operator to \eqref{eq:LinearSistem} described above, and by  $\psi_h$ the final, approximate solution to the non-linear problem then, in its simplest form, the acceleration algorithm that uses $m$ previous iterates can be described as follows:
\begin{center}
\begin{algorithm}[H]

 \Begin{
 $n=0$\, , \quad $change = 1$\;
 $\widetilde \psi^1 =\left(-\Delta^*\right)^{-1}F(r,\psi^0)/r$\;
 $G^1 = \widetilde \psi^1-\psi^0$\;
 $\psi^1 = \widetilde \psi^1$\;
 \While{change $\geq \epsilon$}{
 $ n= n + 1$\;
 $k = \min\{m,n\}$\;
 $\widetilde{\psi}^{n+1} = \left(-\Delta^*\right)^{-1}F(r,\psi^n)/r$\;
 $G^{n+1} = \widetilde{\psi}^{n+1}-\psi^{n}$ \; 
 Find: \, $(\alpha_1,\ldots,\alpha_{k+1})\in \mathbb R^{k+1}$ such that
 \begin{enumerate}
 \item $\sum_{j=1}^{k+1}{\alpha_j}=1$
 \item $(\alpha_1,\ldots,\alpha_{k+1}) = \text{argmin } \|\sum_{j=1}^{k+1}{\alpha_j G^{n+j-k}}\| $
 \end{enumerate}
 $\psi^{n+1} = \sum_{j=1}^{k+1}\alpha_j\widetilde{\psi}^{n+j-k}$\;
 $change = \|\psi^{n+1}-\psi^{n}\|/\|\psi^{n+1}\|$; 
 }
  $\psi_h = \psi^{n+1}$\;
\caption{Anderson-accelerated Picard iterations}
 }
\end{algorithm}
\end{center}

%
\subsection{Non-linear local post-processing}
Following the idea introduced by Stenberg \cite{Stenberg:1991}, once the approximations $\psi_h$ and $\boldsymbol q_h$ have been determined from the solution of \eqref{eq:LinearSistem}, it is possible to define a locally post-processed function $\psi^*_h$ with enhanced convergence properties. There a several different ways of defining the post-processing, but in order to take advantage of the increased accuracy of the post processing as part of a residual estimator we will define $\psi_h^*$ to be the piecewise polynomial function satisfying
\begin{subequations}\label{eq:psistar}
\begin{alignat}{6}
\nonumber
\psi_h^*\in\;& \mathbb P_{k+1}(K) &\quad& \forall\, K\in \mathsf T^{h},\\
\label{eq:psistarA}
(\nabla\psi_h^*,\nabla w_h)_K -( F(\psi^*_h)/r,\nabla w_h)_K=\,& (r\boldsymbol q_h,\nabla w_h)_K -(F(\psi_h)/r,\nabla w_h)_K &\quad& \forall\,w_h\in \mathbb P_{k+1}(K), \\
\label{eq:psistarB}
(\psi_h^*,1)_K =\,& (\psi_h,1)_K. &
\end{alignat}
\end{subequations}
Note that when $F$ is independent of $\psi$, this reduces to the case analyzed in \cite{CoGoSa:2010}, where it was shown that the solution to this auxiliary problem converges towards $\psi$ with order $k+2$ when $k\geq 1$. Numerical evidence suggests that the simpler post processing that arises if the terms involving $F$ in equation \eqref{eq:psistarA} are dropped is also effective; even in the semi-linear case. However for our convergence analysis in \cite{SaSaSo:2018} the effect of the non-linear source term needs to be considered and this leads to the non linear post processing above. The solution to this auxiliary problem will be used in the error estimator described in the next section.

\section{The adaptive algorithm}\label{sec:adaptive}
%
In many situations of physical interest, the solution $\psi$ and its derivatives may vary rapidly in localized regions in $\Omega$ \cite{PaCeFrGrOn:2013}. In such cases, adaptive mesh refinement is an effective way to minimize the number of degrees of freedom for a given target accuracy. Our adaptive strategy follows the standard ``solve $\rightarrow$ estimate $\rightarrow$ mark $\rightarrow$ refine" iterative paradigm. Specifically, starting from an initial triangulation $\mathsf T^{h,0}$, the problem is solved and a suitable error estimator is computed using the obtained approximate solution. Based on the error estimator, a subset of the triangulation is marked for refinement. This generates a new triangulation $\mathsf T^{h,1}$ where the process can be started over until a predetermined number of cycles is reached or the estimator falls below a given threshold.

Our algorithm is based on theoretical work done by Cockburn and Zhang \cite{CoZh:2012,CoZh:2013}, and  Cockburn, Nochetto and Zhang \cite{CoNoZh:2016}. In the first references, the authors proposed and studied the performance of a residual-based error estimator, and in the third one they were able to prove the convergence of the adaptive HDG method assuming that D\"orfler's marking criterion is used (we will return to this later). The problems studied in those cases were linear and the equations were posed in polygonal domains discretized with fitted triangulations. Our focus on the Grad-Shafranov equation poses additional challenges, namely the semi-linearity of the problem and the non-fitting nature of the computational domain---which in turn imposes the requirement that the distance between the boundaries $\Gamma$ and $\Gamma^h$ remains locally  $\mathcal{O}(h_{loc}(\boldsymbol y))$. The non-linearity is dealt with through the accelerated iterative process described above. The refinement of the unfitted grid will require some additional care, as we discuss below.

\subsection{A residual-type estimator}\label{sec:estimator}
For an edge $e$ with length $h_e$ and a function $u$ defined on $e$ (or on a superset containing $e$) we will denote by $\|u\|_e$ its $L^2$ norm on $e$ (or the norm of its restriction to $e$). Considering $K$ to be a generic element of the triangulation $\mathsf T^{h}$, we will adopt the following local error estimator
\begin{align}
\nonumber
\eta_{K}^2   =\, & h_{K}^2\|F(\psi_h^*)/r+\nabla\cdot\boldsymbol q_h\|^2_{K} + \|\boldsymbol q_h-\frac{1}{r}\nabla\psi^*_h\|^2_{K} \\ 
\label{eq:estimator}
&  + \frac{1}{2}\left( \sum_{e\in\mathcal E^\circ\cap\partial K}{h_e\|\jump{\boldsymbol q_h}\|^2_{e}}  + \sum_{e\in\mathcal E^\circ\cap\partial K}{h_e^{-1}\|\jump{\psi^*_h}\|_e^2}\right) + \sum_{e\in\mathcal E\cap \partial K}h_e^{-1} \| \widehat{\psi}_h-\psi^*_h\|_e^2,
\end{align}
where $\psi^*_h$ is the post-processed numerical solution obtained by solving the local auxiliary problem \eqref{eq:psistar}. This estimator is based on the one proposed and analyzed by Cockburn and Zhang \cite{CoZh:2012,CoZh:2013} for linear elliptic equations posed in polygonal domains. The global error estimate is obtained by adding all the local contributions over the computational domain and is therefore defined as
\[
\eta^2(\mathsf T^h) := \sum_{K\in \mathsf T^h}\eta_K^2.
\] 
A detailed analysis of the estimator and its properties for semilinear problems like ours as well as possible improvements for it are the subject of a separate communication \cite{SaSaSo:2018}, but some intuitive understanding can be gained by expressing the estimator in the form $\eta^2 = \eta_1^2 + \eta_2^2 + \eta_3^2 + \eta_4^2 + \eta_5^2$ , where
\begin{alignat*}{6}
\eta_1^2 :=\,& \sum_{K\in\mathsf T}{h_{K}^2\|F(\psi_h^*)/r+\nabla\cdot\boldsymbol q_h\|^2_{K}}, & \qquad \qquad && \eta_2^2 :=\,& \sum_{K\in\mathsf T}{\|\boldsymbol q_h-\frac{1}{r}\nabla\psi^*_h\|^2_{K}}, \\
\eta_3^2 :=\,& \frac{1}{2}\sum_{K\in\mathsf T}{\sum_{e\in\mathcal E^\circ\cap\partial K}{h_e\|\jump{\boldsymbol q_h}\|^2_{e}}}, & \qquad \qquad && \eta_4^2 :=\,& \frac{1}{2}\sum_{K\in\mathsf T}{ \sum_{e\in\mathcal E^\circ\cap\partial K}{h_e^{-1}\|\jump{\psi^*_h}\|_e^2} },\\
\eta_5^2 :=\,& \sum_{K\in\mathsf T}{ \sum_{e\in\mathcal E\cap \partial K}h_e^{-1} \| \widehat{\psi}_h-\psi^*_h\|_e^2  }, & \qquad \qquad &&
\end{alignat*}
and studying each term separately.  The term $\eta_1$ corresponds to the local residual of the strong equation for the flux \eqref{eq:HDGcontinuousSystemB}. Similarly, it would be desirable to consider the residual $q_h-\frac{1}{r}\nabla\psi_h$ of the strong equation \eqref{eq:HDGcontinuousSystemA} as part of the estimator. However, it would converge with reduced order due to the differentiation of the approximate solution $\psi_h$. To overcome this and achieve the desired order of convergence, the term $\eta_2$ makes use of the post processed solution $\psi^*_h$ instead, thus preserving the desired convergence order $k+1$. In this sense, the second term of the estimator is reminiscent of indicators based on gradient recovery, where an improved approximation of the gradient is obtained through post processing and is then used to estimate the error.

Finally, the edge terms $\eta_3$ and $\eta_4$ provide, respectively, a measure of the local loss of conformity of the solution and of its flux, by considering their jumps across element edges in the normal direction. $\eta_5$ estimates the rate of convergence of the hybrid variable and post-processed solution---restricted to the element boundaries---as approximations to the local trace.
%
\subsection{Marking strategies}\label{sec:marking}
%
Given an initial triangulation $\mathsf T^{h,0}$, the discrete problem is solved and post-processed yielding the approximations $(\psi_{h,0},\boldsymbol q_{h,0},\widehat{\psi}_{h,0},\psi^*_{h,0})$ which are then used to compute the local error estimator \eqref{eq:estimator}. One now must choose the elements that will be marked for refinement based on the local values of $\eta_K$. Different marking strategies have been tried in the literature. Here we consider D\"orfler's criterion \cite{Dorfler:1996,Dorfler:1995} and the so-called maximum criterion \cite{BaRh:1978}. In both cases, one must first choose a value of the \textit{marking parameter} $\gamma\in[0,1]$ and then the elements 
\begin{enumerate}[A)] 
\item Either belonging to a minimal set $\mathcal M$ such that
\[
\gamma\sum_{K\in\mathsf T^{h,0}}\eta^2_K \leq \sum_{K\in\mathcal M}\eta^2_K,  \qquad (\textit{D\"orfler Marking} )
\]
\item or for which the local estimate $\eta_{K}$ is such that
\[
\gamma\,\max_{K\in \mathsf T^{h,0}}{\{\eta_K\}} \leq \eta_{K},  \qquad (\text{\textit{Maximum marking}})
\]
\end{enumerate}
are marked and subsequently refined. The choice of the value of the marking parameter $\gamma$ depends on the needs and constraints of the user. It is usually picked based on considerations such as memory availability, desired speed of the computation, etc. In the case of D\"orfler's marking, values of $\gamma$ closer to zero result in fewer elements being marked on each cycle, and very localized refinement; larger values of $\gamma$ tend to produce refinements that are more uniform. This qualitative behavior is reversed when maximum marking is used: refinement is localized for large values of $gamma$ and becomes uniform as the parameter approaches zero.

The convergence of the adaptive refinement loop on fixed polygonal domains was established by Cockburn, Nochetto and Zhang within the context of hybridizable discontinuous Galerkin methods for \textit{linear} problems \cite{CoNoZh:2016}; and assuming D\"orfler's method for marking. Regarding the maximum marking criterion, Mor\'in, Siebert and Veeser \cite{MoSiVe:2008} proved the convergence of the method for a wide class of linear problems discretized with Finite Elements. The analysis of this strategy applied to HDG is still an ongoing task, even for the linear case, but the strategy seems to be robust, as suggested by our numerical experiments.
%
\subsection{Local mesh refinement}\label{sec:meshrefine}
%
Once some elements of the triangulation have been marked by one of the criteria presented above, triangle refinement can be carried out through standard methods such as \textit{newest vertex bisection} (NVB) \cite{Rivara:1984a,Rivara:1984b} or a \textit{red-green} procedure \cite{BaShWe:1983}. In the standard setting where the computational domain $\Omega^{h}$ coincides with the domain of definition of the PDE, each of the refined meshes $\{\mathsf T^{h,n}\}_{n\geq0}$ produced in such fashion will remain a triangulation of the original domain $\Omega_h$. 

\begin{wrapfigure}{R}{0.4\textwidth}\centering
\begin{tabular}{ccc}
{\small Level 0} & {\small Level 2} & {\small Level 4} \\ 
\includegraphics[height=0.45\linewidth]{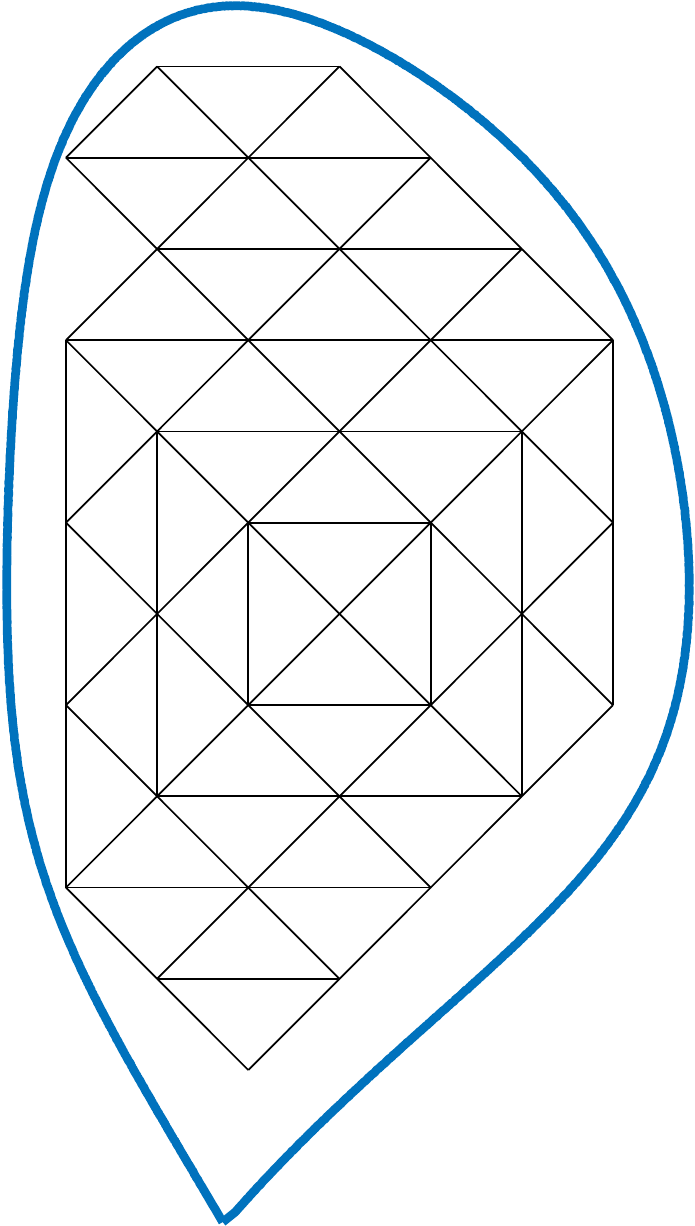} \; & \includegraphics[height=0.45\linewidth]{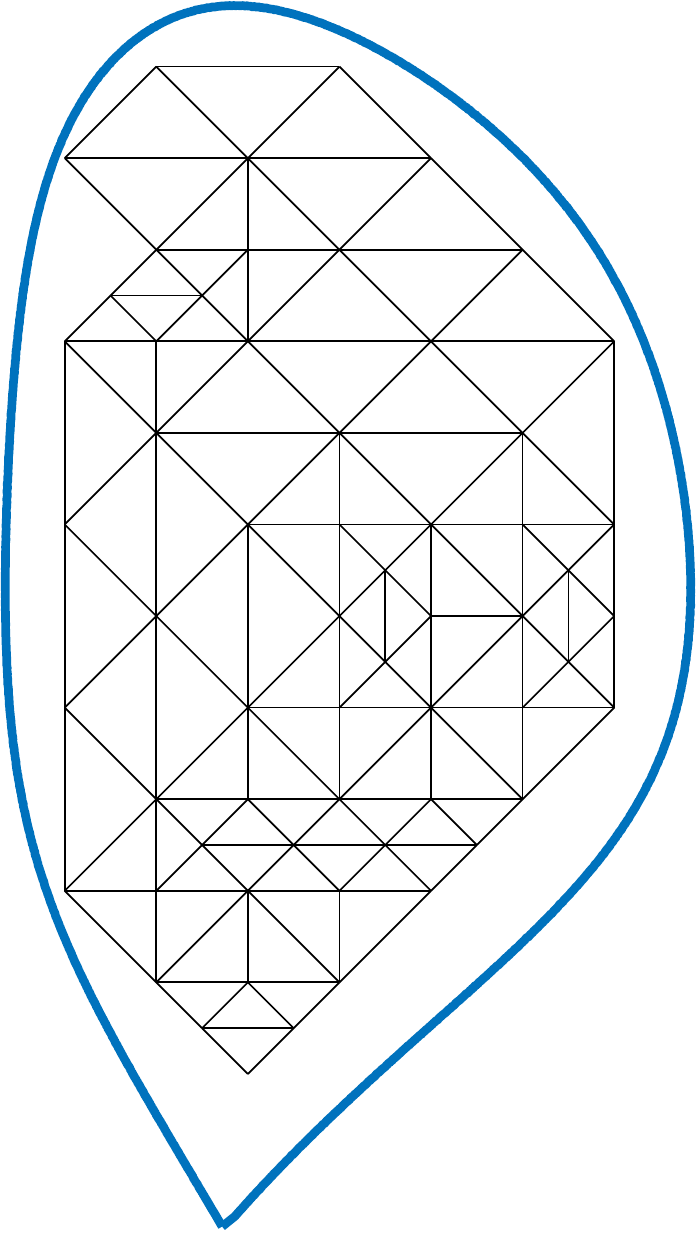} \; & \includegraphics[height=0.45\linewidth]{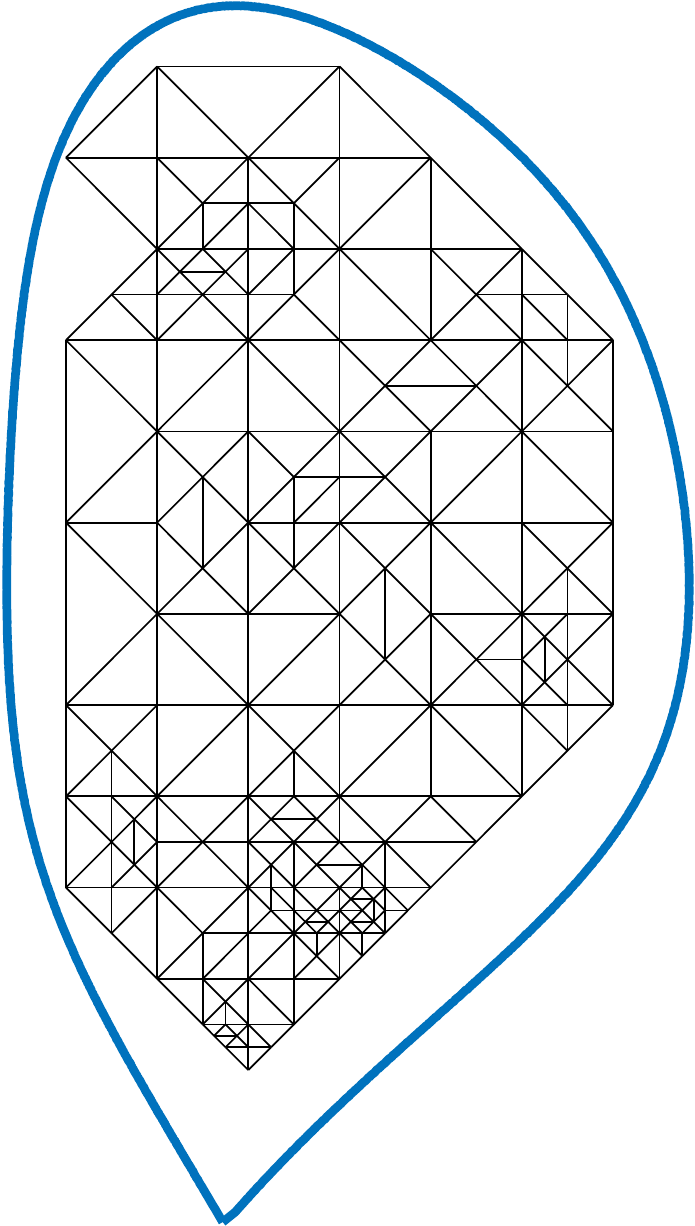} 
\end{tabular}
\caption{If the computational domain $\Omega_h$ (left) is kept fixed as the mesh is locally refined, the successive triangulations (center and right) will fail to keep the distance between $\partial\Omega_h$ and $\partial\Omega$ of the order of the local mesh diameter $h_{loc}$. This effect can be observed, for instance, in the lower part of the domain.}\label{fig:LocalDistance}
\end{wrapfigure}

In the present situation however, the computational domain is in fact a strict subdomain of $\Omega$. Thus, if $\mathsf T^{h,0}$ is a triangulation of a \textit{fixed} computational domain $\Omega_h$ built as described in Section \ref{sec:ComputationalDomain}, with every subsequent refinement step the mesh will drift farther away from satisfying the condition that the distance between the computational boundary $\partial\Omega_h$ and the actual boundary $\partial \Omega$ remains locally of the order of the triangle diameter $h_{loc}$, as depicted in Figure \ref{fig:LocalDistance}. As a result, the transfer procedure would not yield satisfactory results.

\begin{figure}\centering
\begin{tabular}{cccccc}
 & Solve \& Estimate & Mark & Refine & Select & Restart \\
\rotatebox{90}{{\small \; Computational Mesh}} & \includegraphics[scale=.2]{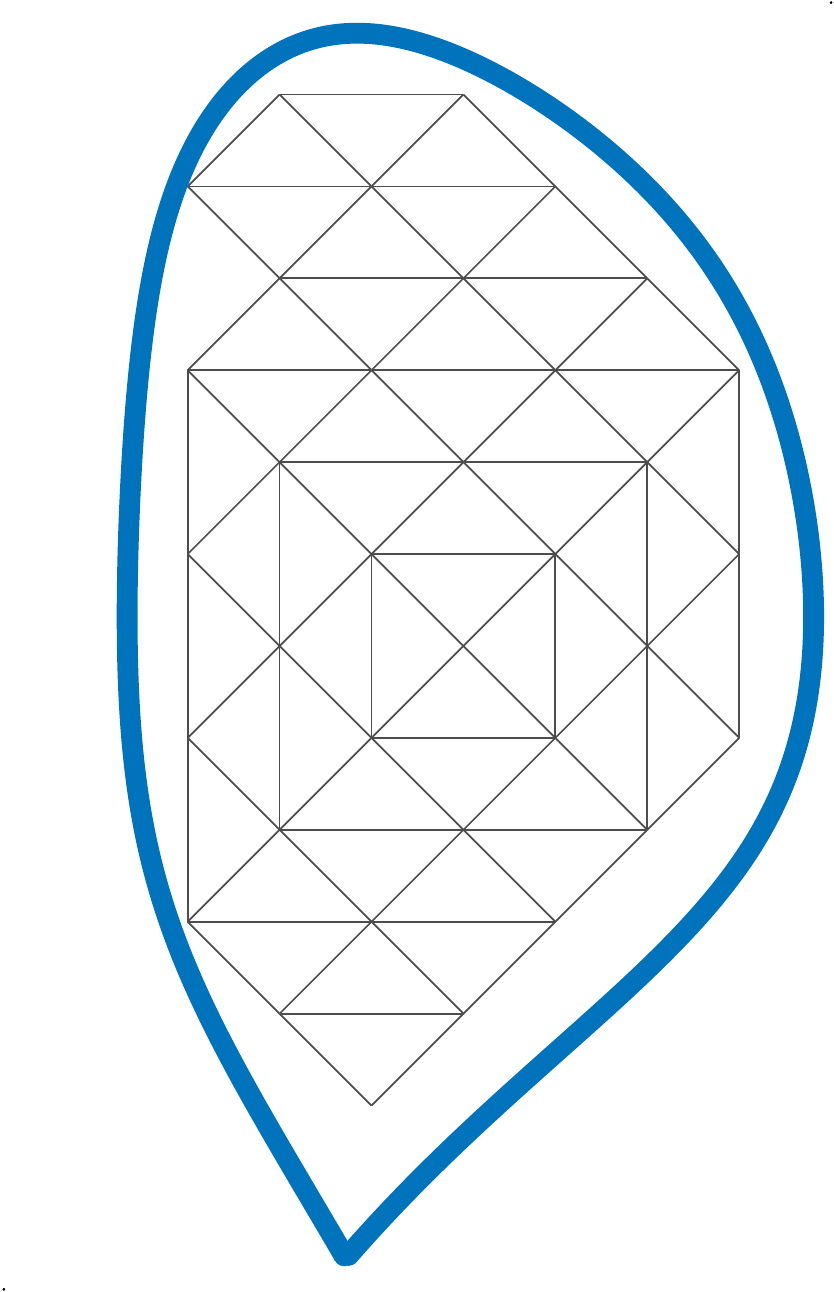} \quad & \includegraphics[scale=.2]{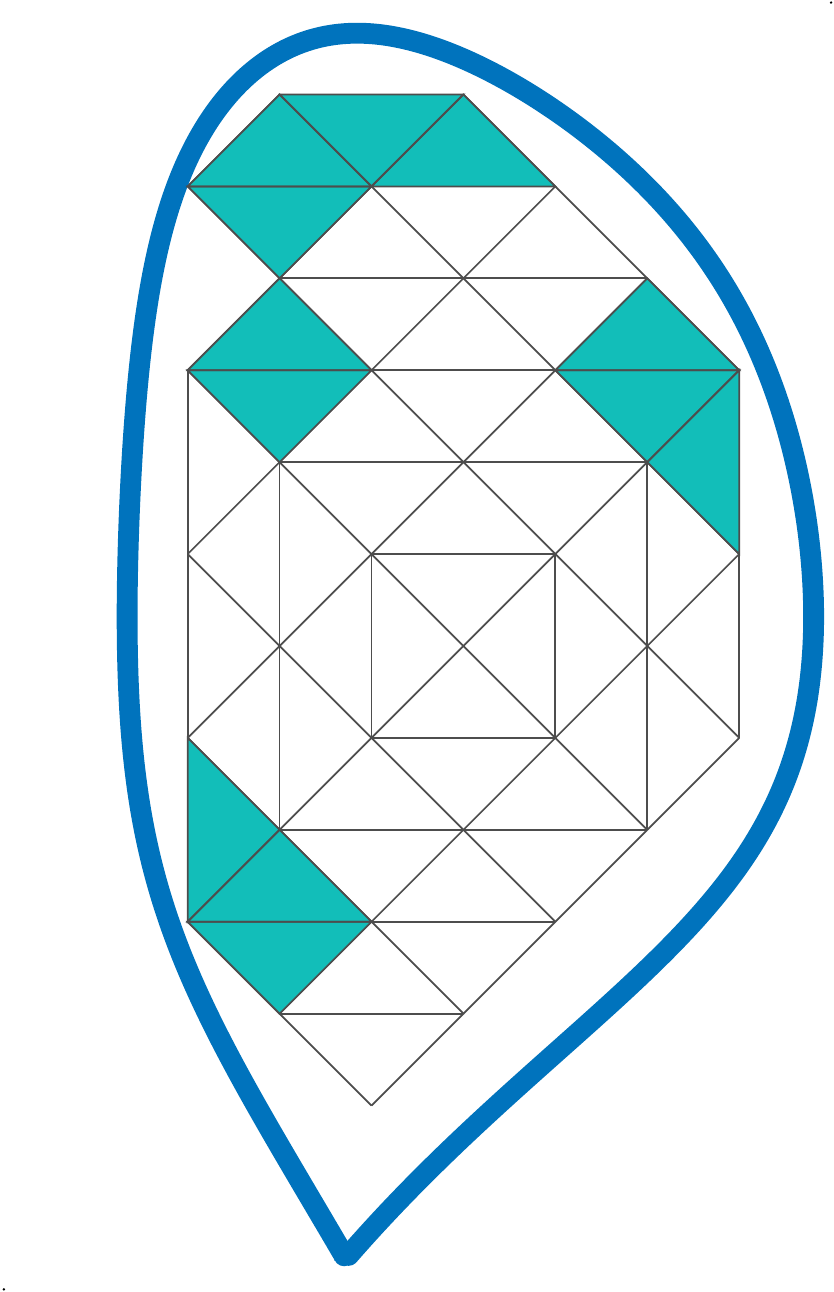} \quad & \multirow{2}{*}[\dimexpr 0.65in]{\includegraphics[scale=.28]{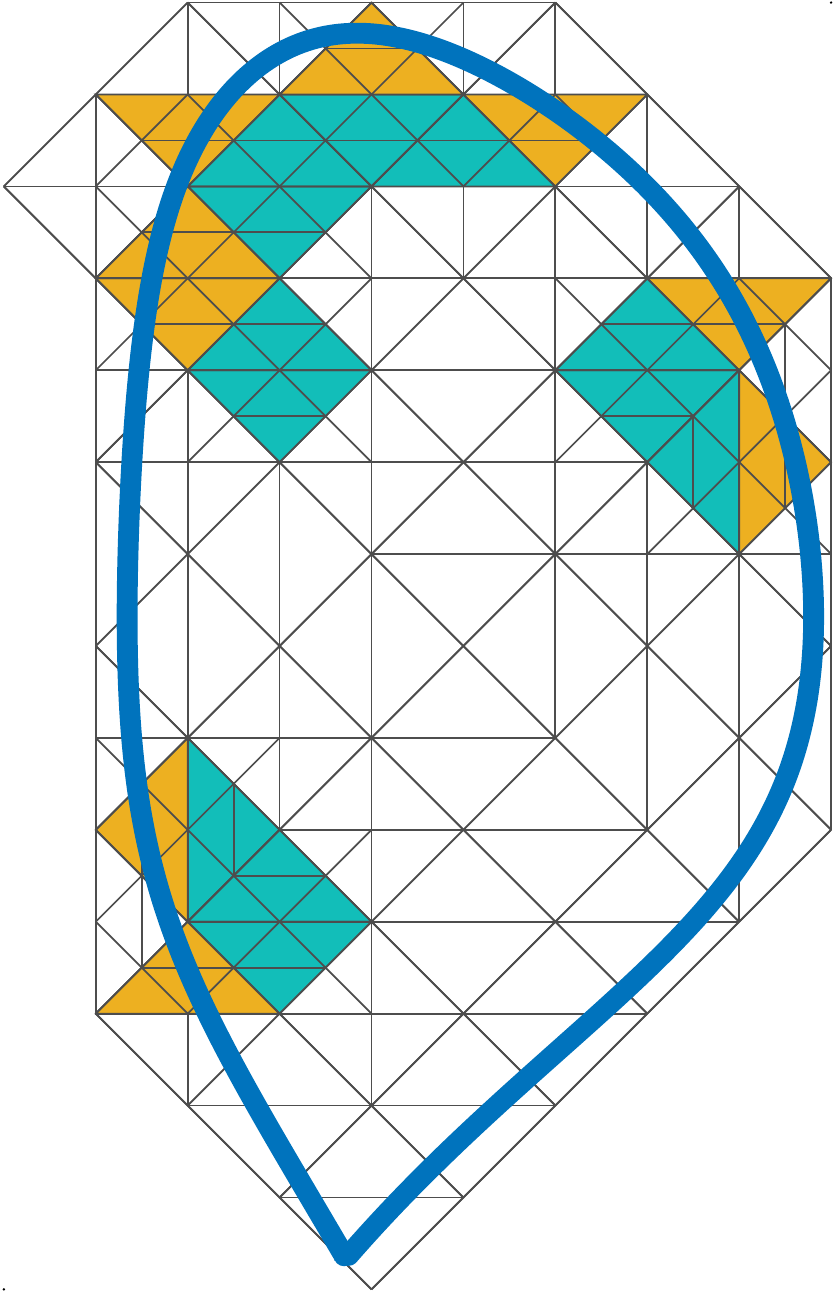}} \quad & \includegraphics[scale=.2]{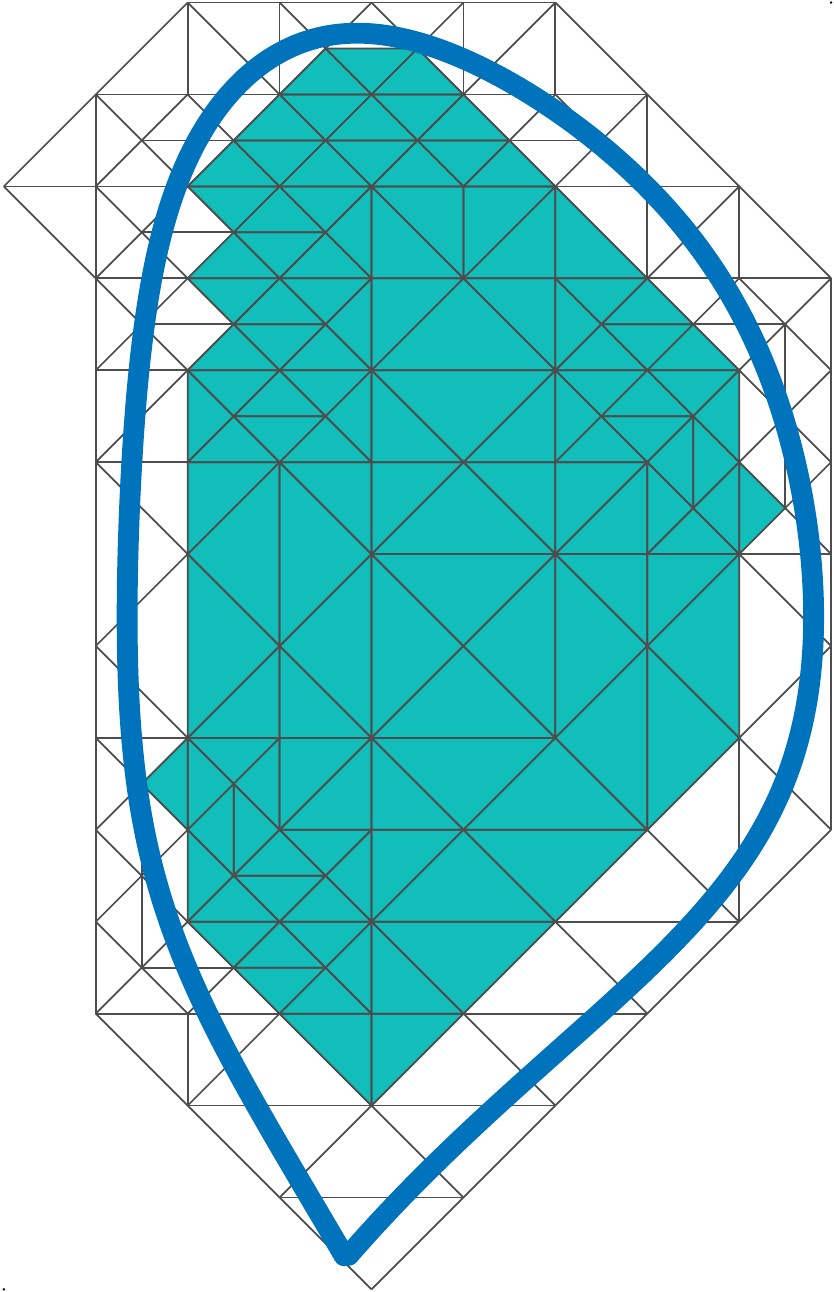} \quad & \includegraphics[scale=.2]{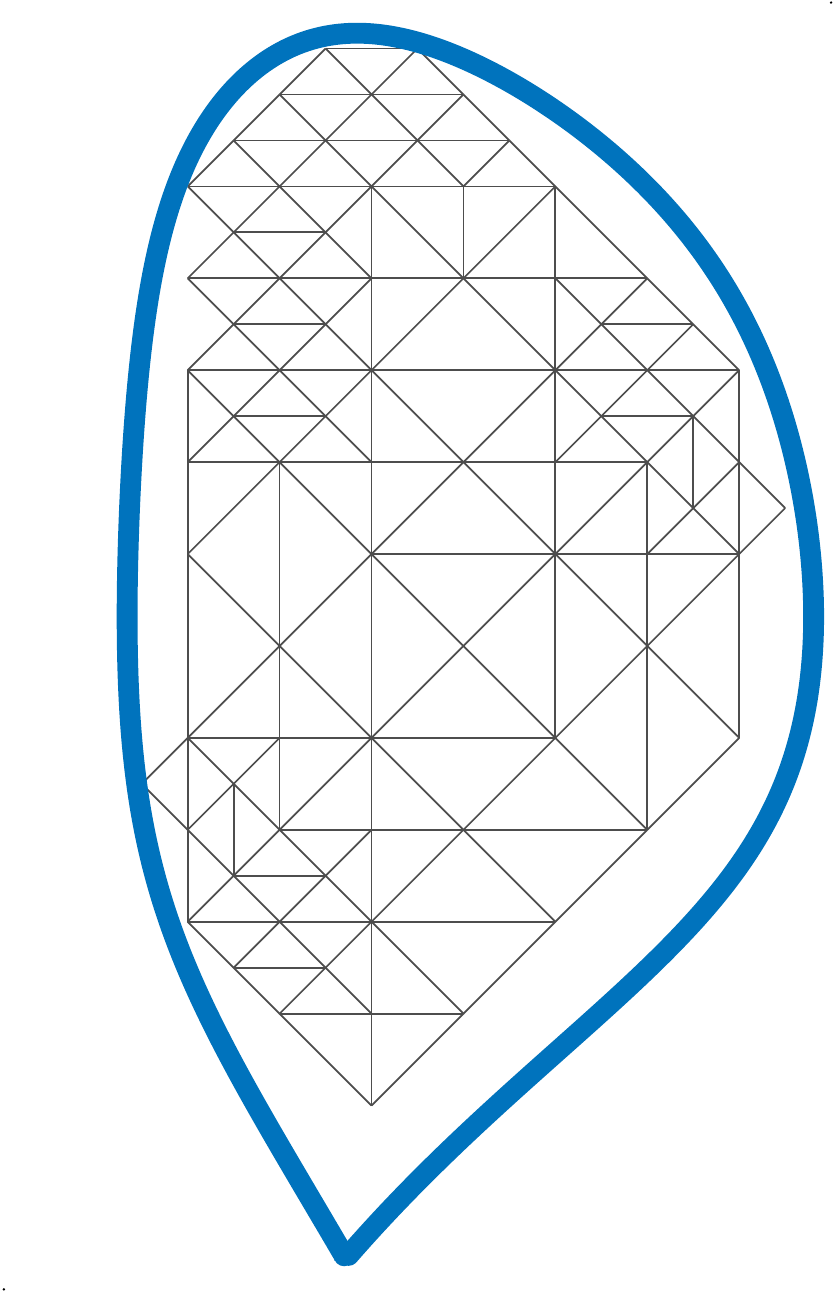} \\
\rotatebox{90}{{\small \; Companion Mesh}} &  \includegraphics[scale=.2]{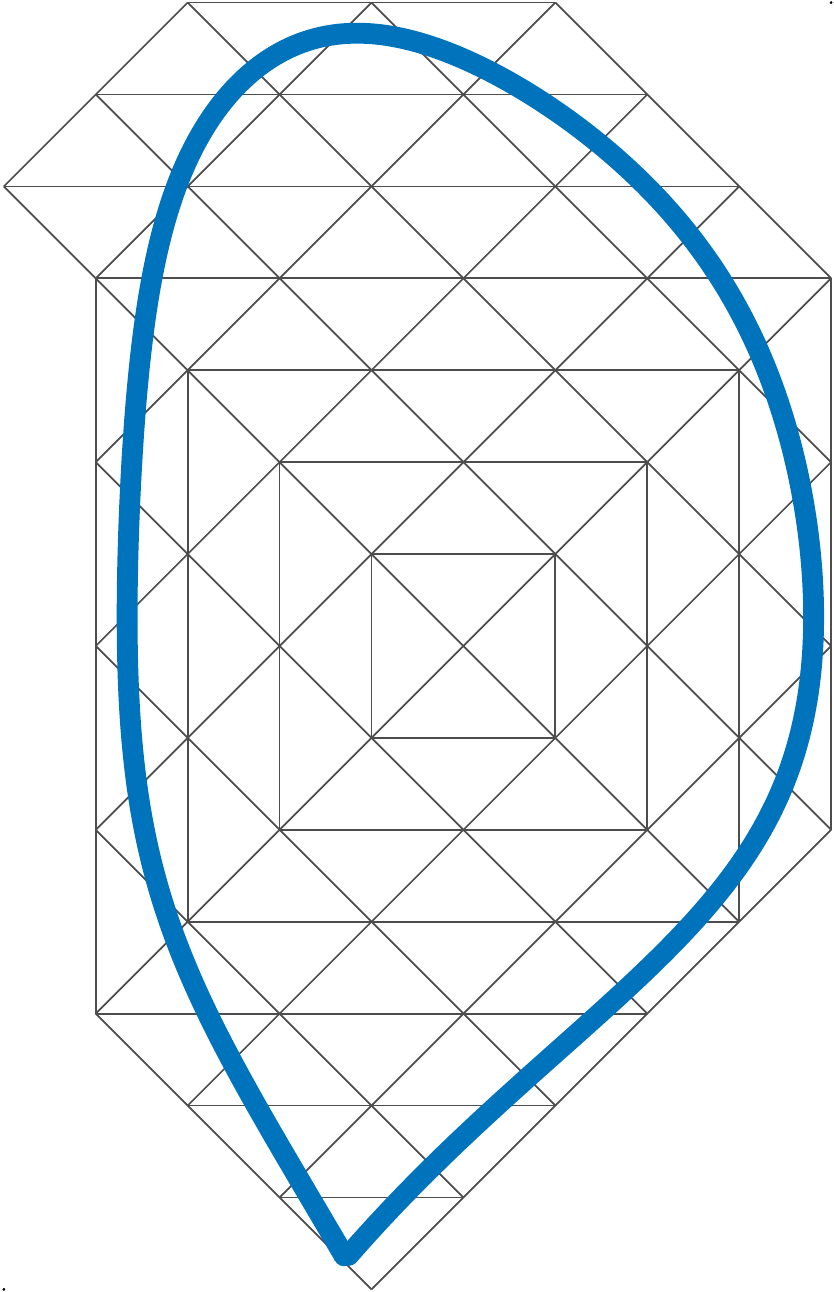} \quad & \includegraphics[scale=.2]{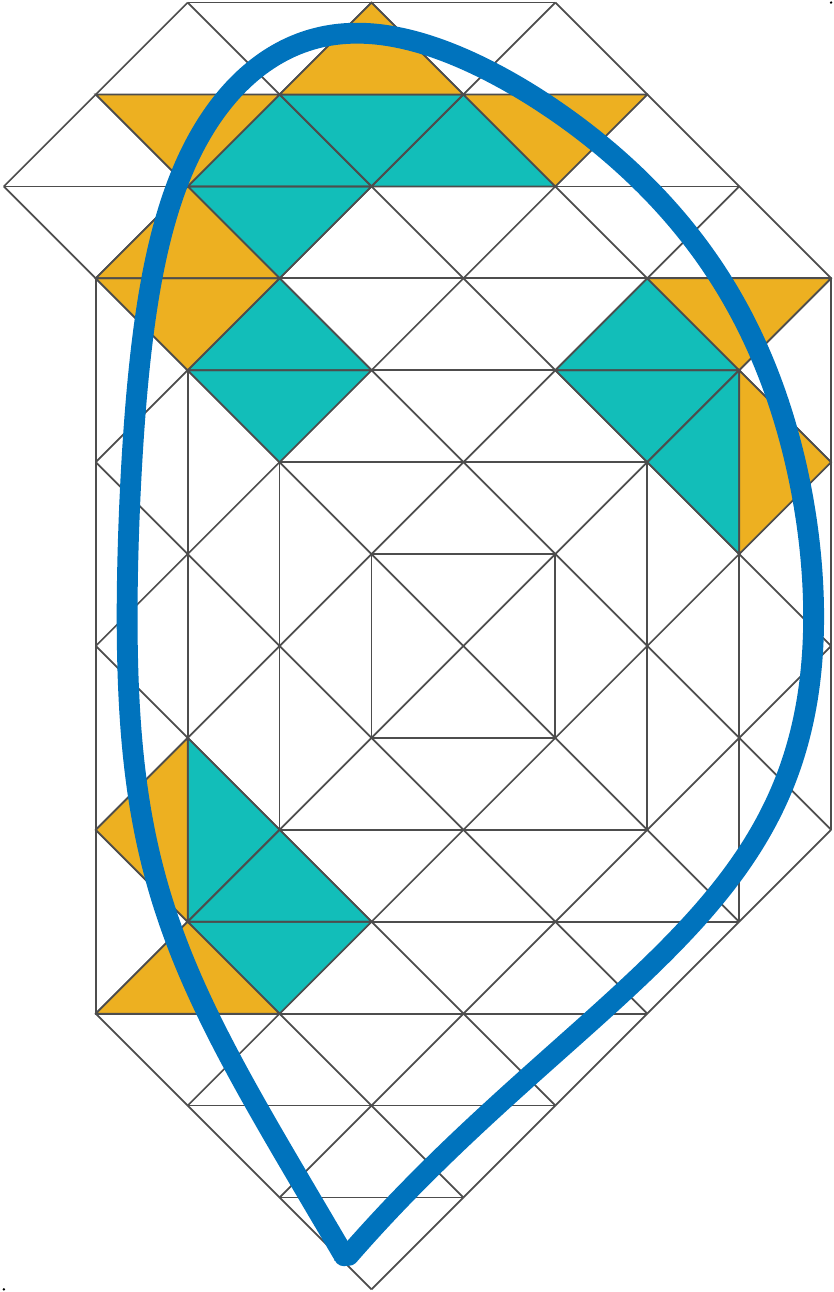} \quad &  & \includegraphics[scale=.2]{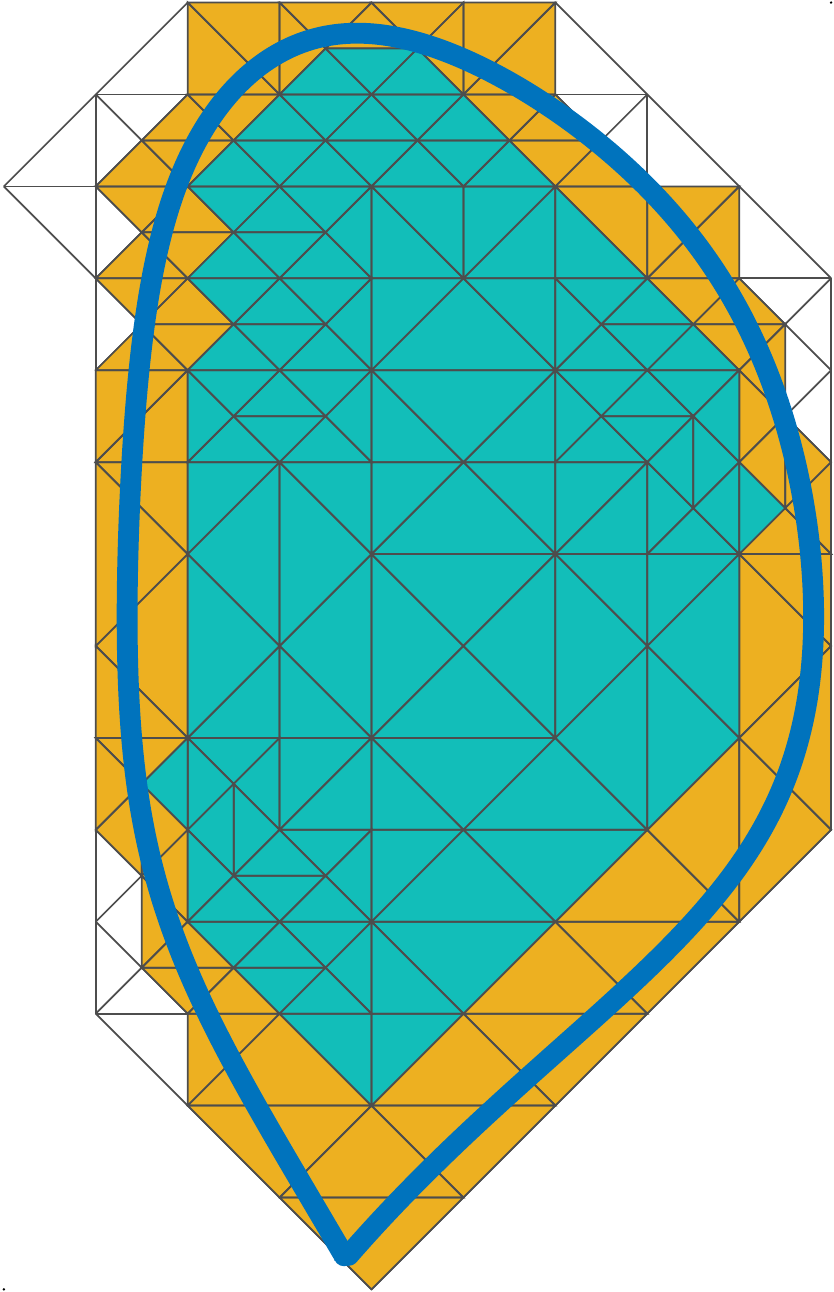} \quad & \includegraphics[scale=.2]{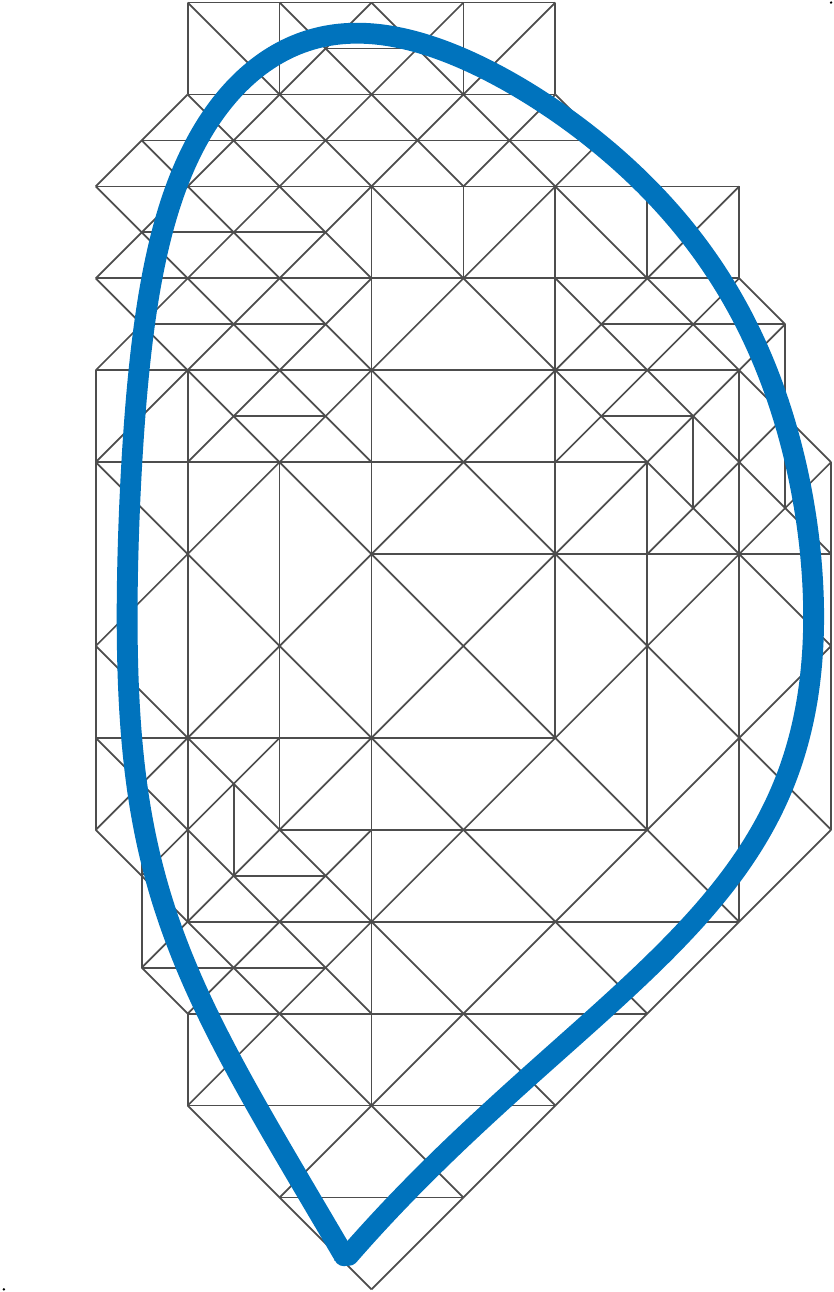}
\end{tabular}
\caption{An iteration of the adaptive algorithm is illustrated. The equation is solved on the computational mesh on the top left. The error is estimated on the initial mesh and a few elements are marked for refinement (top center left). The corresponding elements in the companion mesh are marked for refinement as well as the elements which both intersect the boundary and share an edge with the ones marked (bottom center left). The elements are then refined avoiding the creation of hanging nodes (center). From the refined companion mesh the elements completely contained within $\Omega$ are selected as a new computational mesh (top center right) while those forming a minimal cover of $\overline{\Omega}$ are selected as the updated companion mesh (bottom center right). The data structures are then updated eliminating the unnecessary elements and the process can be restarted (right). }\label{fig:refinement}
\end{figure}

In order to avoid such situations, we propose a strategy to update the computational domain consistent with the mesh refinement in such a way that the local distance condition is always satisfied. The method, showed schematically in Figure \ref{fig:refinement}, can be described as follows: 

1) Starting from the computational domain $\Omega_{h,0}$, a pair of computational and companion meshes $\mathsf T^{h,0}$ and $\mathsf T^{h,0}_c$ are built following the process detailed in Section \ref{sec:ComputationalDomain}.

2) The problem is solved and the error is estimated on the mesh $\mathsf T^{h,0}$, which results in a list $\mathcal M$ of elements marked for refinement (green triangles in the second column of Figure \ref{fig:refinement}).

3) The elements in the companion mesh $\mathsf T^{h,0}_c$ which correspond to those marked on the computational mesh in the previous step are marked for refinement. In addition, all the elements in the companion mesh which intersect the true boundary $\partial\Omega$ and share an edge with any triangle in $\mathcal M$ are marked for refinement as well (yellow triangles in the second column of Figure \ref{fig:refinement}). This results in an augmented list $\mathcal M_c$ of elements in the companion mesh.

4) The companion mesh $\mathsf T^{h,0}$ is updated by performing triangle refinement on all elements in $\mathcal M_c$, yielding a temporary background mesh $\widetilde{\mathsf T}$ that completely contains  $\Omega$, as depicted in the central column of Figure \ref{fig:refinement}. Note that since the new elements may have smaller diameter, some of them may now in fact be completely contained in $\Omega$ even if the parent triangle was not. In a similar fashion, some of them may neither be contained in $\Omega$ nor intersect $\partial\Omega$.

5) A new computational domain $\Omega_{h,1}$ and its corresponding triangulation $\mathsf T^{h,1}$ are defined by selecting the triangles in $\widetilde{\mathsf T}$ that are completely contained
in $\Omega$. Analogously, a new companion mesh $\mathsf T^{h,1}_c$ is defined by selecting all the elements of $\widetilde{\mathsf T}$ that are either completely contained in $\Omega$ or intersect the boundary $\partial\Omega$. The remaining triangles are discarded. The resulting level of refinement will then use the new computational and companion meshes and the process will continue until the predetermined stopping criterion is met.

\begin{wrapfigure}[21]{R}{0.4\textwidth}\centering
\begin{tabular}{cccc}
\kern-1em  & \kern-1em {\small Level 0} & \kern-1em  {\small Level 3} & \kern-1em  {\small Level 6}\\
\kern-1em \rotatebox{90}{{\small \; Computational Mesh}} & 
\kern-1em \includegraphics[height=0.43\linewidth]{IntAdaptiveL0} &  
\kern-1em \includegraphics[height=0.43\linewidth]{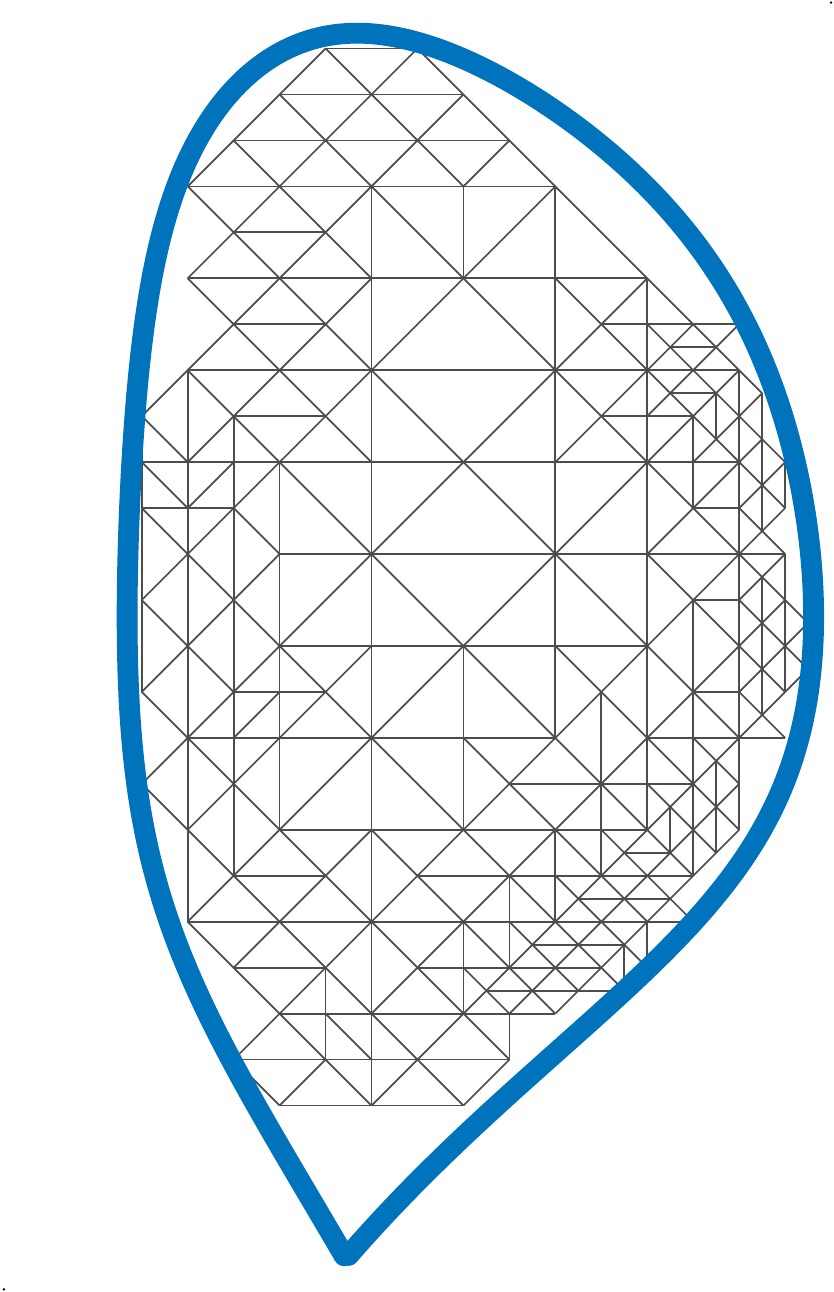} &  
\kern-1em \includegraphics[height=0.43\linewidth]{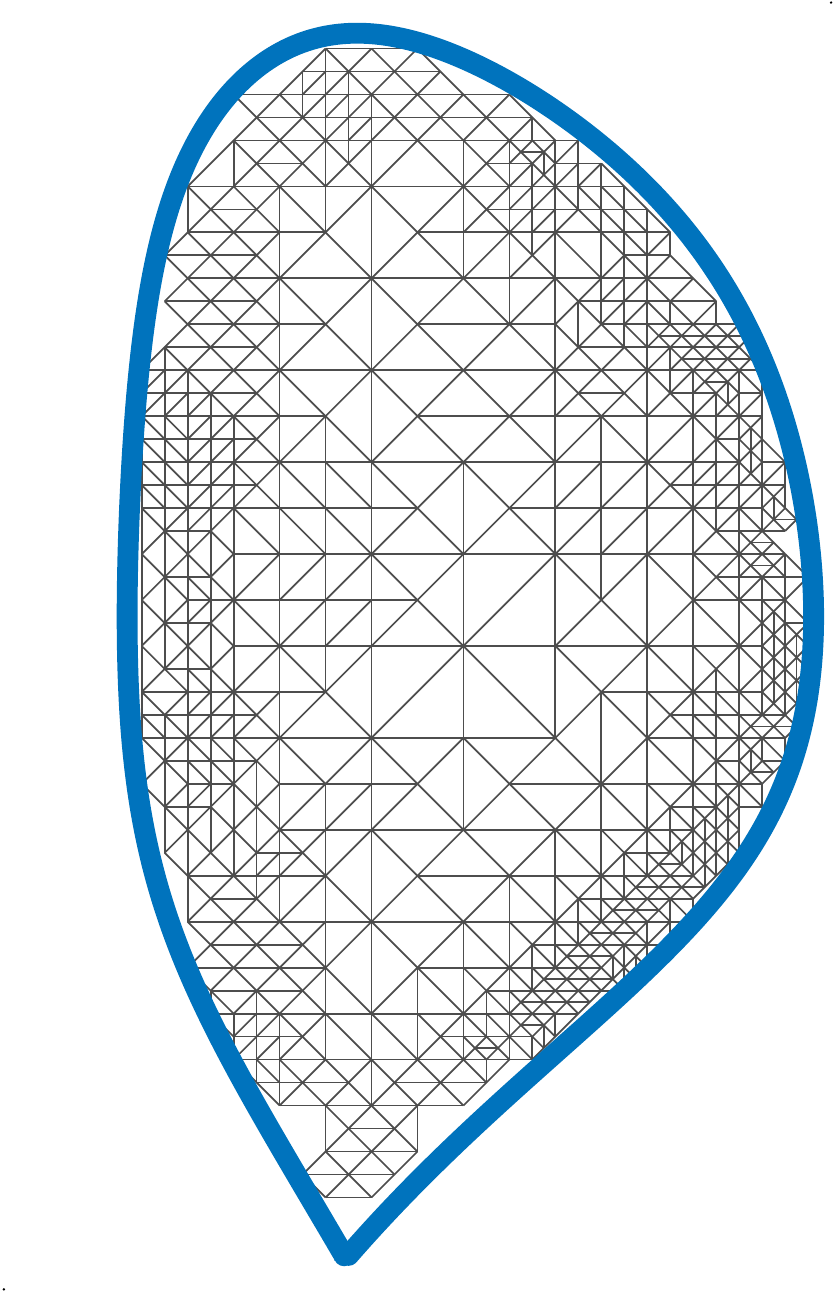} \\
\kern-1em \rotatebox{90}{{\small \; Companion Mesh}} &  
\kern-1em \includegraphics[height=0.43\linewidth]{ExtAdaptiveL0} &  
\kern-1em \includegraphics[height=0.43\linewidth]{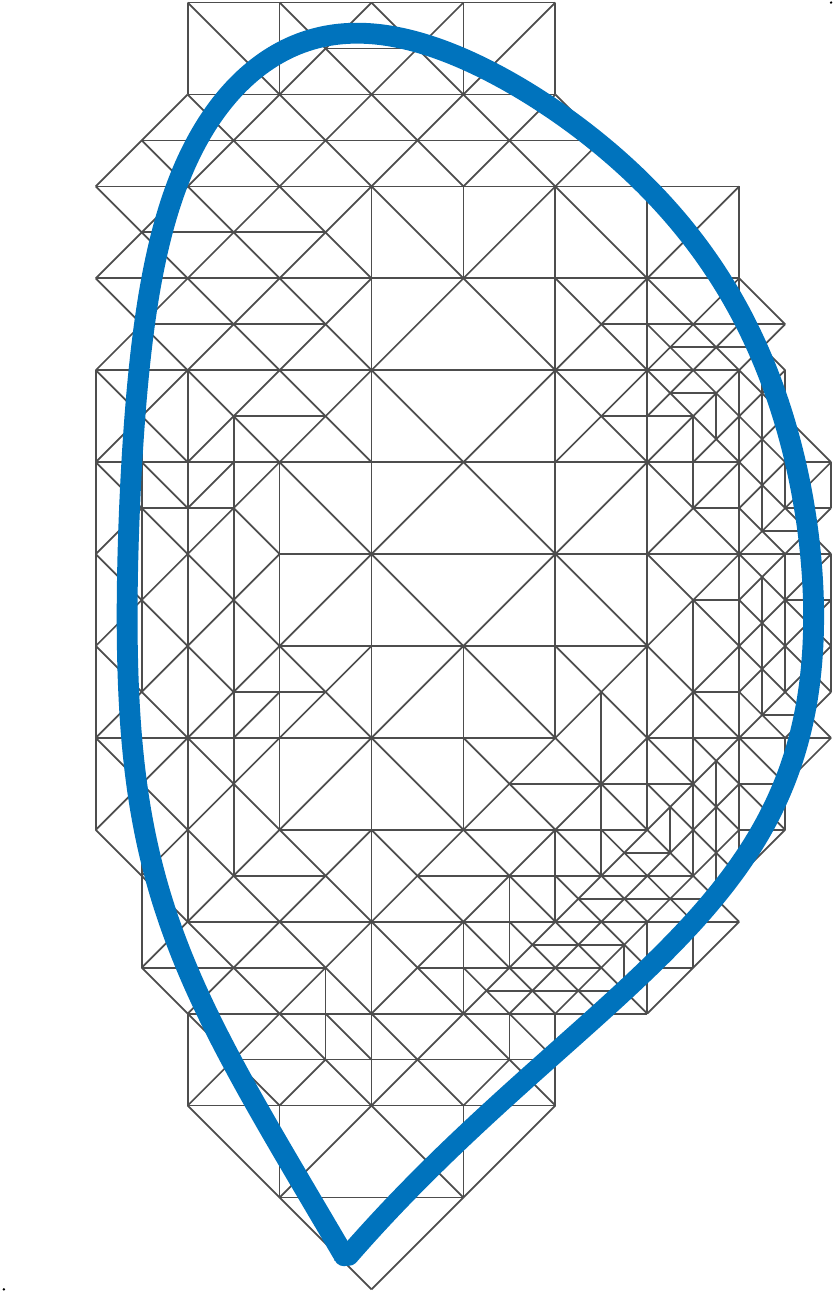} &  
\kern-1em \includegraphics[height=0.43\linewidth]{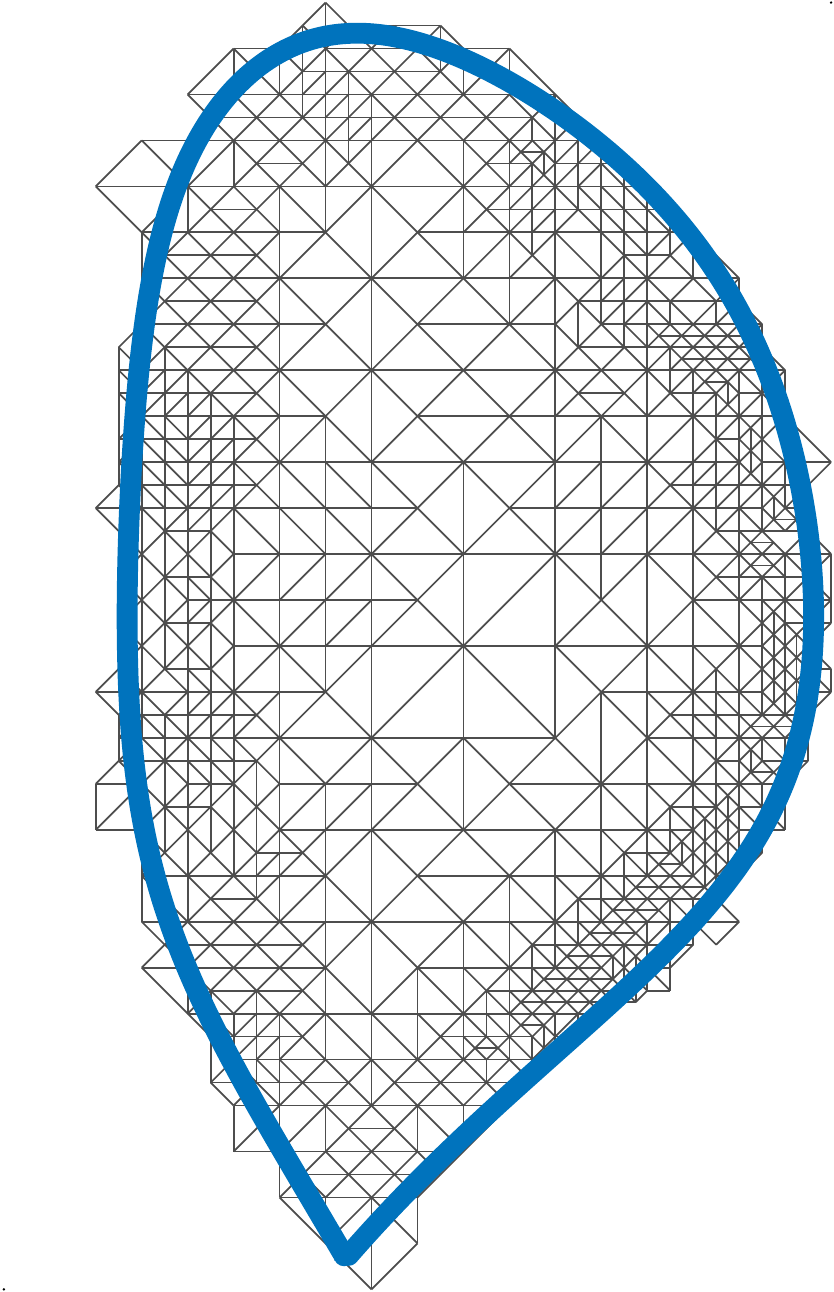}
\end{tabular}
\caption{A sequence of adaptively refined computational (top) and companion meshes (bottom). As the refinement progresses, the triangulations approximate the computational domain by exhaustion. The computations are carried out using only the meshes in the top row. }\label{fig:MeshSequence}
\end{wrapfigure}

Note that no computations are ever carried out using the companion grid, which is needed only to update the background triangulation in a way that gives rise to a refined computational domain $\Omega_{c,n}$ satisfying the separation condition $d(\boldsymbol y,\partial \Omega) = \mathcal O(h_{loc}(\boldsymbol y))$ for every $\boldsymbol y \in \partial \Omega_{c,n}$. Moreover, since at every level the computational domain $\Omega_{h,n}$ is defined as a subdomain of an increasingly finer background triangulation, the algorithm yields a sequence of computational domains that effectively ``exhaust" $\Omega$ as $n\to\infty$ so long as the local estimator $\eta_K$ remains nonzero on the elements having an edge on $\partial\Omega_{h,n}$. This is illustrated in Figure \ref{fig:MeshSequence}. 

\section{Numerical Experiments}\label{sec:numerical}
We present five examples to showcase the performance of the adaptive algorithm. To establish the correct behavior of the error estimator and the refinement strategy  we begin with a linear test case where the analytic solution is known. The chosen analytic equilibrium has the advantage of providing the desired geometric and parametric flexibility, but corresponds to solutions that vary smoothly across the confinement region and thus tend to favour a uniform refinement strategy. Nevertheless, the availability of an exact solution allows us to verify the overall behavior of the estimator. 

More challenging and physically relevant situations arise when the source term of the equation is non-linear and has large gradients. The solutions for these cases develop features that would be hard to resolve accurately while simultaneously keeping a small number of elements and using the boundary transfer technique. It was precisely to address cases like these that we included adaptive refinement capabilities in our solver, and the last four examples in this section belong to this category. In all the experiments, the maximum criterion was used for marking.\\
\newpage
\subsection{A Solov'ev equilibrium}
In order to test the performance of the error estimator and the adaptive algorithm, we start with a simple linear setting for which an exact solution is available. The example corresponds to a so-called Solov'ev profile \cite{Solov'ev:1968} where the source term of the equation is taken to be of the form \cite{CeFr:2010}
\begin{equation}\label{eq:solovevsource}
F(r,\psi):=(1-A)r^2 + A.
\end{equation}
The free parameter $A$ determines the ratio of plasma pressure to magnetic pressure in the equilibrium of interest.   With such a source term, the Grad-Shafranov equation is linear and exact solutions can be constructed by imposing physical or geometrical constraints. The case that we will consider here corresponds to a geometry similar to that of the National Spherical Toroidal Experiment (NSTX) in the high beta regime. 

\begin{figure}\centering\scalebox{1}{
\begin{tabular}{cccccc}
\kern-2em $\psi$ & \kern-2em $\partial_r\psi$ & \kern-2em $\partial_z\psi$ & \kern-2em $\partial_{rr}\psi$ & \kern-2em $\partial_{zz}\psi$ & \kern-2em $J_\phi$ \\
\kern-1em \includegraphics[height=0.2\linewidth]{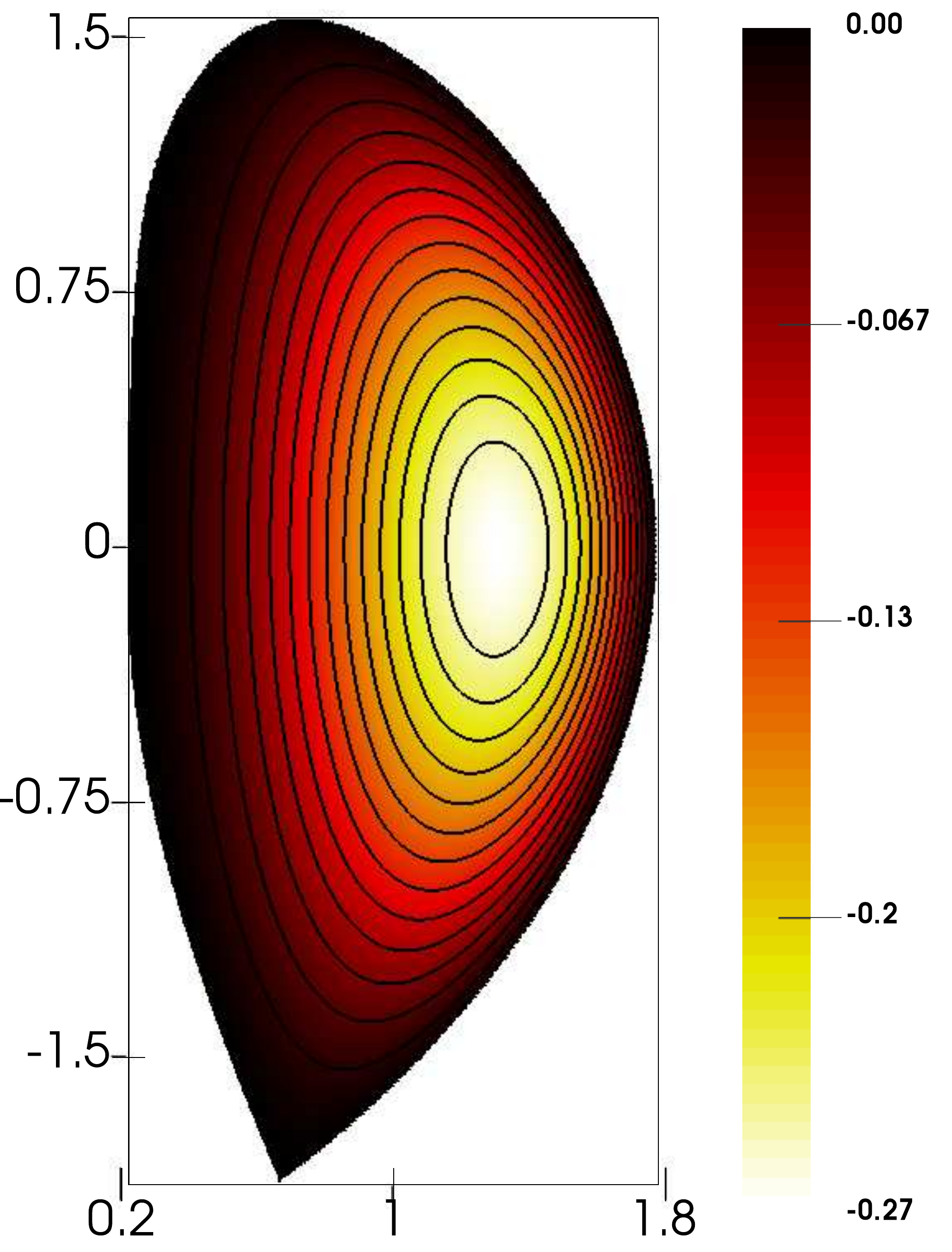} &
\kern-1em \includegraphics[height=0.2\linewidth]{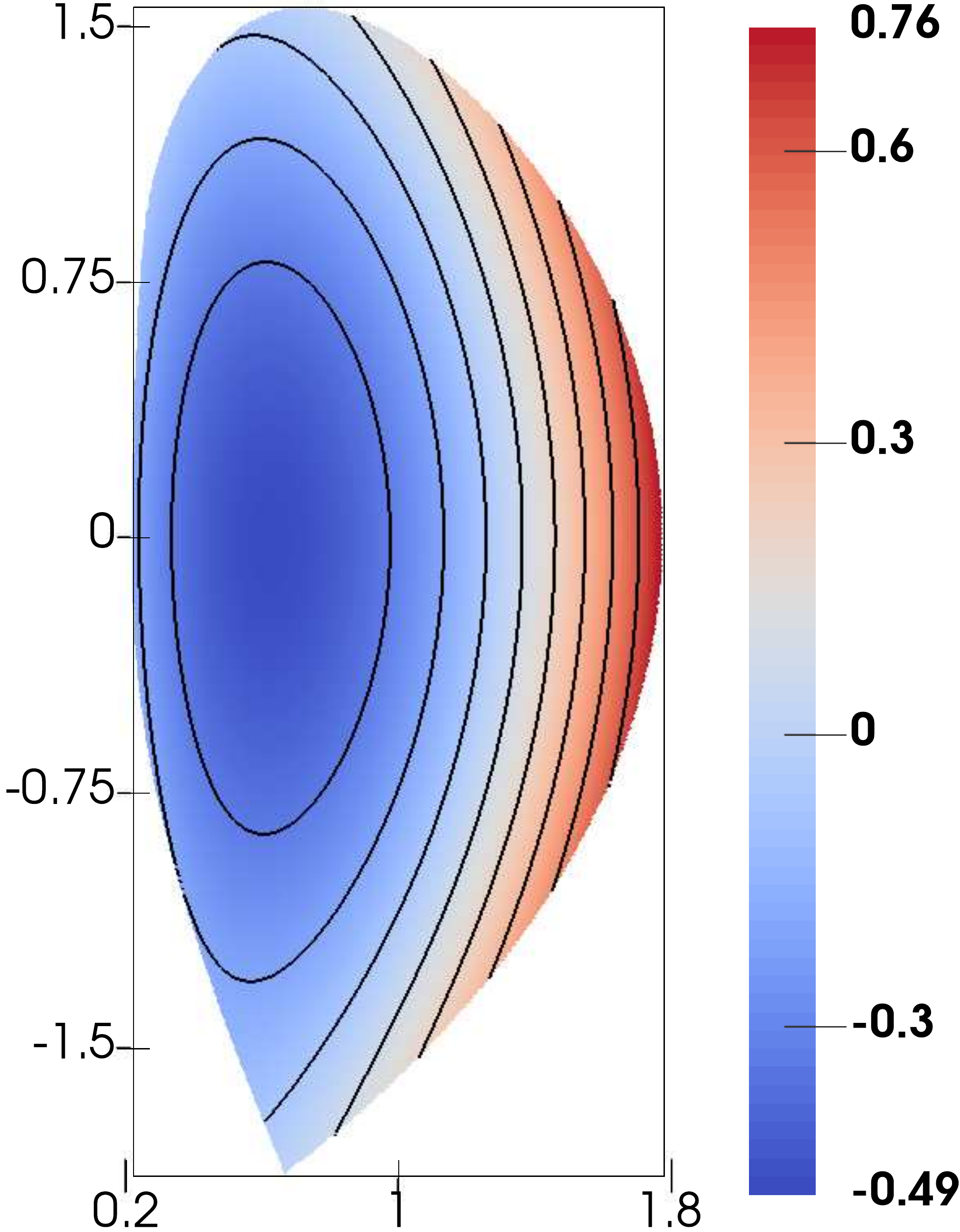} & 
\kern-1em \includegraphics[height=0.2\linewidth]{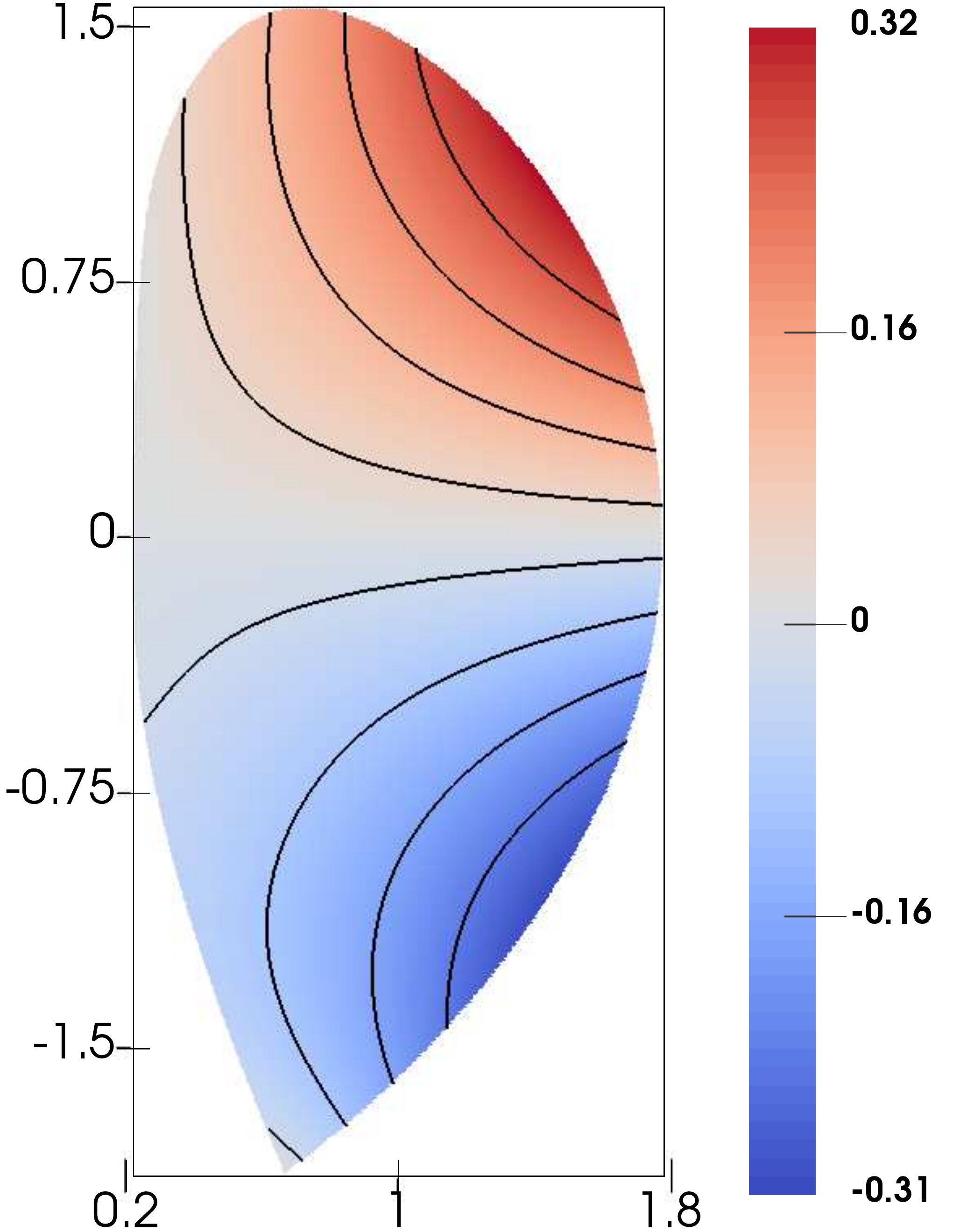} & 
\kern-1em \includegraphics[height=0.2\linewidth]{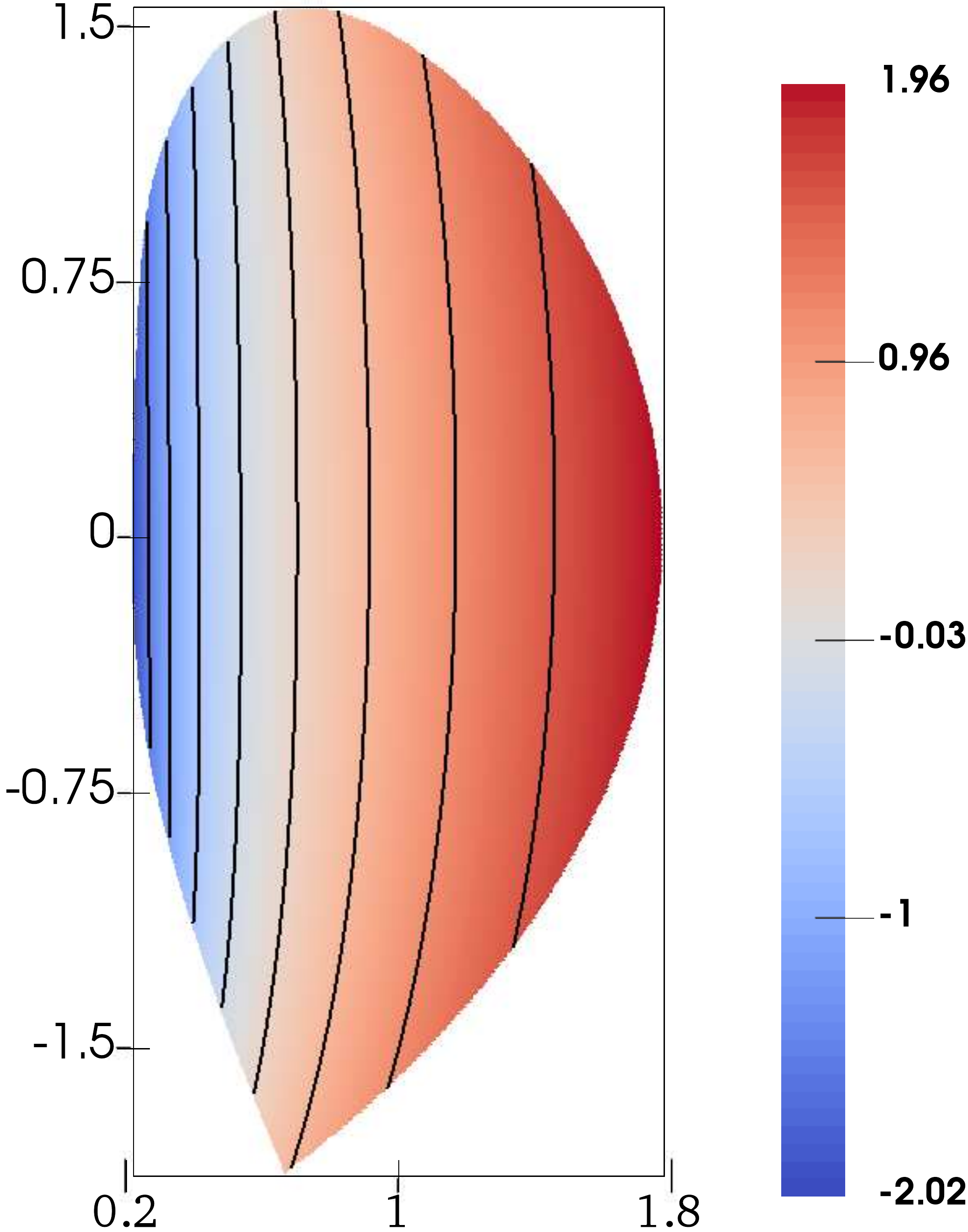}& 
\kern-1em \includegraphics[height=0.2\linewidth]{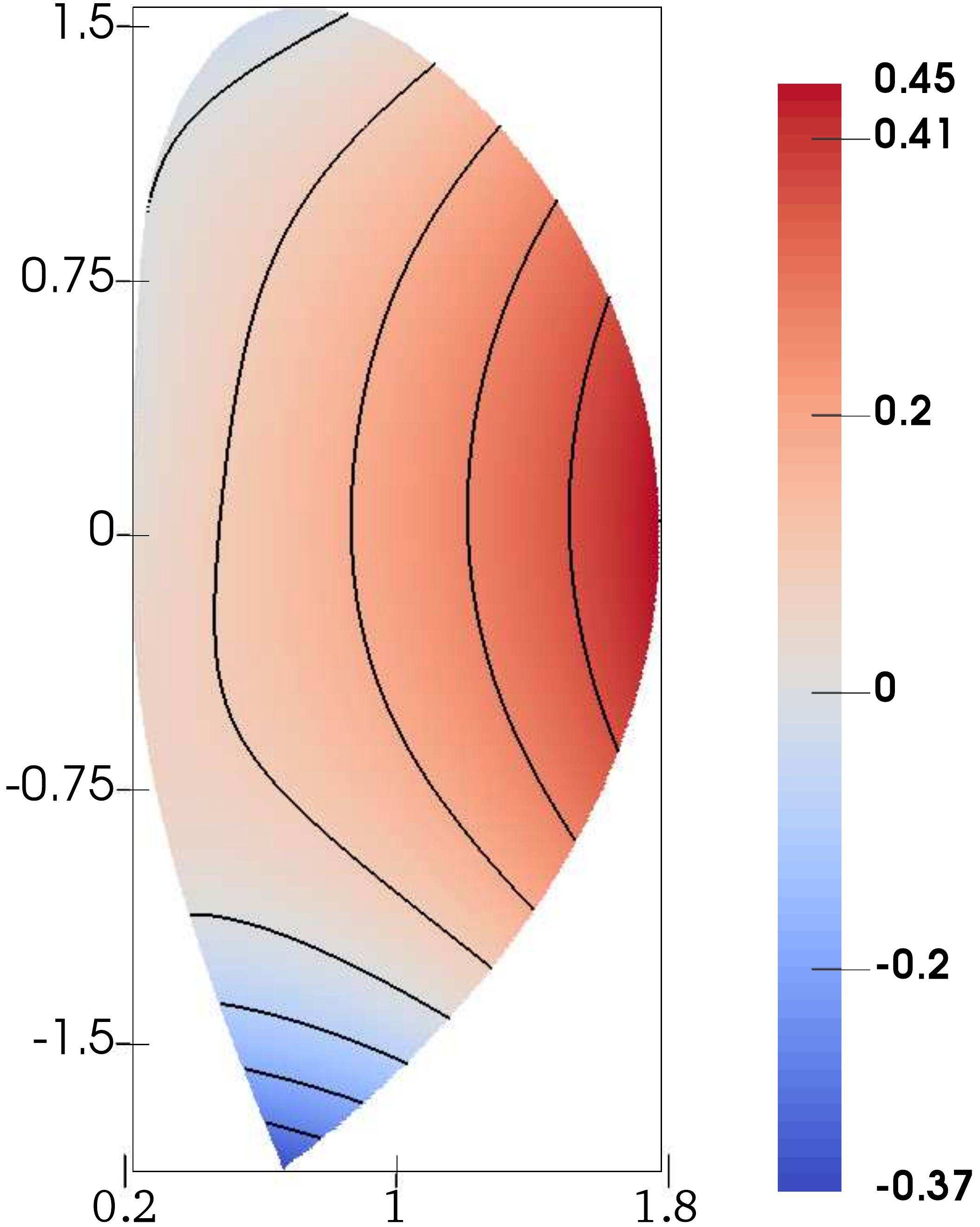}&
\kern-1em \includegraphics[height=0.2\linewidth]{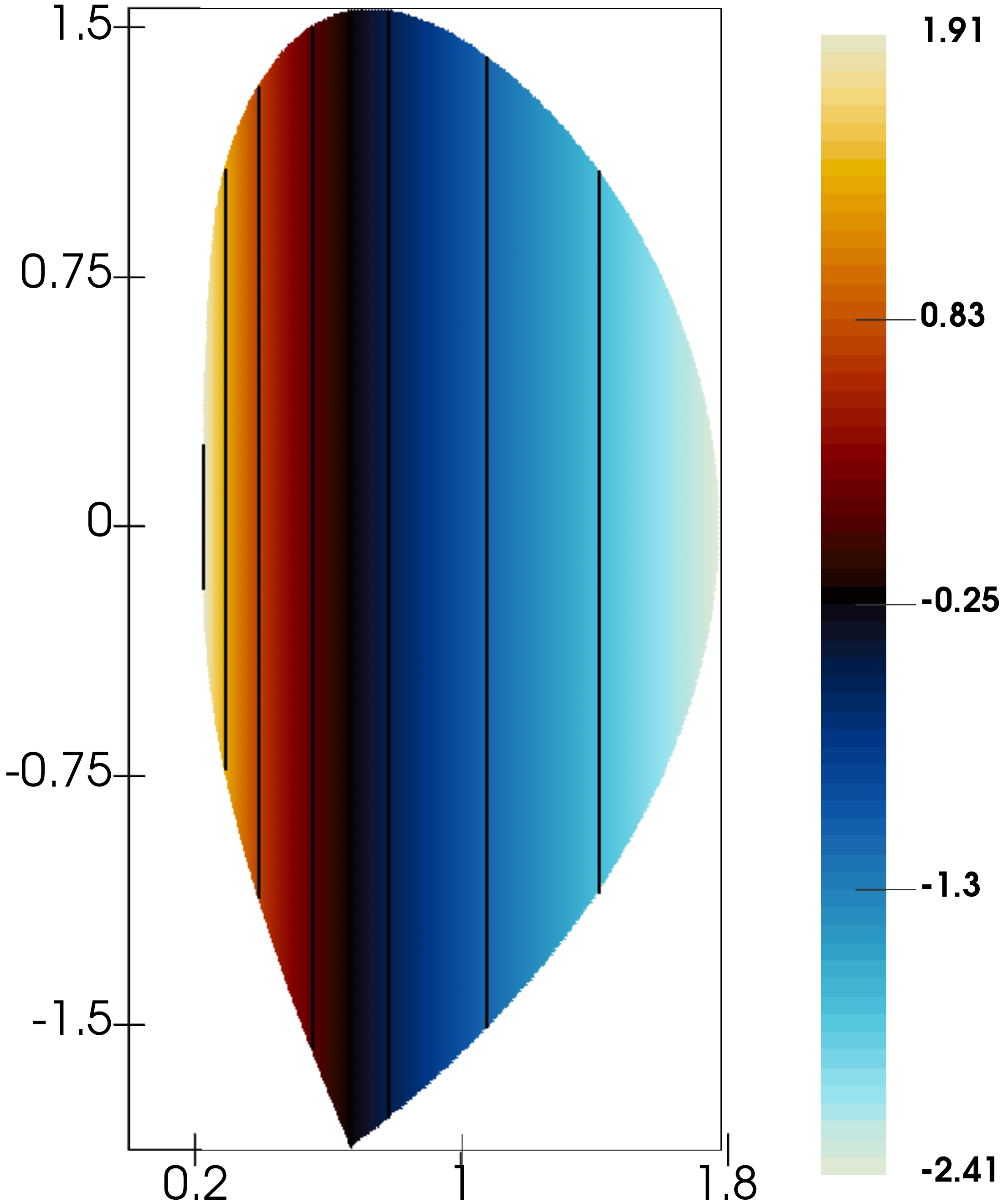}
\end{tabular}}
\caption{Exact Solov'ev solution to the Grad-Shafranov equation with source term $F(r,\psi)=(1-A)r^2+A$ and the parameter values $A, c_1,\ldots,c_{12}$ specified in \eqref{eq:HighBetaSol} (left), and its partial derivatives with respect to $r$ and $z$ (center). The toroidal current density is displayed on the right.}\label{fig:HighBeta}
\end{figure}

The exact solution is of the form
\begin{subequations}\label{eq:HighBetaSol}
\begin{equation}
\label{eq:HighBetaSolA}
\psi =  \frac{r^4}{8} + A\left(\frac{1}{2}r^2\ln r - \frac{r^4}{8}\right) + \sum_{i=1}^{12}c_i\,\psi_i,
\end{equation}
where $A=-0.52$ and
\begin{alignat}{6}
\nonumber
\psi_1 =\,& 1, & \qquad & \psi_7 = \,&& 8z^6-140z^4r^2+75z^2r^4-15r^6\ln{r} \\
\nonumber
\psi_2 =\,& r^2, & \qquad &  &&+180r^4z^2\ln{r}-120r^2z^4\ln{r},\\ 
\nonumber
\psi_3 =\,& z^2-r^2\ln{r}, &\qquad &  \psi_8 =\,&& z,\\ 
\nonumber
\psi_4 =\,& r^4-4r^2z^2, &\qquad &  \psi_9 =\,&& zr^2,\\ 
\nonumber
\psi_5 =\,& 2z^4-9z^2r^2+3r^4\ln{r} &\qquad & \psi_{10} =\,&& z^3-3zr^2\ln{r},\\ 
\nonumber
 &  -12r^2z^2\ln{r}, & \qquad &  \psi_{11} =\,&& 3zr^4-4z^3r^2,\\ 
\label{eq:HighBetaSolB}
\psi_6 =\,& r^6-12r^4z^2+8r^2z^4, &\qquad& \psi_{12} =\,&& 8z^5-45zr^4 -80z^3r^2\ln{r} +60zr^4\ln{r}. 
\end{alignat}
Following the process presented in detail in \cite{CeFr:2010} and using the geometric parameters corresponding to NSTX in that article, the undetermined constants $c_1,\dots,c_{12}$ for the case at hand can be easily computed, and we find
\begin{equation}\label{eq:HighBetaSolC}
\begin{array}{cccc}
c_1 = -0.001479661575325, & c_2=-0.366568333204813, & c_3=0.002409406149732, \\
c_4=-0.023957517168316, & c_5=0.000692888519765, & c_6=-0.001768712177298, \\
c_7=-0.000044132956899, & c_8=0.000433522611526, & c_9=0.008286849573230, \\
c_{10}=-0.000044132956899, & c_{11}=-0.001299619729855, & c_{12}=0.000072050578303. 
\end{array}
\end{equation}
\end{subequations}
Graphs of the solution and its partial derivatives with these parameter values on the target geometry can be seen in Figure \ref{fig:HighBeta}.

\begin{figure}\centering\scalebox{1}{
\begin{tabular}{ccccc}
\rotatebox{90}{\quad Uniform refinement} &   \includegraphics[height=0.2\linewidth]{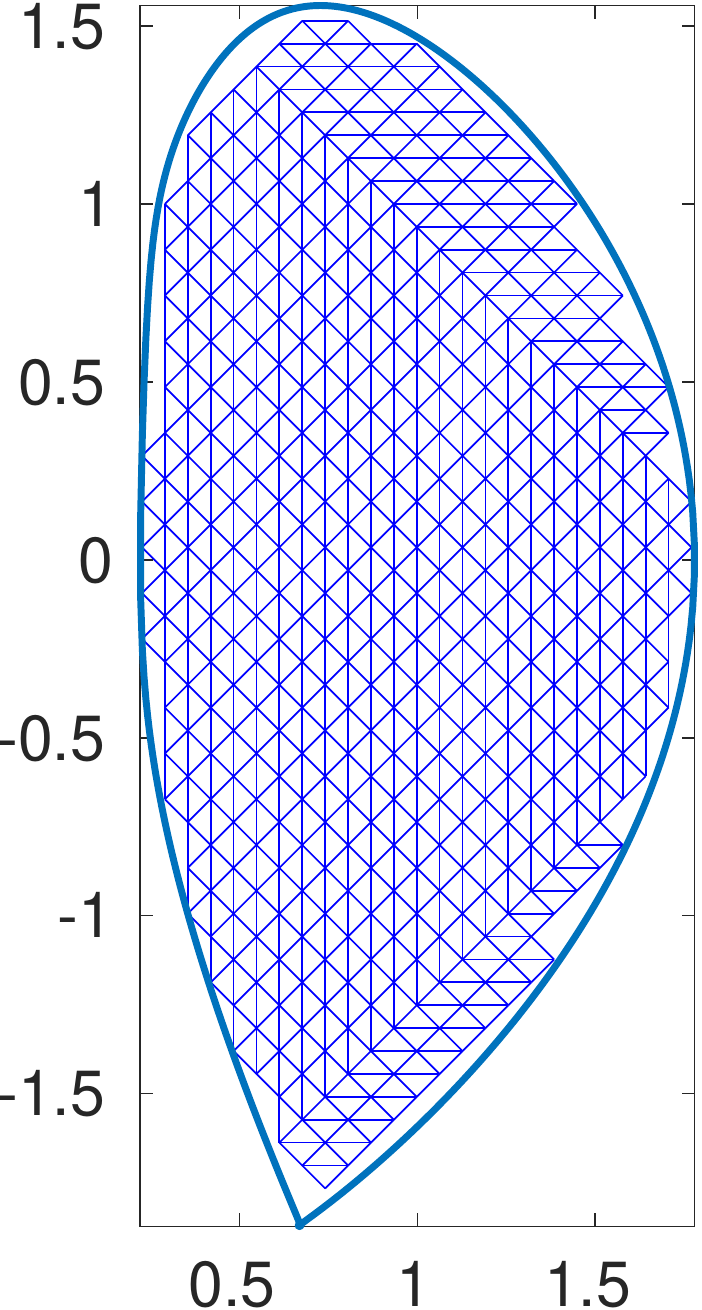} & \kern-1em 
\includegraphics[height=0.2\linewidth]{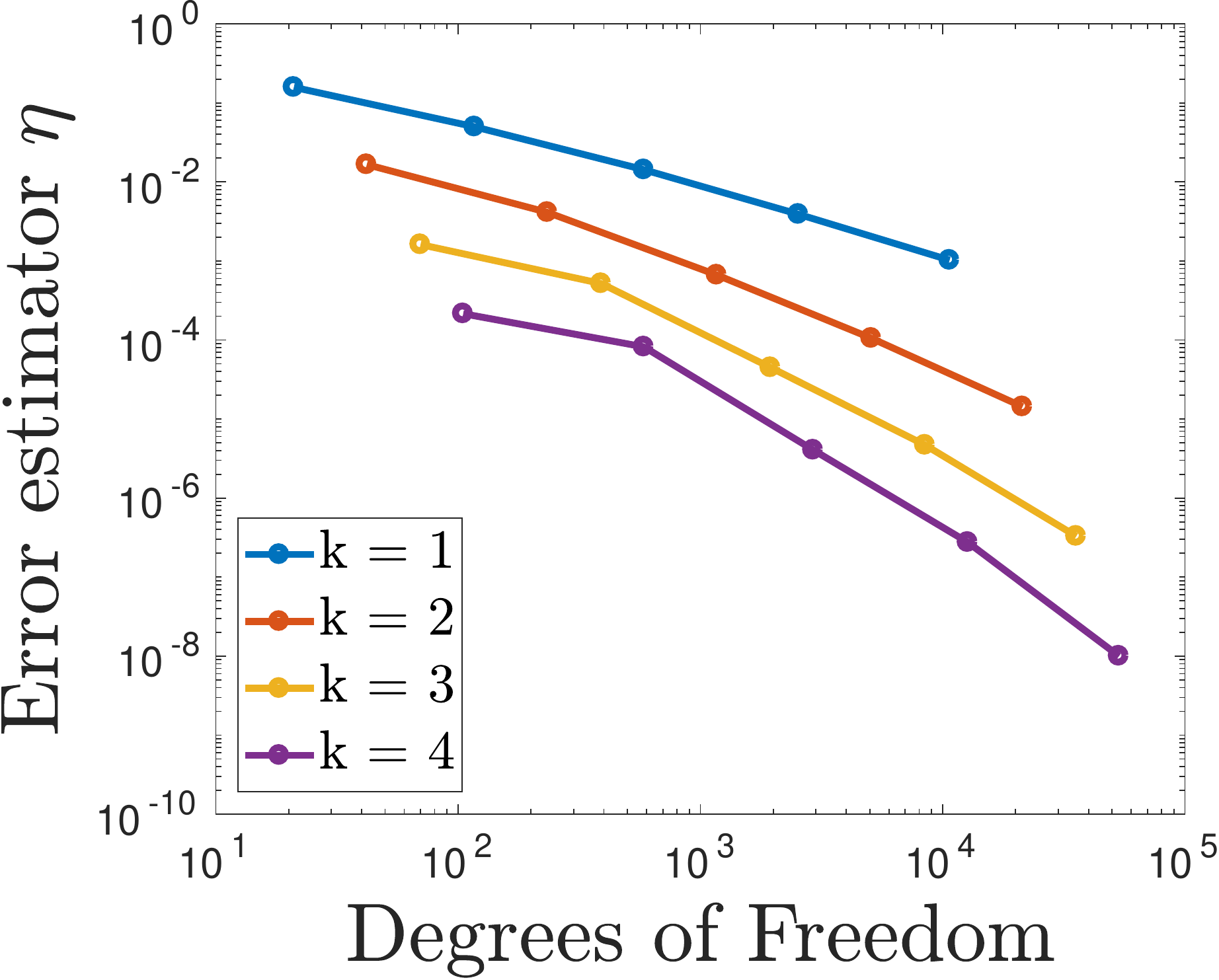}  & \kern-1em  \includegraphics[height=0.2\linewidth]{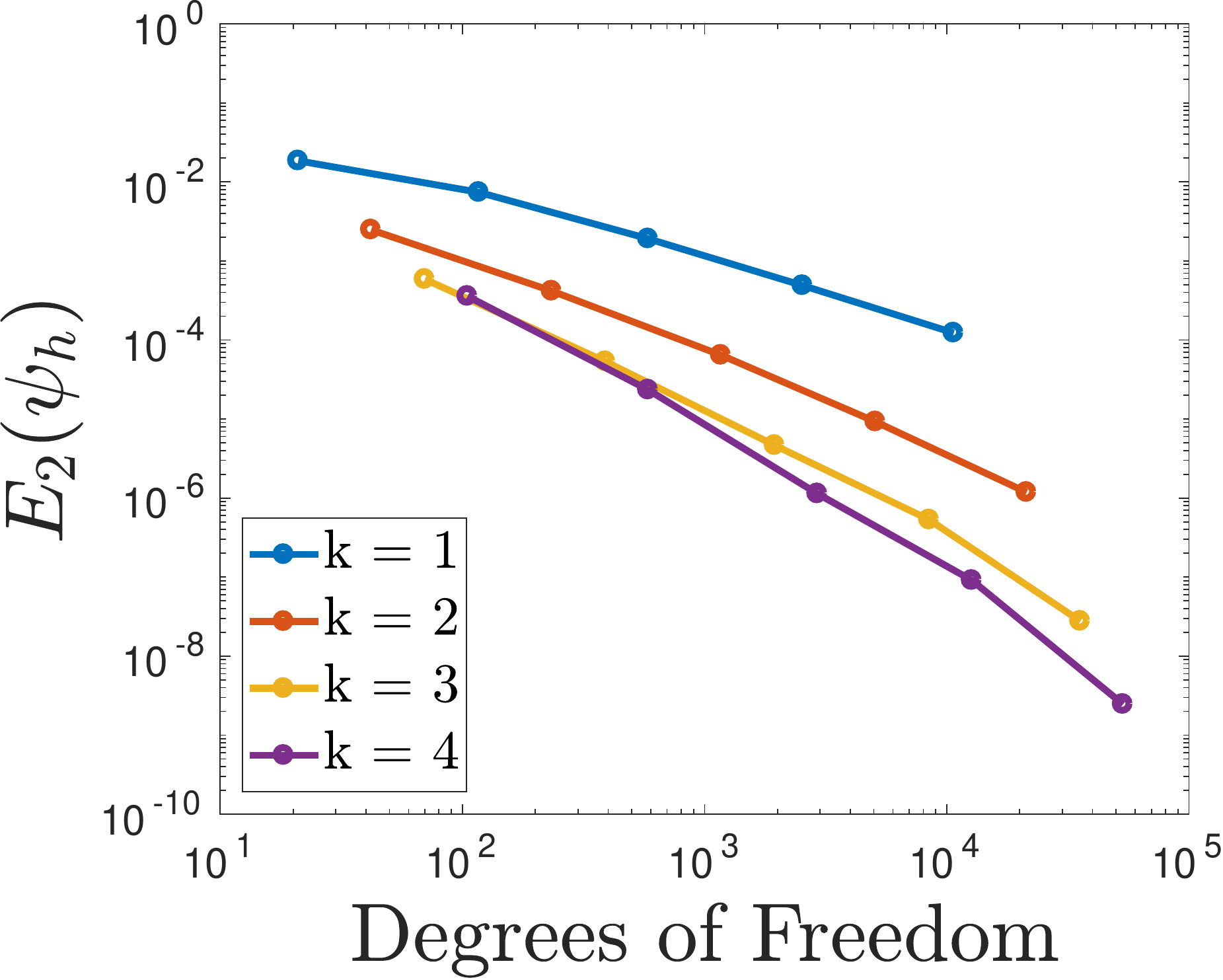}  & \kern-1em  \includegraphics[height=0.2\linewidth]{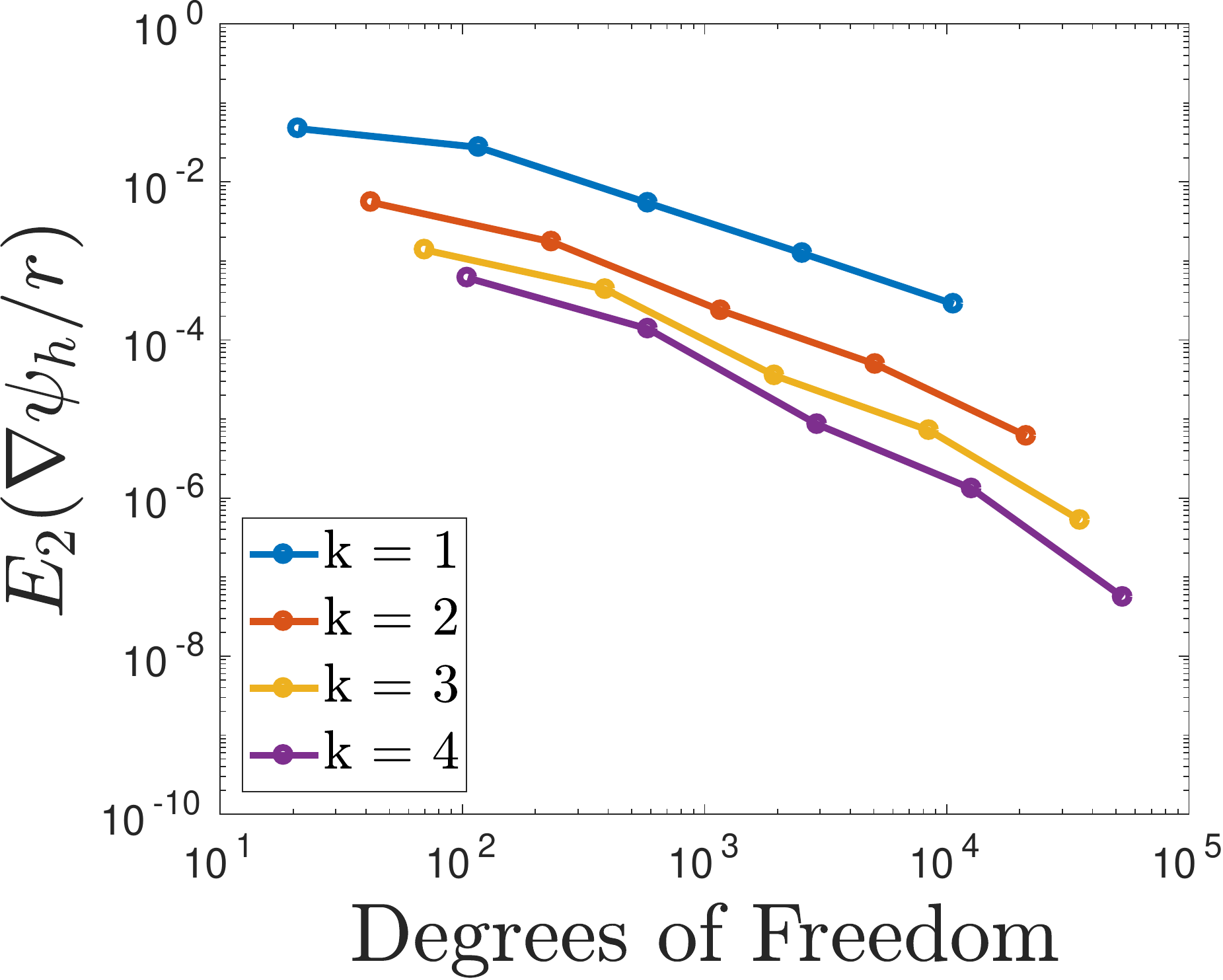} \\
\rotatebox{90}{\quad Adaptive refinement} & 
\includegraphics[height=0.2\linewidth]{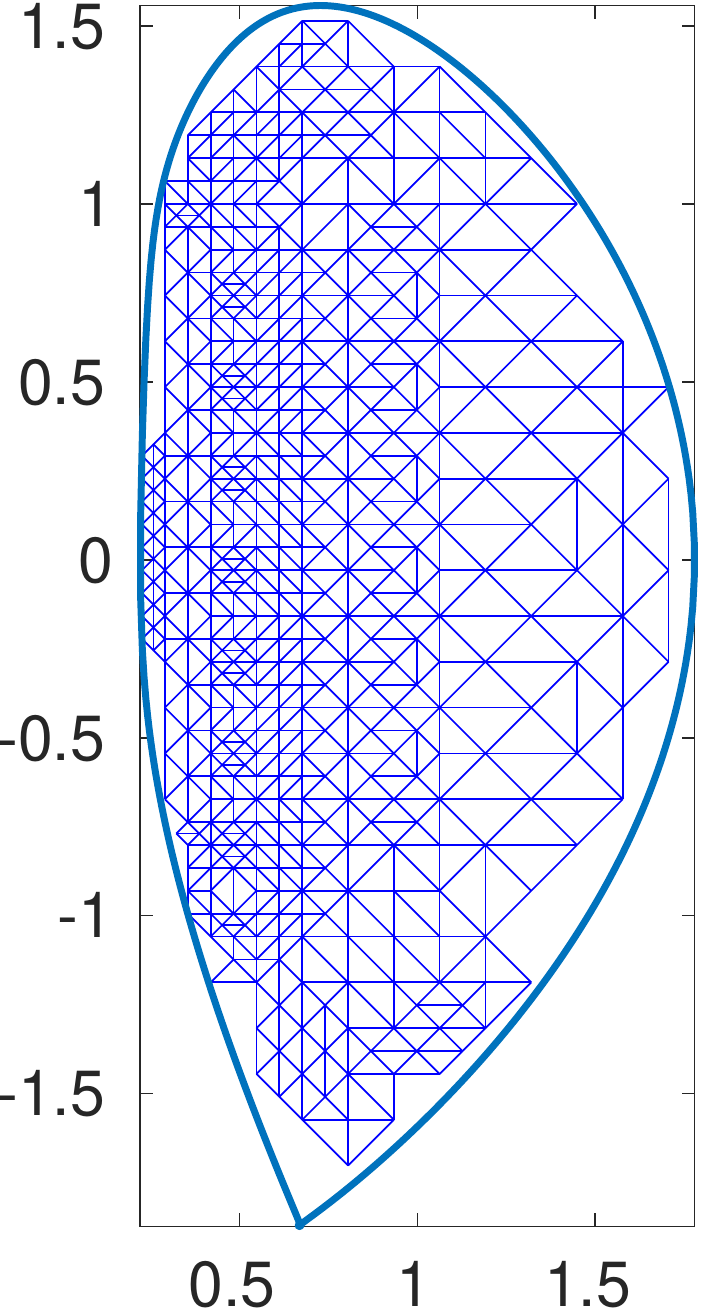} &  \kern-1em  \includegraphics[height=0.2\linewidth]{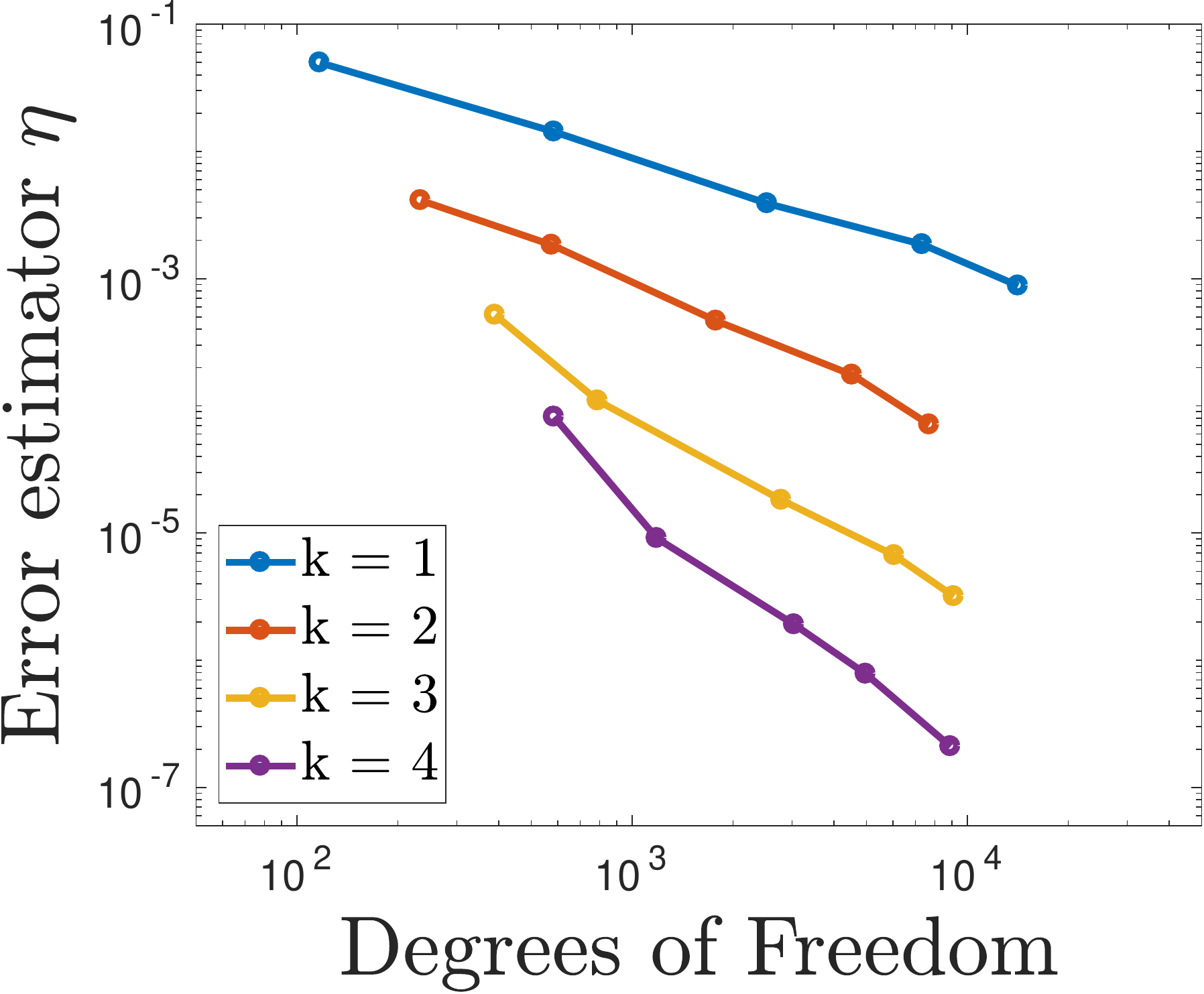} & \kern-1em  \includegraphics[height=0.2\linewidth]{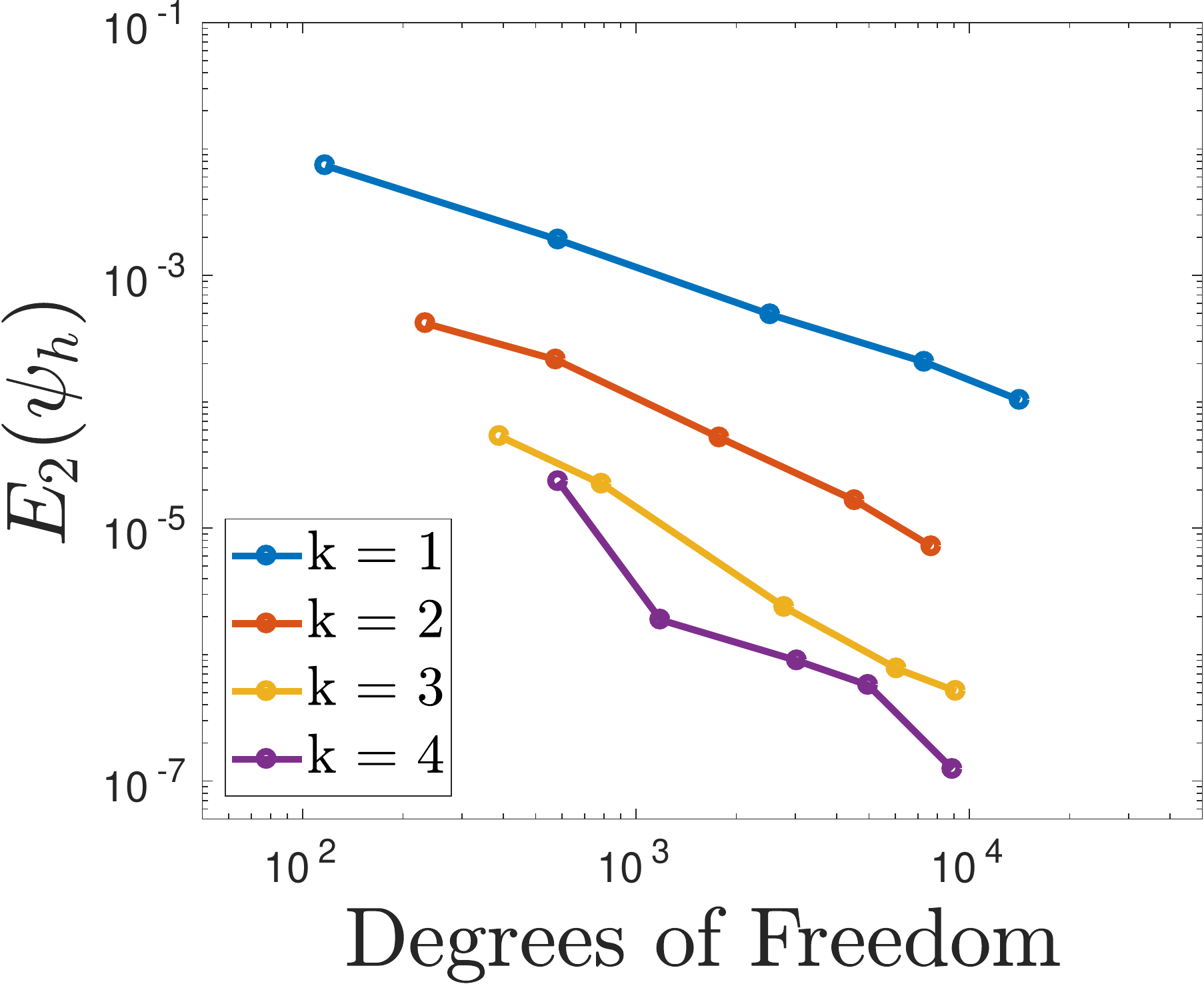}  & \kern-1em  \includegraphics[height=0.2\linewidth]{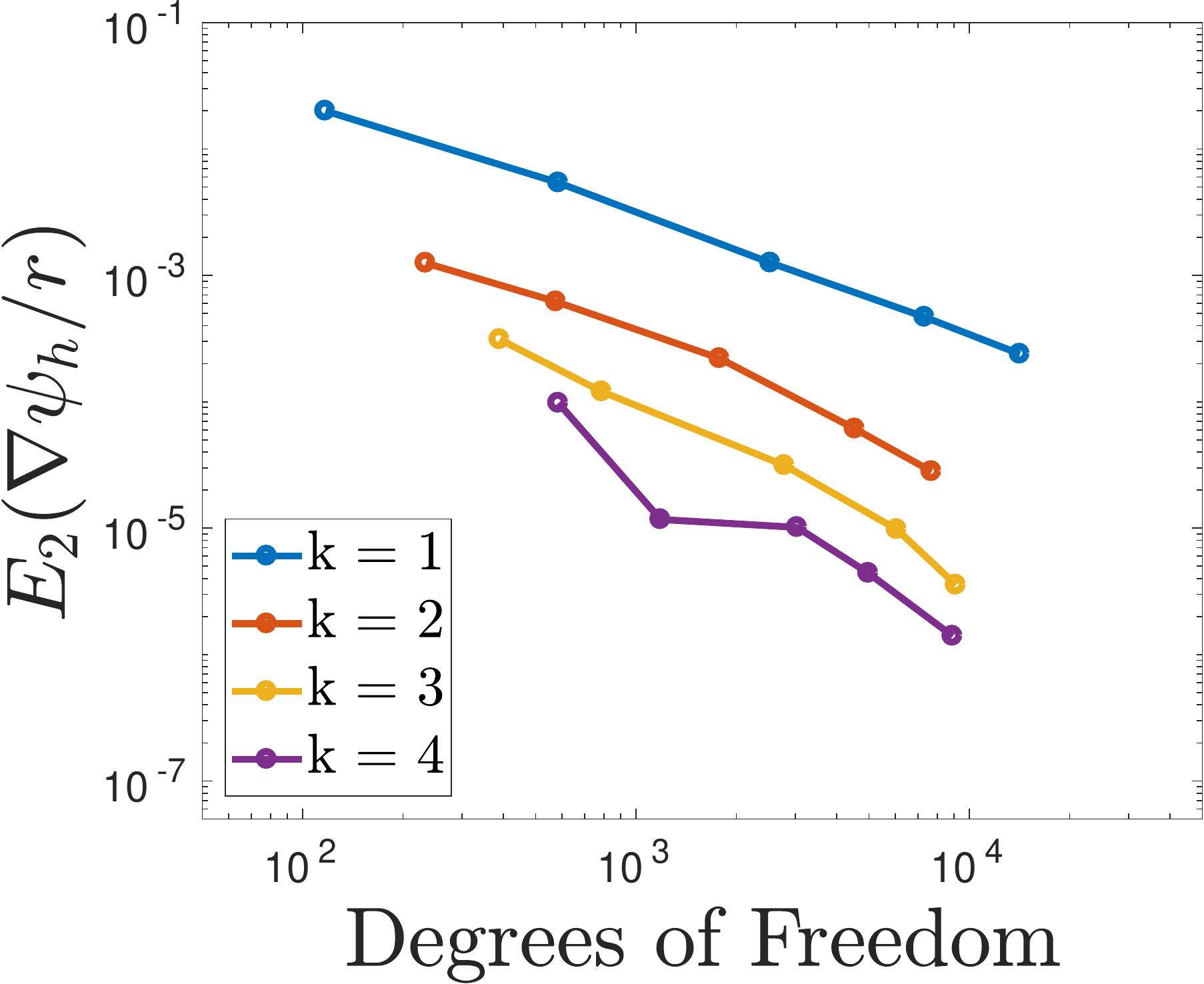}
\end{tabular}}
\caption{Convergence plots for the global error estimator $\eta$ (left), and the mean square errors for $\psi_h$ (center), and $\nabla\psi_h/r$ (right) for the Solov'ev equilibrium arising from the source term \eqref{eq:solovevsource}. The case of uniform refinement is shown in the top row, while adaptive refinement with maximum marking and $\gamma=0.3$ is displayed in the bottom row. The final mesh for $k=4$ in each case is displayed on the left.}\label{fig:ConvergenceHighBeta}
\end{figure}

To test the behavior of the error estimator with respect to the true error, the equation was solved with polynomial basis with degrees $k=1,\ldots,4$ and uniform mesh refinement (i.e. for any subsequent levels $h_{n+1}= h_n/2$). The initial mesh diameter was $h=1.03$. The convergence plots for this experiment are shown in the top row of Figure \ref{fig:ConvergenceHighBeta}, where it can be seen that the estimator accurately captures the qualitative behavior of the error. Moreover, as can be seen in Table \ref{tab:convergencerates}, the rates of convergence of the numerical solutions $\psi_h$ and $\boldsymbol q_h$, as well as those of the error estimator $\eta$ and all its component terms $\eta_1,\ldots,\eta_5$ are nearly optimal.

For comparison, five levels of the adaptive algorithm were ran on the same problem for polynomial degrees from 1 to 4 on the same initial grid. Marking was done using the maximum criterion with parameter $\gamma=0.3$: elements whose estimator is at least 30\% of the maximum local estimate are marked. The convergence history can be seen in the bottom row of Figure \ref{fig:ConvergenceHighBeta}. As can be seen in the same figure, for $k=4$ the number of degrees of freedom after four levels of uniform refinement is about the same order of magnitude as that of five levels of adaptive refinement, but the adaptive algorithm places most of the computational effort on the left side of the domain. Comparing with the plot of $\partial_{rr}\psi$ in Figure \ref{fig:HighBeta} it is clear that the refinement is focusing on the region where the magnitude of the second derivative in the horizontal direction peaks.

\begin{table}\centering\resizebox{\textwidth}{!}{
\begin{tabular}{c|cccccccc|cccccccc|}
\cline{2-17}
& \multicolumn{8}{c}{Polynomial degree $k=1$} & \multicolumn{8}{|c|}{Polynomial degree $k=2$} \\
\cline{2-17}
 & $\psi_h$ & $\boldsymbol q_h$ & $\eta$ & $\eta_1$ & $\eta_2$ & $\eta_3$ & $\eta_4$ & $\eta_5$  & $\psi_h$ & $\boldsymbol q_h$ & $\eta$ & $\eta_1$ & $\eta_2$ & $\eta_3$ & $\eta_4$ & $\eta_5$ \\ \hline 
\multicolumn{1}{|c|}{$h\rightarrow h/2$}   & 1.32 & 0.79 & 1.66 & 1.68 & 1.64 & 1.49 & 2.1  & 1.81 & 2.57 & 1.66 & 2.01 & 2.04 & 2.02 & 1.63 & 1.52 & 2.78  \\
\multicolumn{1}{|c|}{$h/2\rightarrow h/4$} & 1.95 & 2.34 & 1.8  & 1.84 & 1.62 & 1.59 & 2.05 & 1.5  & 2.7  & 2.89 & 2.63 & 2.64 & 2.81 & 2.55 & 2.93   & 3.06 \\
\multicolumn{1}{|c|}{$h/4\rightarrow h/8$} & 1.97 & 2.11 & 1.87 & 1.89 & 1.82 & 1.78 & 2.28 & 1.81 & 2.8  & 2.26 & 2.66 & 2.66 & 2.75 & 2.62 & 2.56 & 2.81 \\
\multicolumn{1}{|c|}{$h/8\rightarrow h/16$} & 1.99 & 2.13 & 1.93 & 1.95 & 1.93 & 1.86 & 2.33 & 1.94 & 2.97 & 3.02 & 2.88 & 2.88 & 2.91 & 2.83 & 2.74 & 2.85 \\
\hline
& \multicolumn{8}{c}{Polynomial degree $k=3$} & \multicolumn{8}{|c|}{Polynomial degree $k=4$} \\
\cline{2-17}
 & $\psi_h$ & $\boldsymbol q_h$ & $\eta$ & $\eta_1$ & $\eta_2$ & $\eta_3$ & $\eta_4$ & $\eta_5$  & $\psi_h$ & $\boldsymbol q_h$ & $\eta$ & $\eta_1$ & $\eta_2$ & $\eta_3$ & $\eta_4$ & $\eta_5$ \\ \hline 
\multicolumn{1}{|c|}{$h\rightarrow h/2$} & 3.46 & 1.66 & 1.64 & 1.61 & 1.82 & 1.99 & 1.96 & 2.82 & 3.94 & 2.14 & 1.39 & 1.4 & 1.65 & 1.28 & 1.64 & 2.47 \\
\multicolumn{1}{|c|}{$h/2\rightarrow h/4$} & 3.52 & 3.62 & 3.53 & 3.53 & 3.58 & 3.53 & 3.91 & 3.93 & 4.37 & 4.02 & 4.34 & 4.34 & 4.29 & 4.21 & 4.53 & 4.56 \\
\multicolumn{1}{|c|}{$h/4\rightarrow h/8$}  & 3.13 & 2.3  & 3.26 & 3.26 & 3.2  & 3.34 & 3.44 & 3.48 & 3.64 & 2.7 & 3.88 & 3.89 & 3.63 & 3.81 & 3.92 & 3.96  \\
\multicolumn{1}{|c|}{$h/8\rightarrow h/16$}  & 4.26 & 3.78 & 3.83 & 3.83 & 3.81 & 3.85 & 3.99 & 3.71 & 5.21 & 4.56 & 4.78 & 4.79 & 4.68 & 4.67 & 4.79 & 4.79 \\
\hline
\end{tabular} }
\caption{Estimated convergence rates (e.c.r.) for the approximate solutions $\psi_h$ and $\boldsymbol q_h$ as well as for the error estimator $\eta$ and all its separate components in the case of uniform refinement. The exact Solov'ev solution is given in Equation \eqref{eq:HighBetaSol}. For a given quantity $U$ we denote the $L^2$ error at the $k-$th refinement level by $E(U)^k_2 $. Then, the estimated convergence rate is computed through the formula $\text{e.c.r} = \log_{2}{\left(E(U)^k_2/E(U)^{k+1}_2\right)}$. }\label{tab:convergencerates}
\end{table}

\subsection{A pressure pedestal}
\begin{figure}\centering\scalebox{.925}{
\begin{tabular}{cccc}
Pressure profile & & Source term & \\
\includegraphics[height=0.2\linewidth]{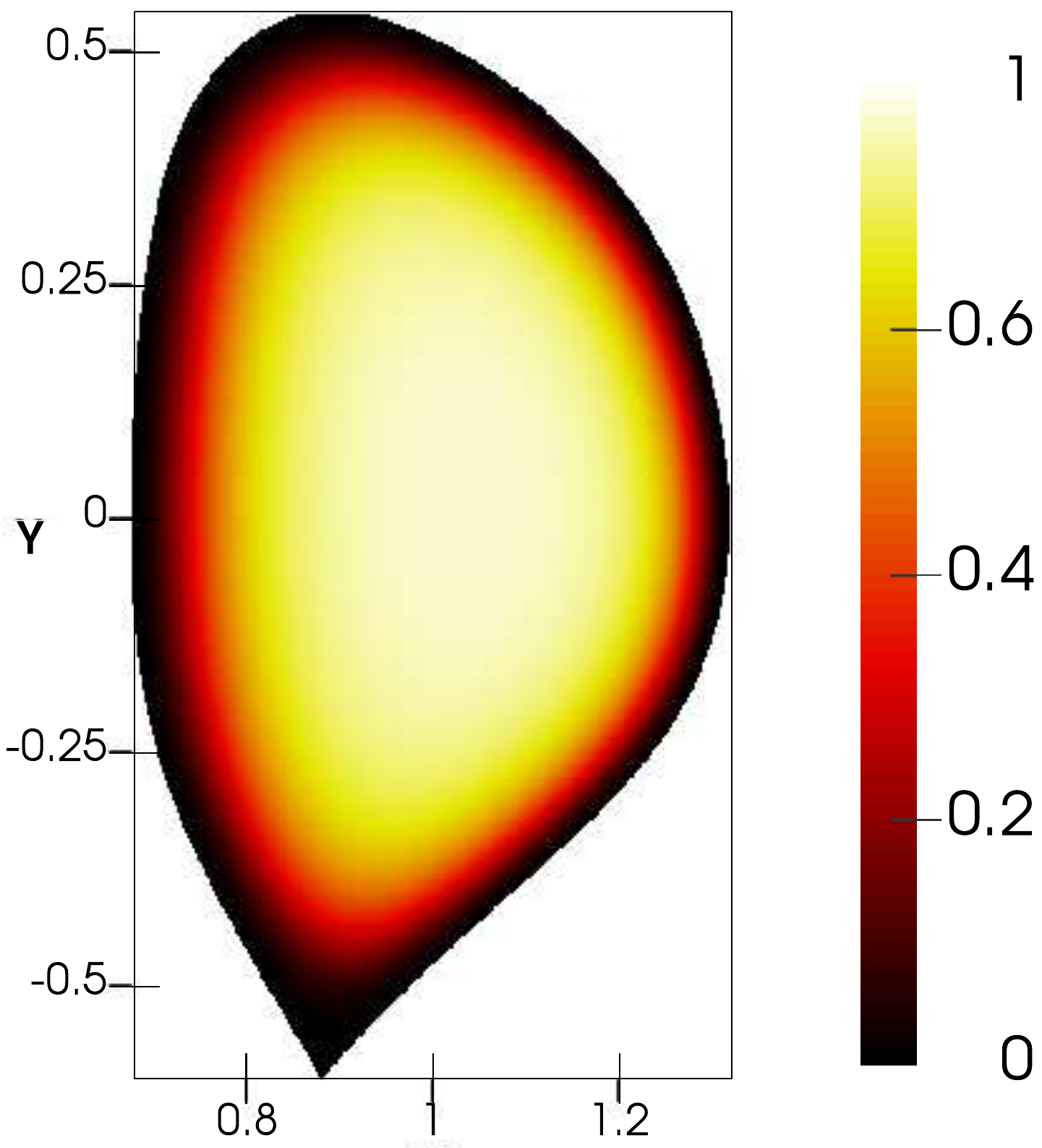} \quad & \includegraphics[height=0.2\linewidth]{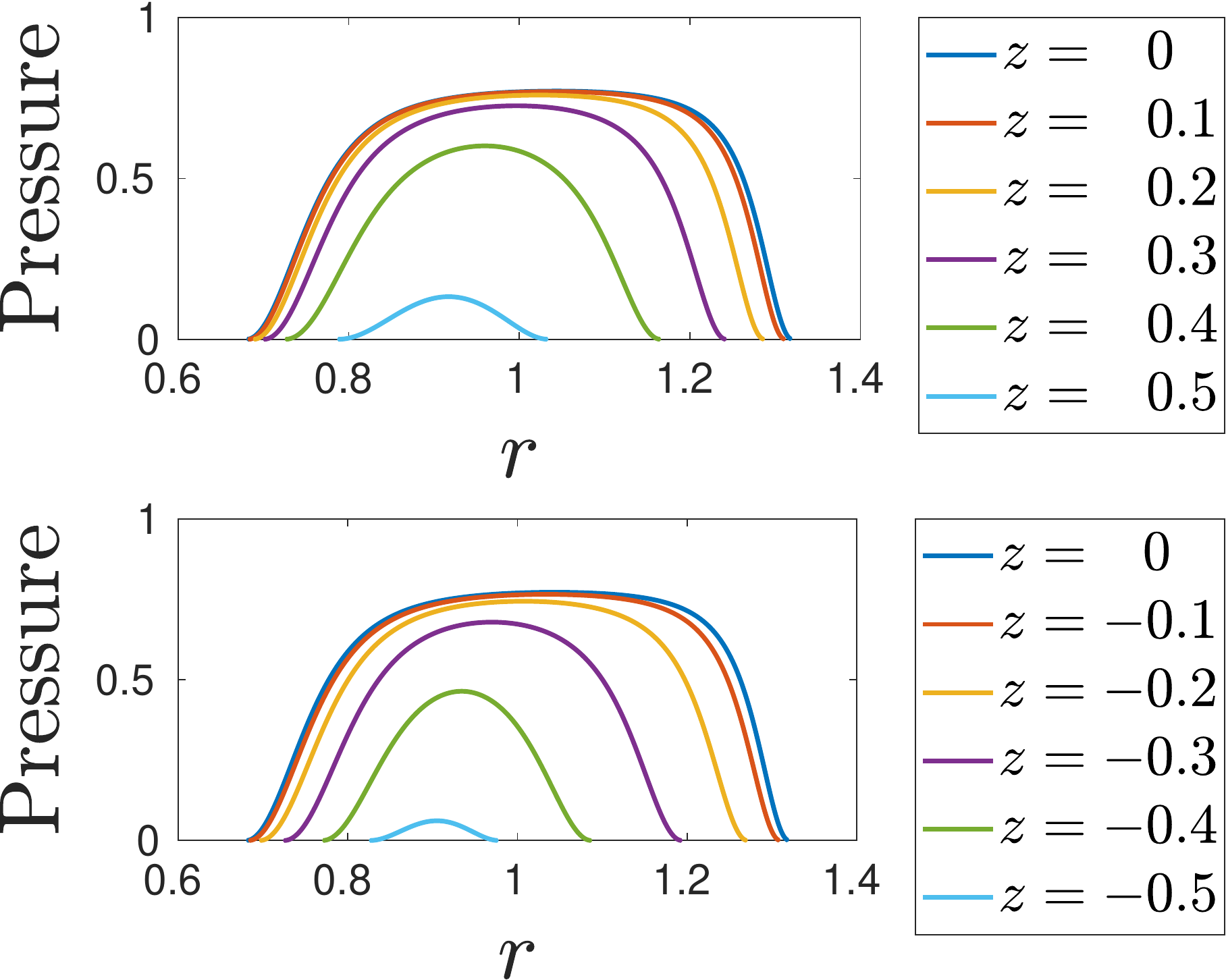} \quad & \includegraphics[height=0.2\linewidth]{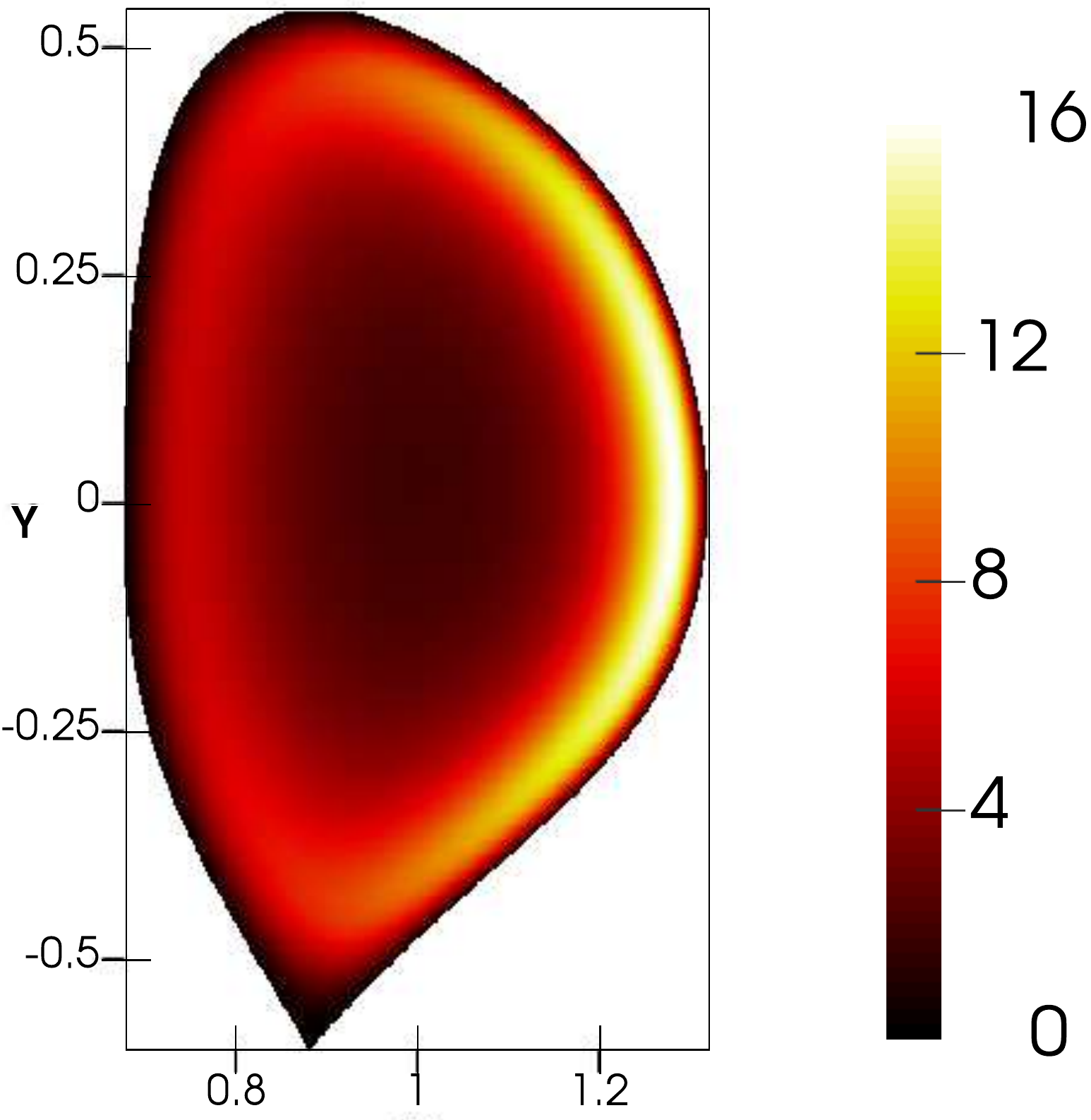} \quad & \includegraphics[height=0.2\linewidth]{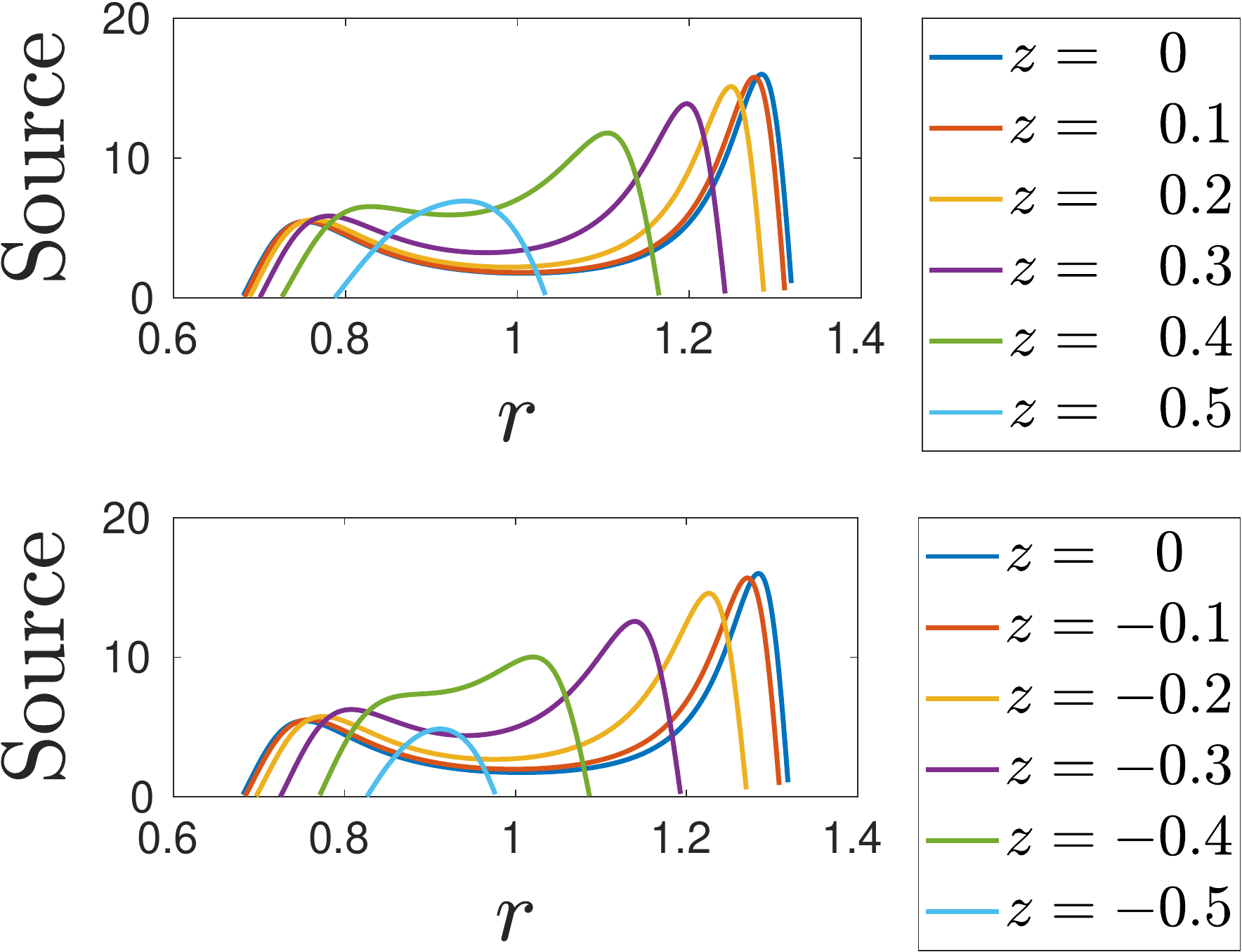}
\end{tabular}}
\caption{A pressure pedestal of the form given by equation \eqref{eq:PressurePedestal} for $\sigma^2 = 0.005$ in an ITER-like geometry (left). Cross sections for values of $z$ ranging from $-0.5$ to $0.5$ are shown (center left). The corresponding source term of the Grad-Shafranov equation presents sharp gradients on the outer edge (center right). Cross sections of the source for the same values of $z$ as for the pressure are shown ( right).}\label{fig:PressurePedestalIterX}
\end{figure}

The following example features a pressure profile that remains almost flat throughout the confinement region and drops abruptly in the vicinity of the boundary:
\begin{equation}\label{eq:PressurePedestal}
p(\psi) = (c_1 + c_2\psi^2)(1-e^{-(\psi/\sigma)^2}).
\end{equation}
with $c_1=0.8$, $c_2=0.2$, and $\sigma^2=0.005$. This kind of profile is quite frequent in magnetic confinement fusion experiments, where the narrow region of fast decrease of the pressure is known as a \textit{pressure pedestal}. If the equilibrium is assumed to be neither paramagnetic nor diamagnetic, $g(\psi)=constant$, and the source term of the Grad-Shafranov equation is
\begin{equation}\label{eq:srcpedestal}
F(r,\psi) = 2r^2\psi\left(c_2(1-e^{-(\psi/\sigma)^2}) + \frac{1}{\sigma^2}(c_1+c_2\psi^2)e^{-(\psi/\sigma)^2}\right),
\end{equation}
which has very strong gradients close to the edge of the confinement region since $\sigma$ is small. Figure \ref{fig:PressurePedestalIterX} shows both the pressure and the source profiles in an ITER-like geometry with an magnetic X-point (as described in \cite{CeFr:2010}). As can be seen from the cross sections at constant values of $z$ in the same figure, the source has large gradients close to the boundaries, especially in the ``outer" region.

\begin{figure}\centering\scalebox{.925}{
\begin{tabular}{ccccc}
{\small Computational Mesh} & $\psi^*_h$ & $\psi^*_h$ {\small (cross section)} & $\partial_r\psi_h$ & $\partial_z\psi_h$ \\
\includegraphics[height=0.2\linewidth]{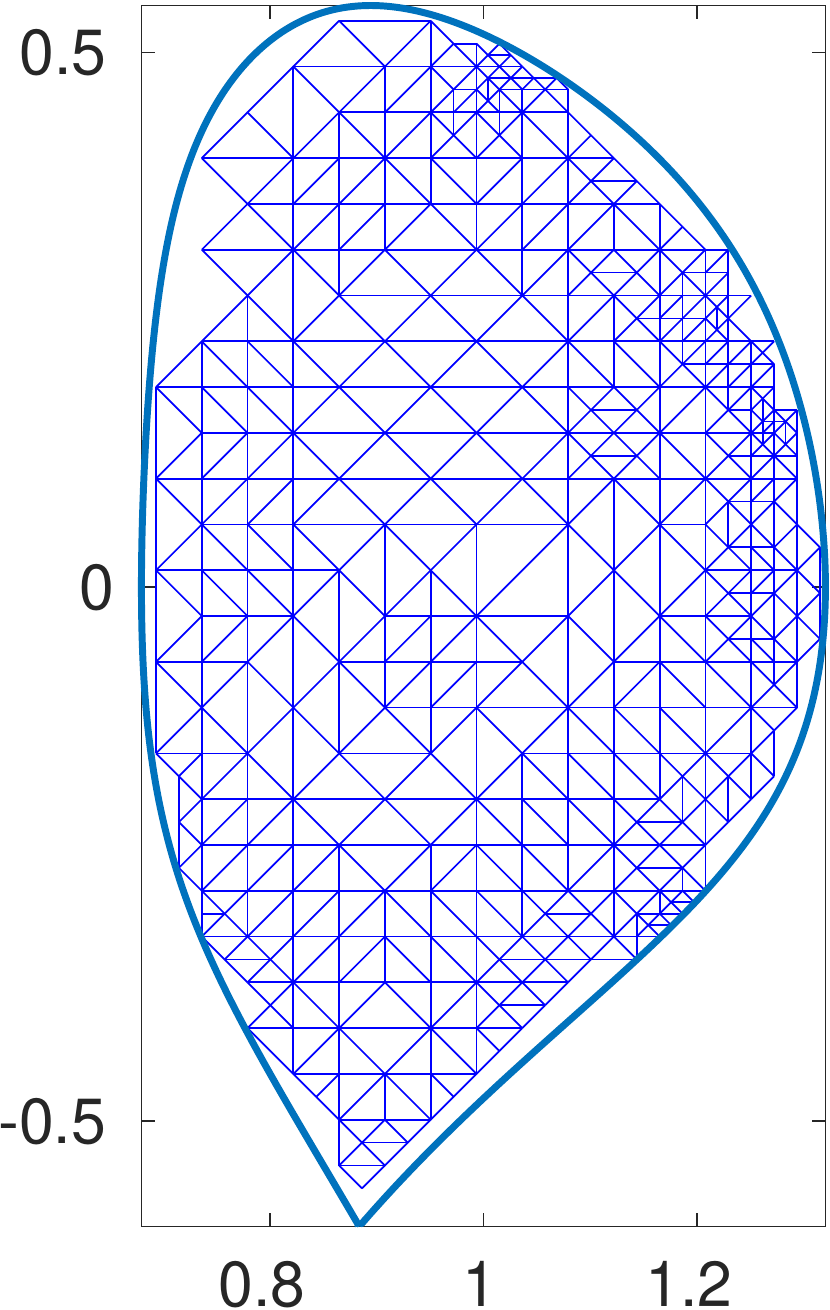} \qquad & \includegraphics[height=0.2\linewidth]{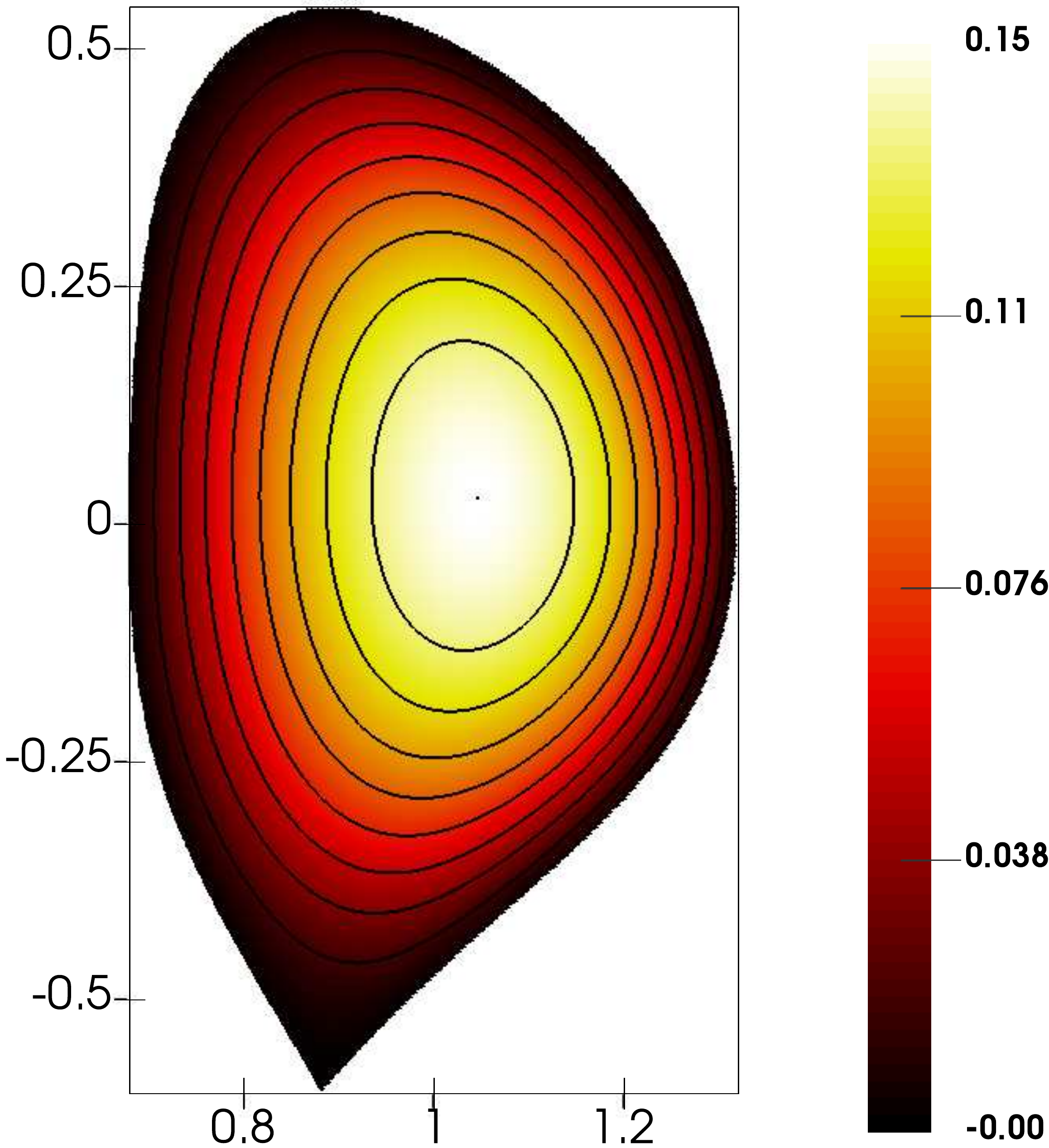} \quad & \includegraphics[height=0.2\linewidth]{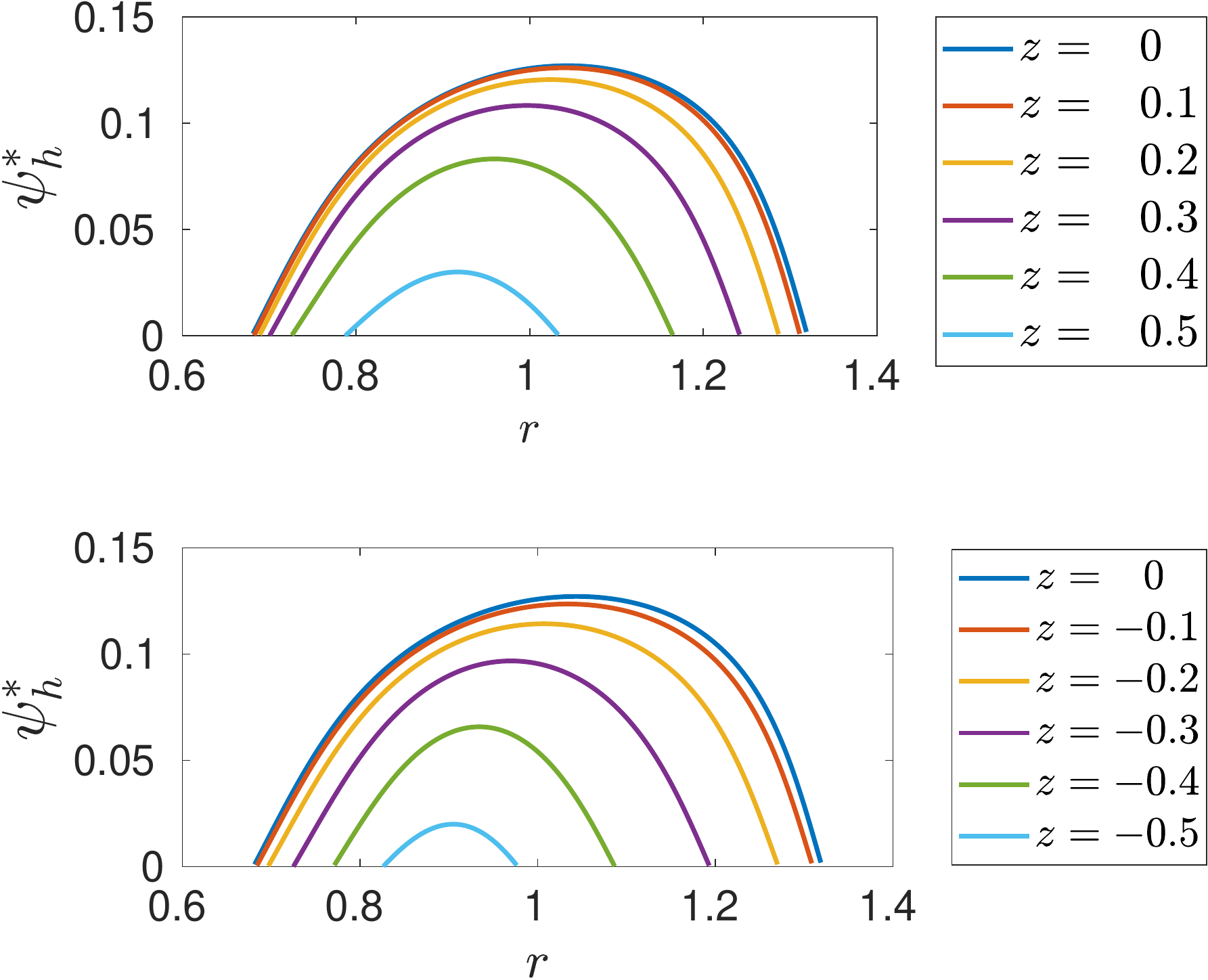} \quad & \includegraphics[height=0.2\linewidth]{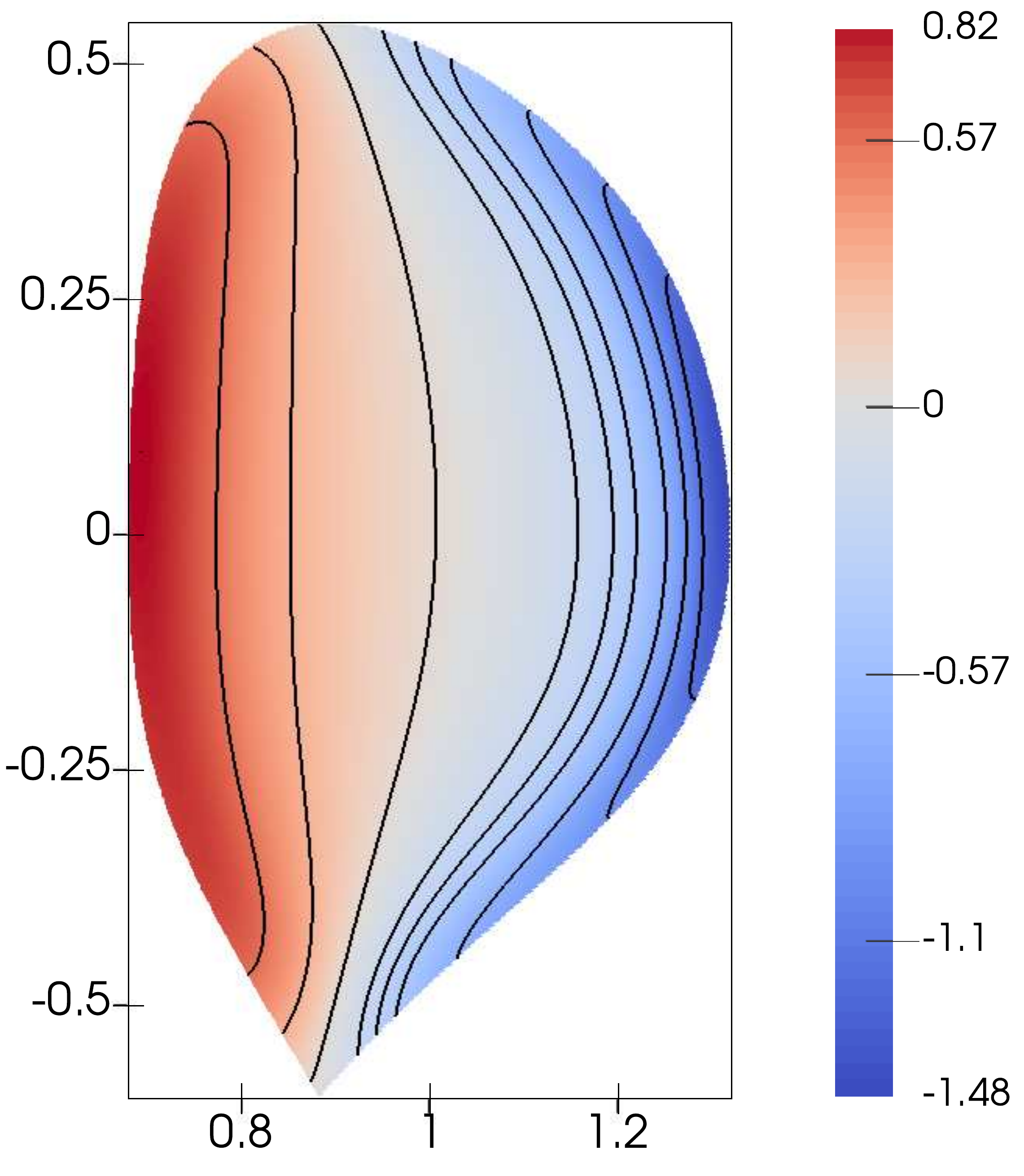} \quad & \includegraphics[height=0.2\linewidth]{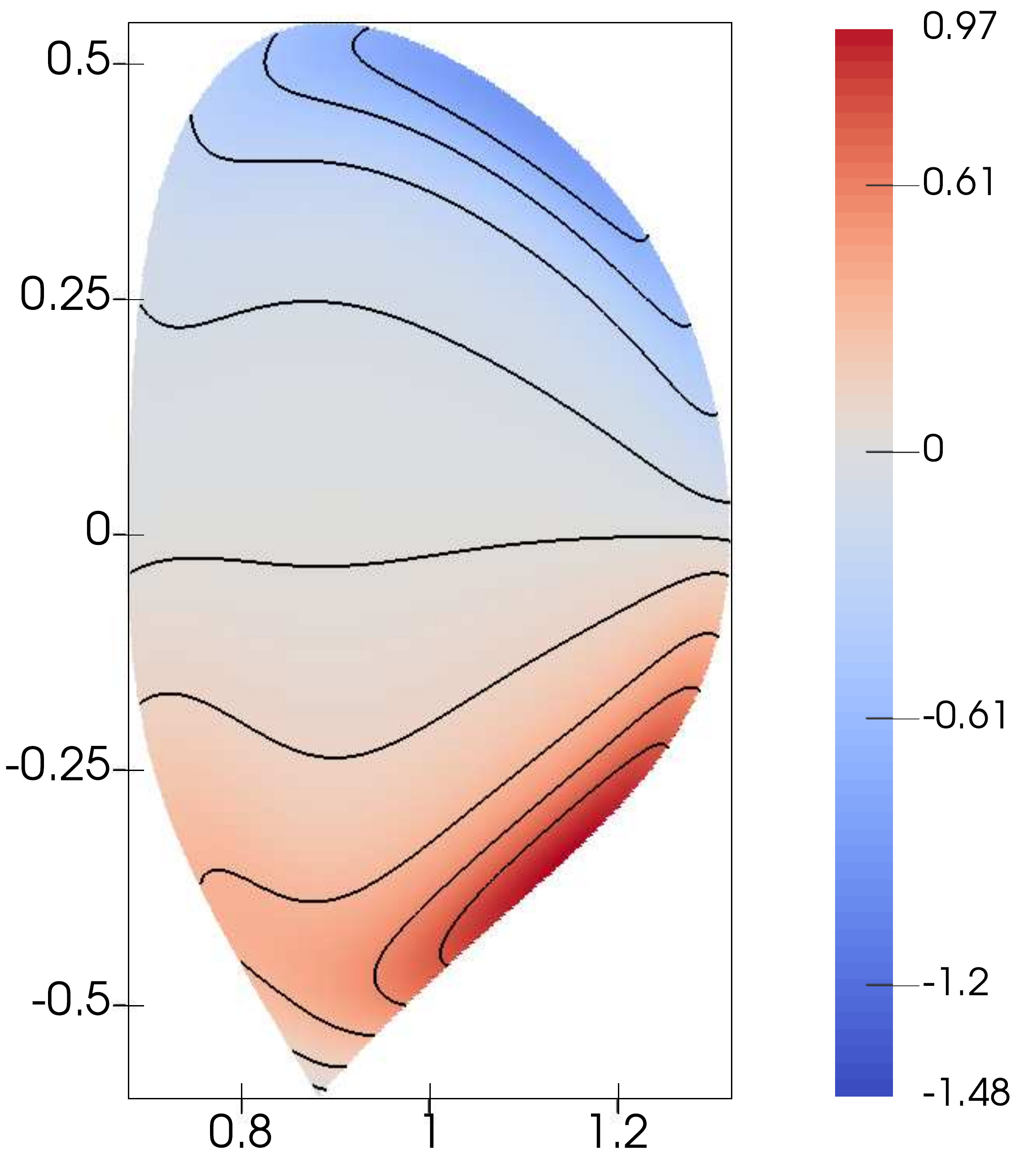}
\end{tabular}}
\caption{Numerical solution for the equilibrium with a pressure pedestal, Equation \eqref{eq:PressurePedestal}, with parameter values $c_1=0.8, c_2=0.2, \sigma^2 = 0.005$. The computation was carried out with a polynomial basis of degree $k=4$ and six levels of adaptive refinement with $\gamma=0.3$. The post processed scalar potential $\psi^*$ and the partial derivatives of $\psi$ are shown. Cross sections for $z=-0.5,-0.25,0,0.25,0.5$ are displayed in the center. The computational grid at the final refinement is shown on the left.}\label{fig:PressurePedestalApproxIterX}
\end{figure}

The equation was solved using \eqref{eq:srcpedestal} as the source term and the latter parameter values. Figure \ref{fig:PressurePedestalApproxIterX} shows the post-processed numerical solution obtained when the polynomial basis was chosen to have degree four, and six levels of refinement were used with marking parameter $\gamma=0.3$. The computational mesh had initial mesh parameter $h=1.71\times10^{-1}$, whereas the final mesh, displayed in the first block of the figure,  consists of 636 elements with maximum diameter $h_{max}=1.71\times10^{-1}$ and minimum diameter $h_{min}=1.07 \times 10^{-2}$. Our adaptive refinement scheme clearly focuses degrees of freedom in the region corresponding to the pressure pedestal, where $\psi$ and its derivatives vary strongly.

%
\subsection{A transport barrier}
\begin{wrapfigure}{l}{0.4\textwidth}\centering
\begin{tabular}{cc}
Pressure profile & Source term \\
 \includegraphics[height=0.35\linewidth]{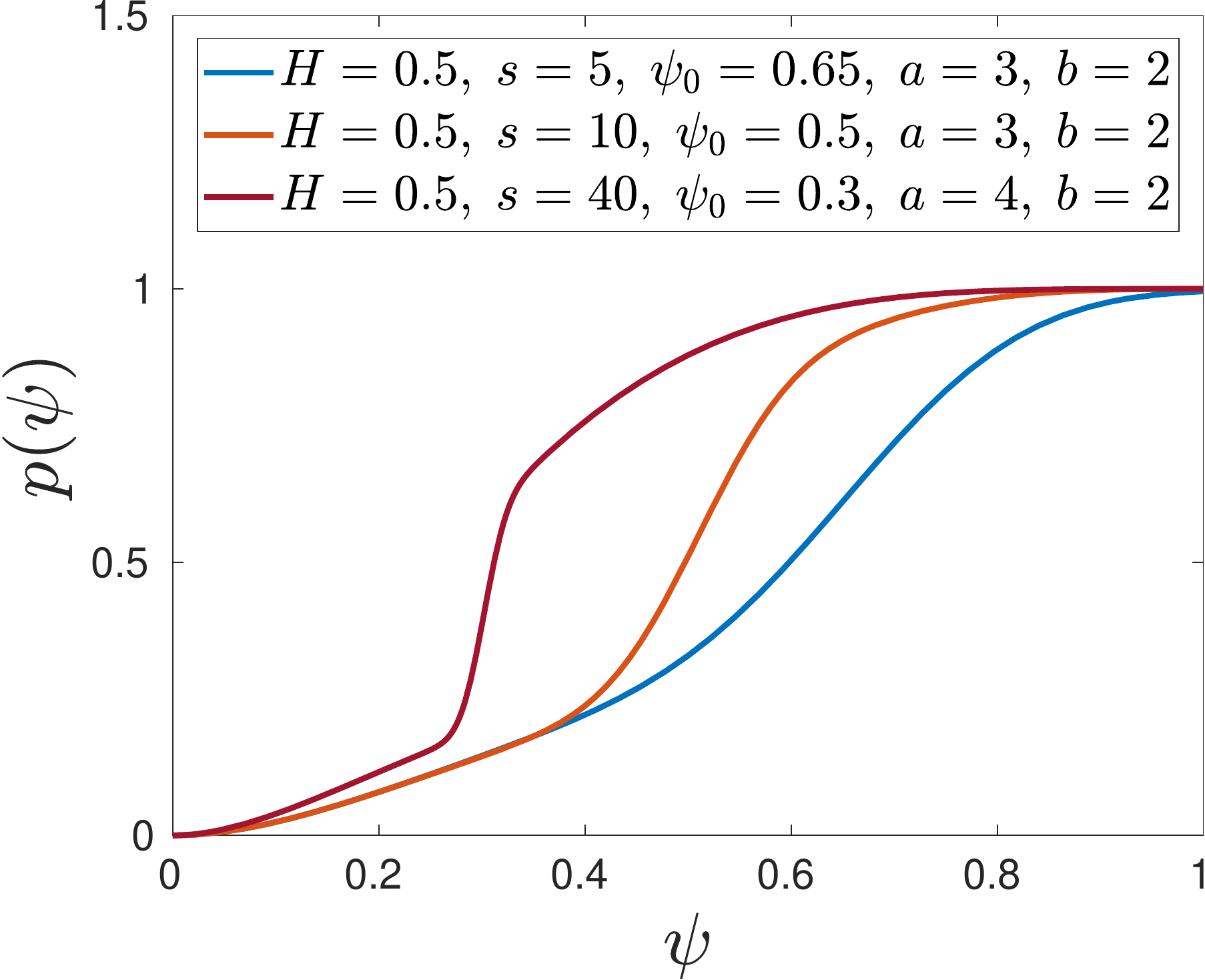} &
 \includegraphics[height=0.35\linewidth]{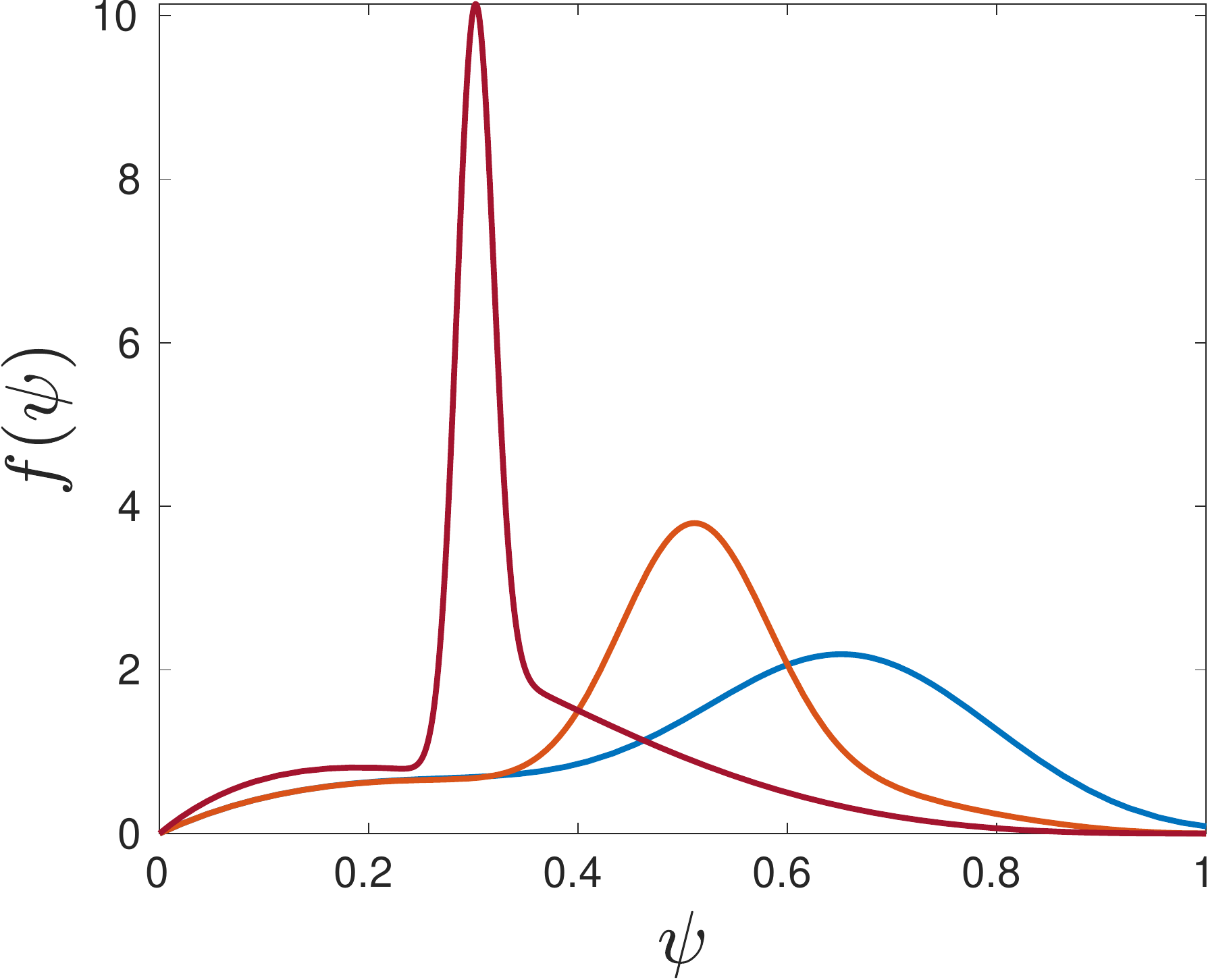}   
\end{tabular}
\caption{Left: Different parameter values in \eqref{eq:transportbarier} give rise to pressure profiles with increasingly steeper gradients. Center: In the case $g=const.$ (see equation\eqref{eq:GradShafranov}) the corresponding source term for the Grad-Shafranov equation has a highly localized structure in the neighborhood of the transport barrier.}\label{fig:TransportBarrier}
\end{wrapfigure}

In magnetic confinement fusion experiments, large pressure gradients may also be observed closer to the core of the discharge, and correspond to internal transport barriers \cite{Wolf:2003}. To model such situations, we consider a pressure profile of the form
\begin{equation}\label{eq:transportbarier}
p(\psi) = \frac{1+H\,\text{erf}(s(\psi-\psi_0))}{1+H}\left(1-(1-\psi)^a\right)^b,
\end{equation}
where erf$(\cdot)$ is the error function, $a$ and $b$ are natural numbers, and the parameters $H$  and $s$ control the height and the steepness of the barrier respectively. The parameter $\psi_0$ gives the location of the transport barrier with respect to a common normalization where $\psi\in[0,1]$ \cite{Cedres:2015,LeCe:2015}. In Figure \ref{fig:TransportBarrier} we present the behavior of the pressure as a function of $\psi$ for different parameter values, and thus verify that we are able to model experimentally relevant situations \cite[Figure 3]{Wolf:2003}.

\begin{figure}[b]\centering\scalebox{1}{
\begin{tabular}{cccc}
$\psi$  &  {\small Pressure }  &  {\small Horizontal Cross Sections }   & {\small Source }\\
\includegraphics[height=0.22\linewidth]{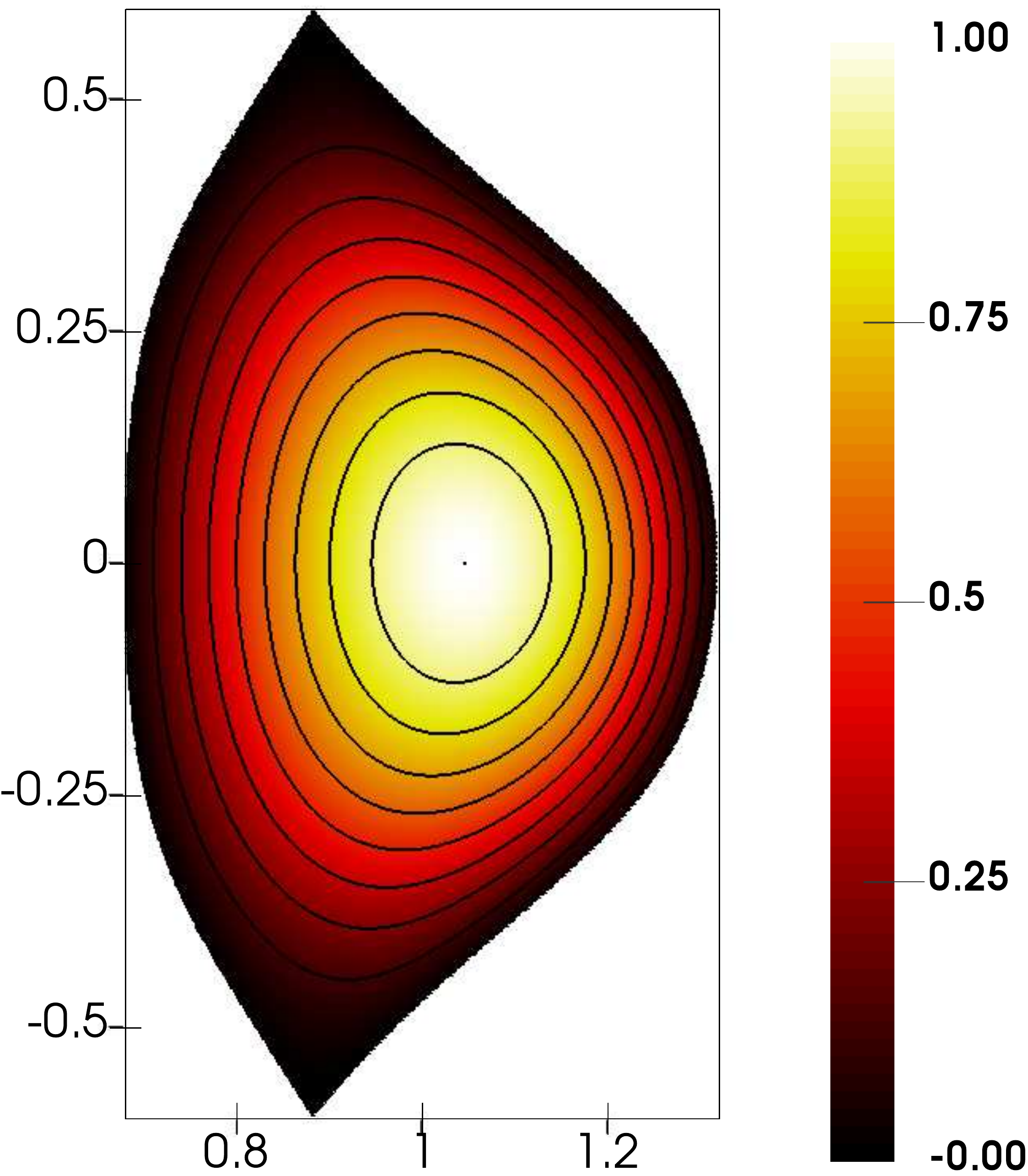} & 
\includegraphics[height=0.22\linewidth]{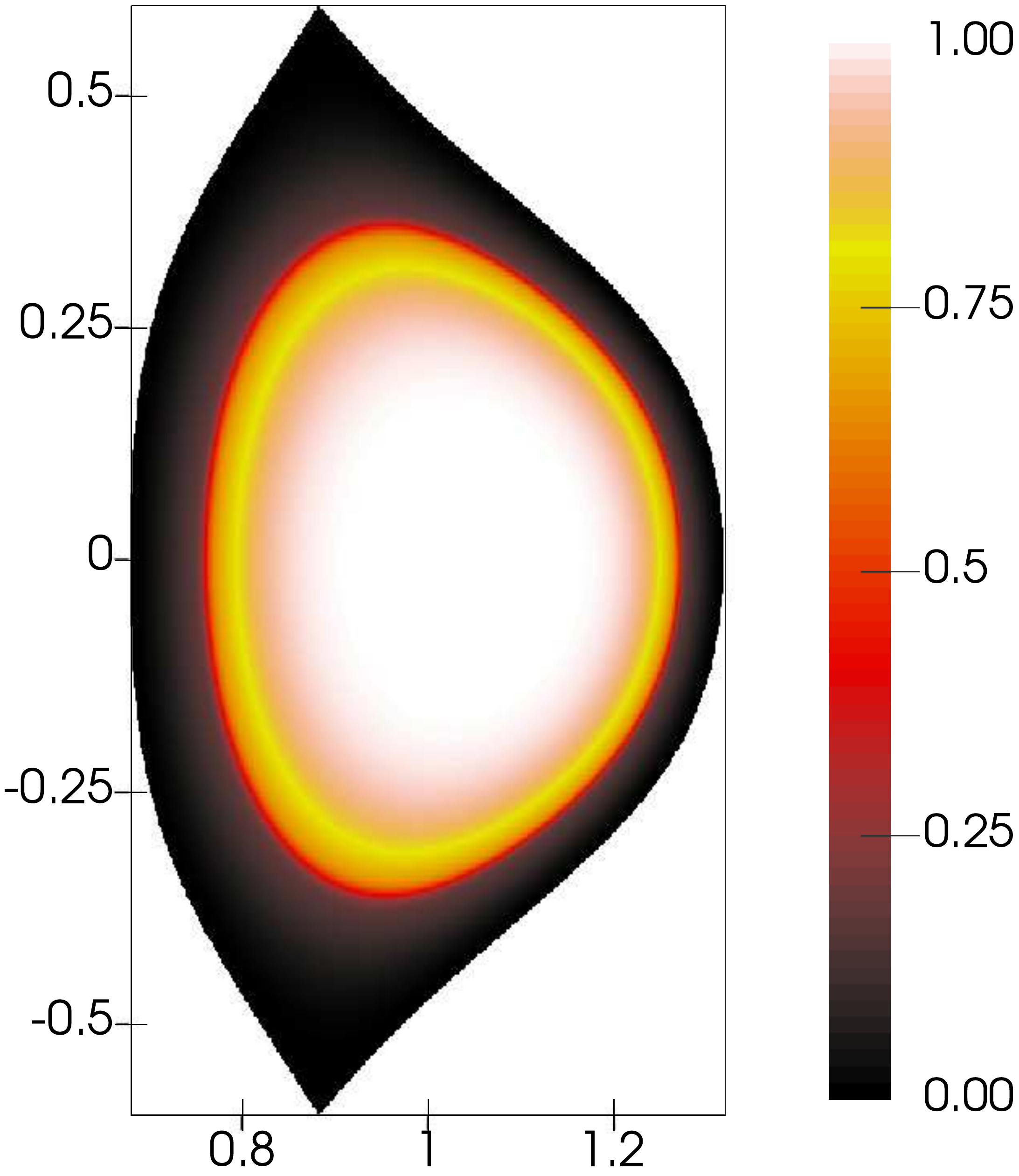} & 
\includegraphics[height=0.22\linewidth] {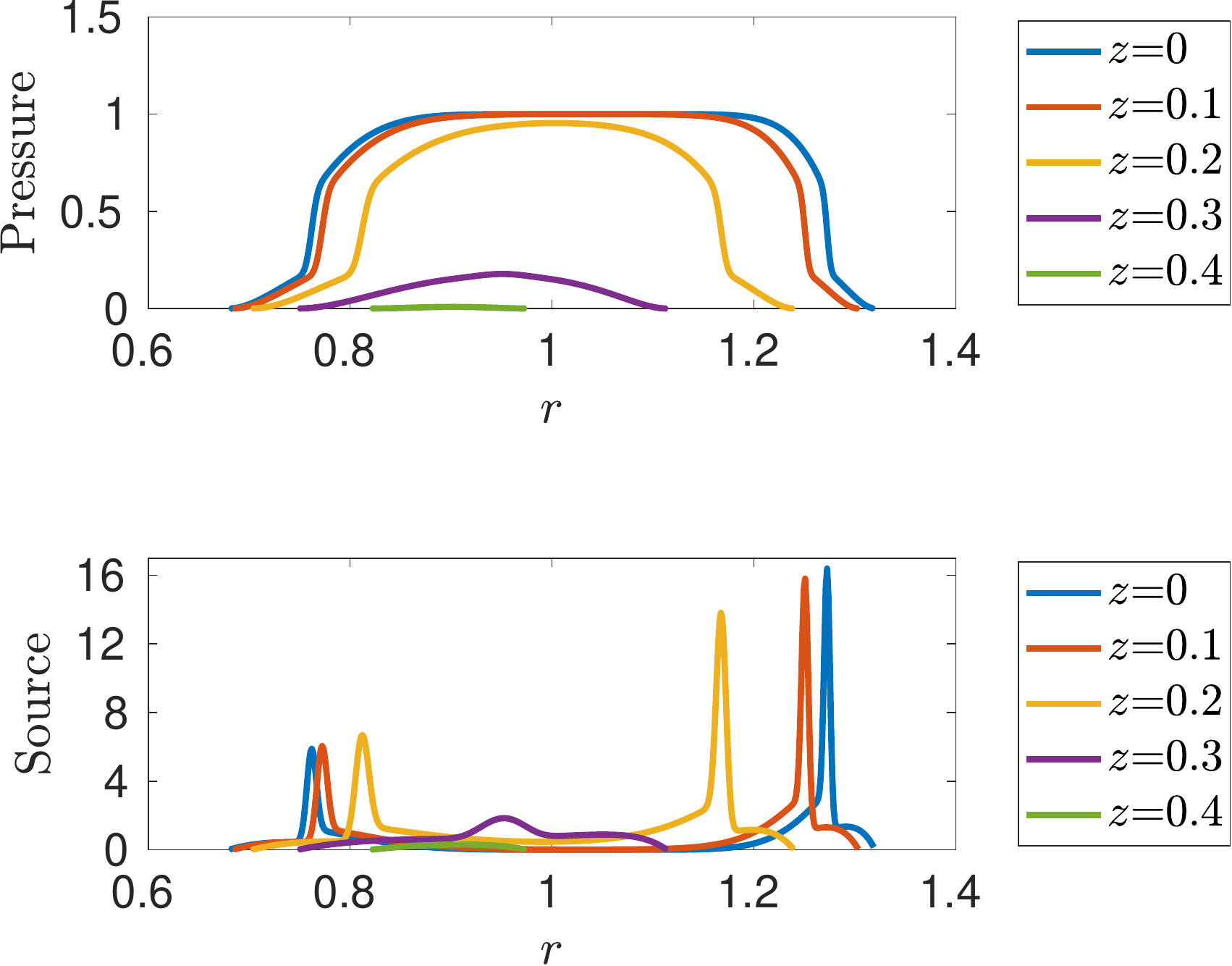} &
\includegraphics[height=0.22\linewidth]{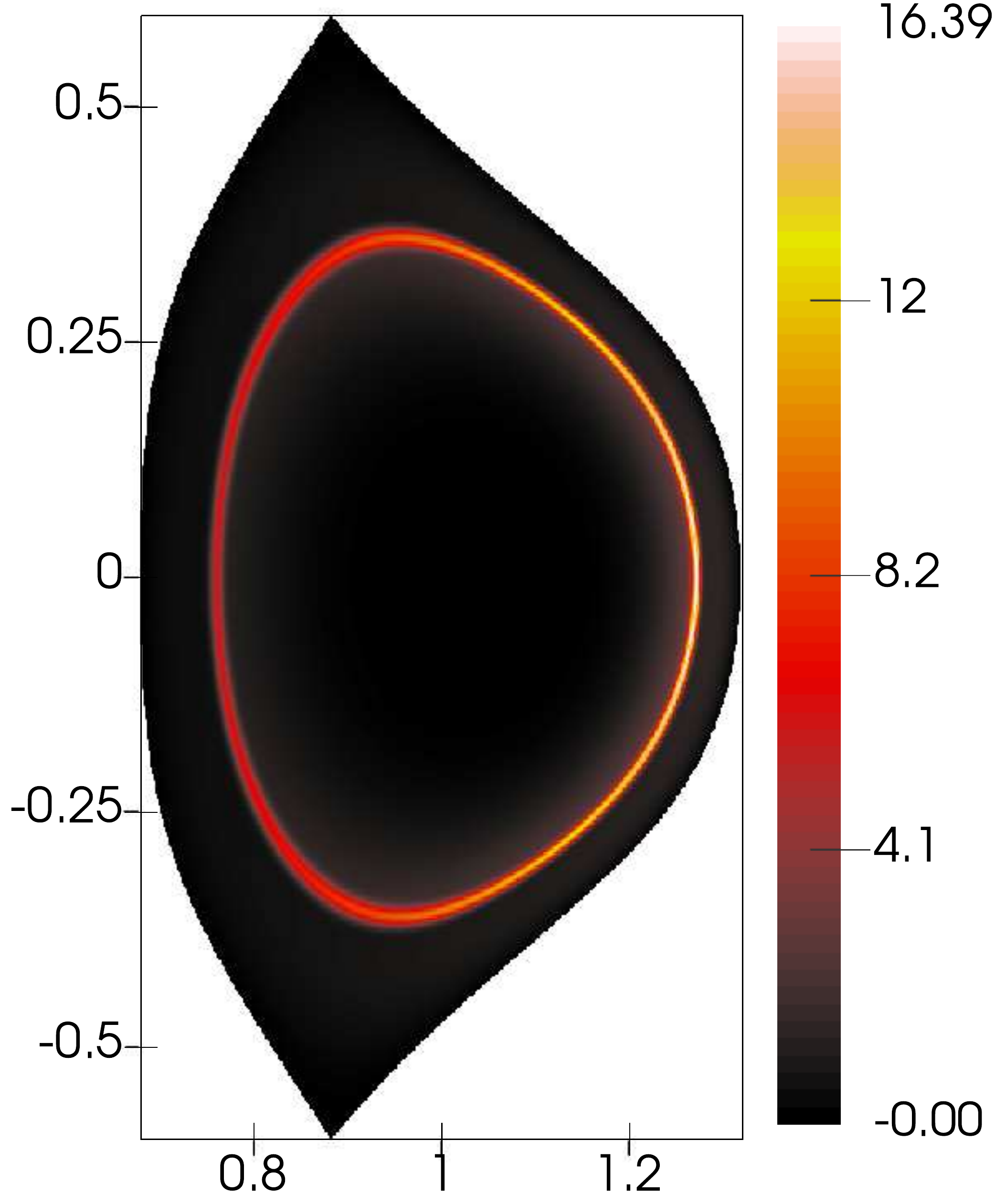}
\end{tabular}}
\caption{Left: Poloidal flux corresponding to the source term associated with the transport barrier. Center left: Pressure profile corresponding to Equation \eqref{eq:transportbarier} and with parameter values $H=0.5$, $s = 40$, $\psi_0=0.3$, $a=4$, and $b=2$. Center right: Horizontal cross sections for the pressure and the source at constant values of $z$. Right: plot of the source term giving rise to the transport barrier. The geometry corresponds to a ITER-like configuration obtained using the process described in \cite{CeFr:2010}.}\label{fig:TransportSimulations}
\end{figure}

\begin{figure}\centering\scalebox{1}{
\begin{tabular}{cccccc}
 & \quad {\small Initial mesh } & {\small Final mesh } & {\small Final mesh } & {\small Final mesh } & {\small Final mesh}\\
\kern-1em  &  & \quad {\small $k=1$ } & \quad {\small $k=2$ }  &  \quad {\small $k=3$}   & \quad {\small $k=4$}\\
\kern-1em \includegraphics[height=0.2\linewidth]{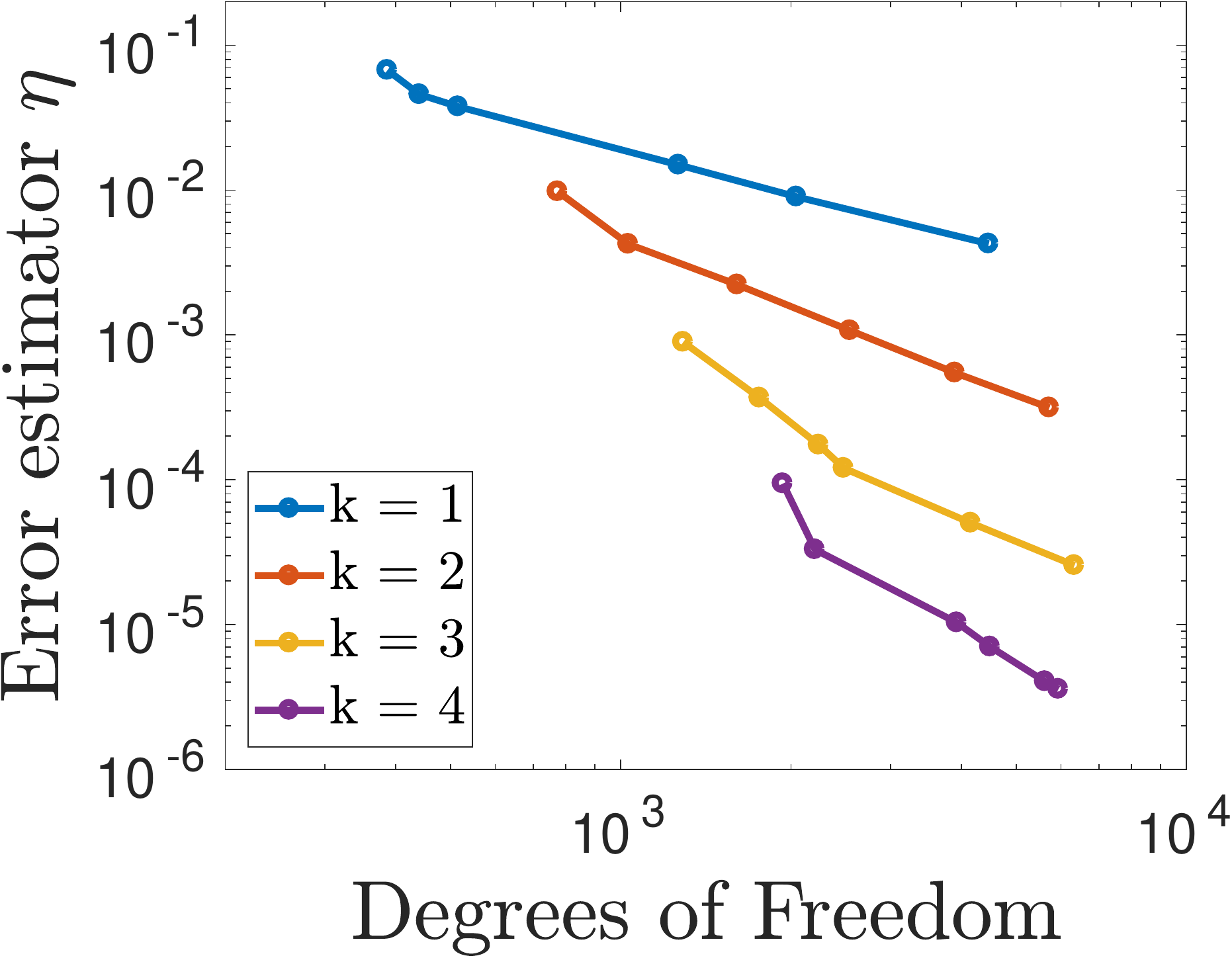} & \!
\includegraphics[height=0.2\linewidth]{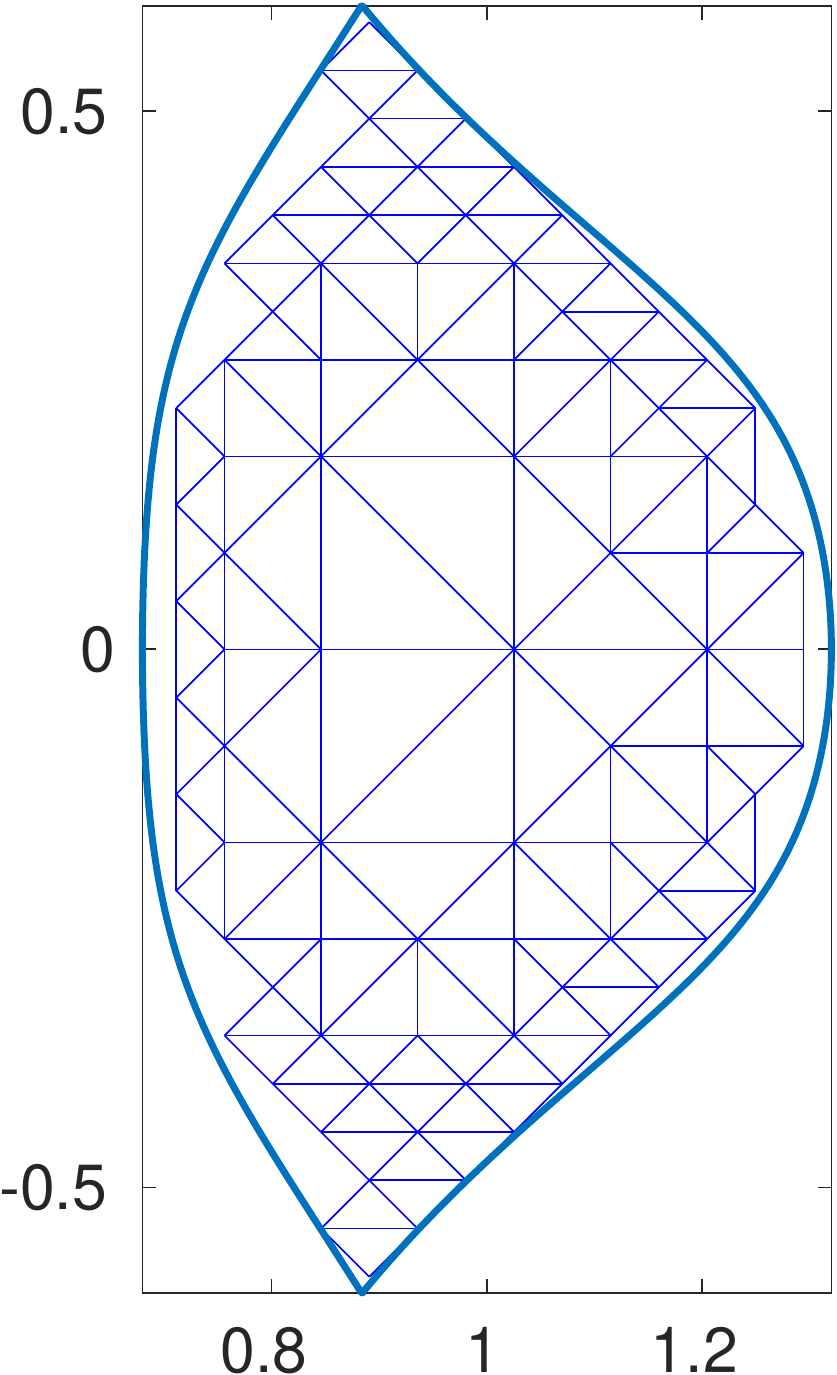} &
\includegraphics[height=0.2\linewidth]{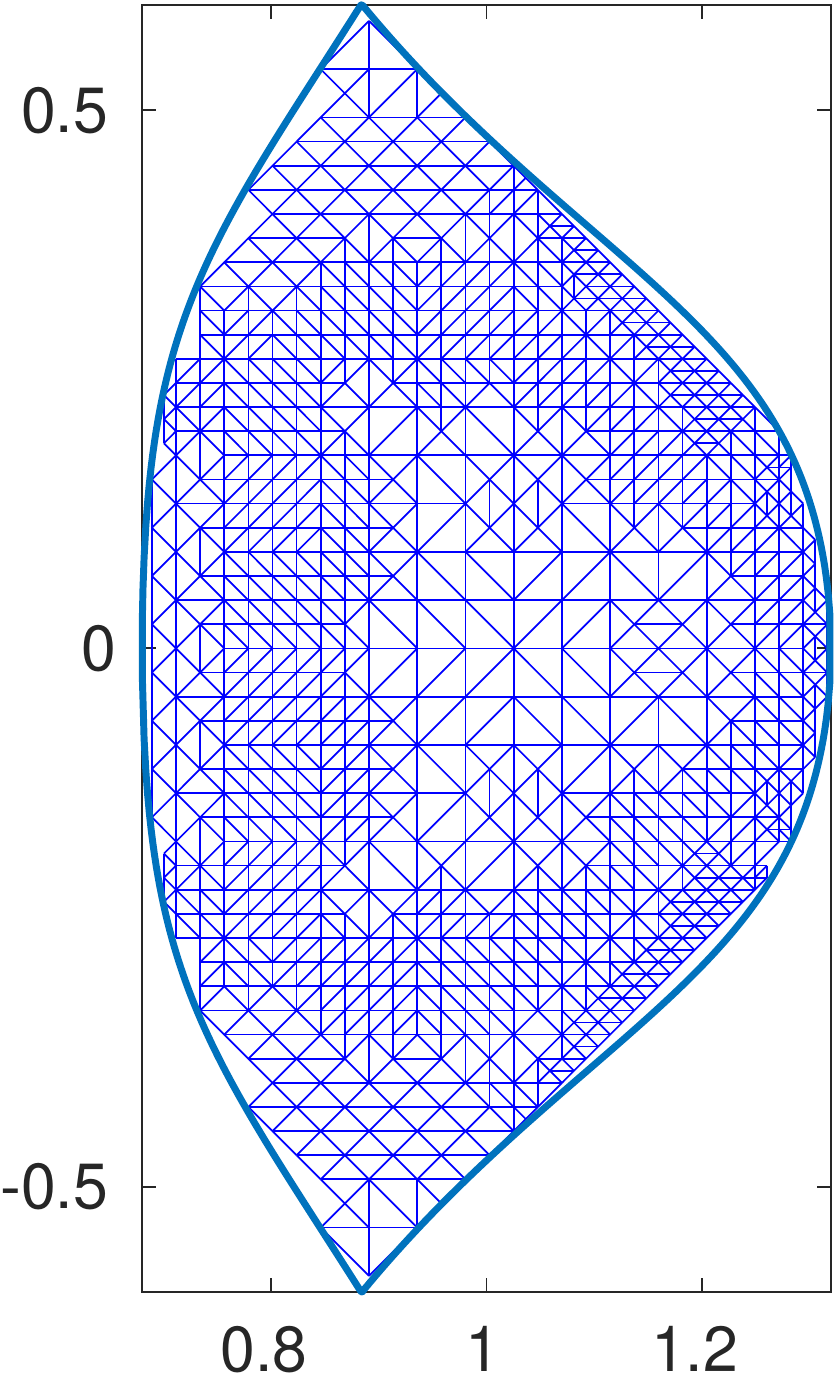} &
\includegraphics[height=0.2\linewidth]{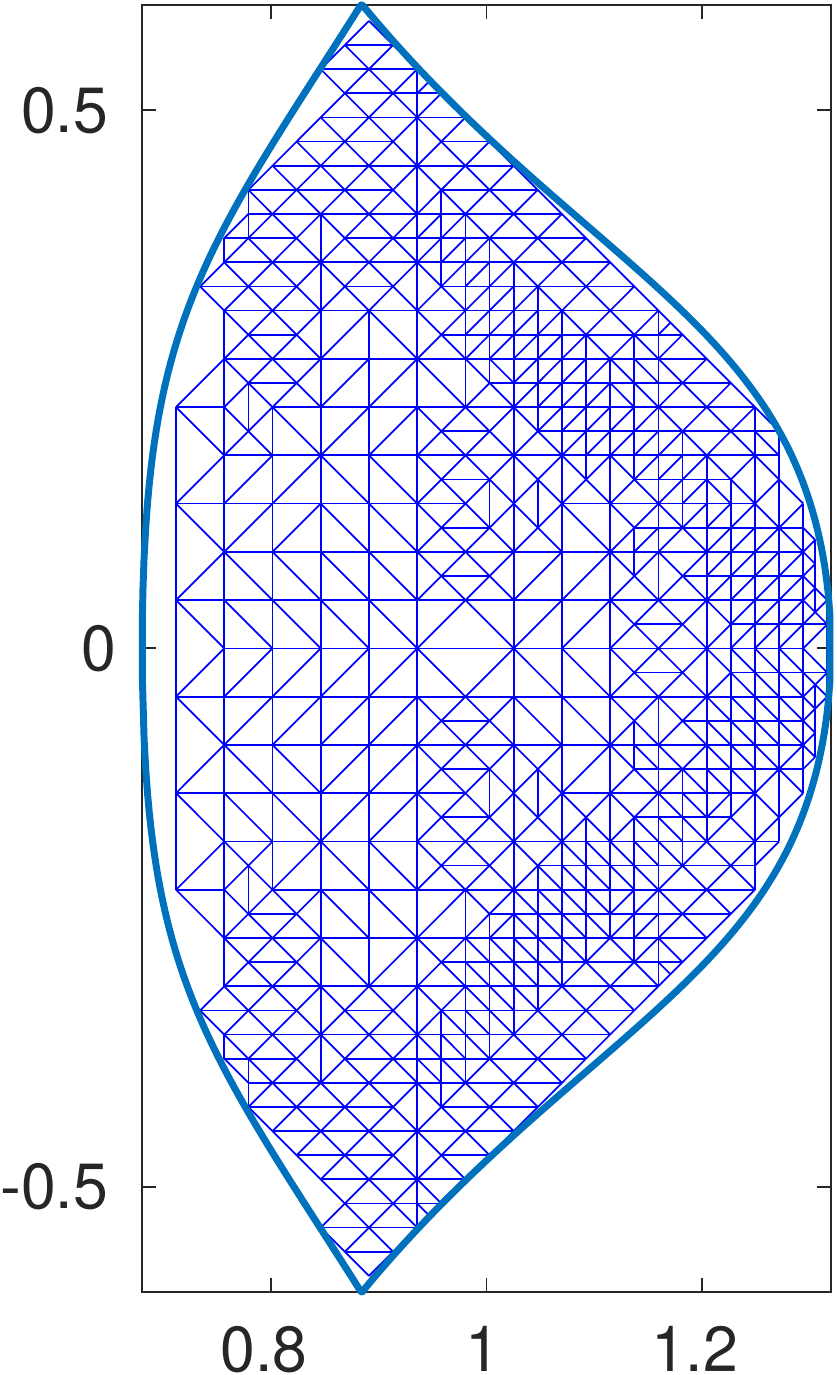} & 
\includegraphics[height=0.2\linewidth] {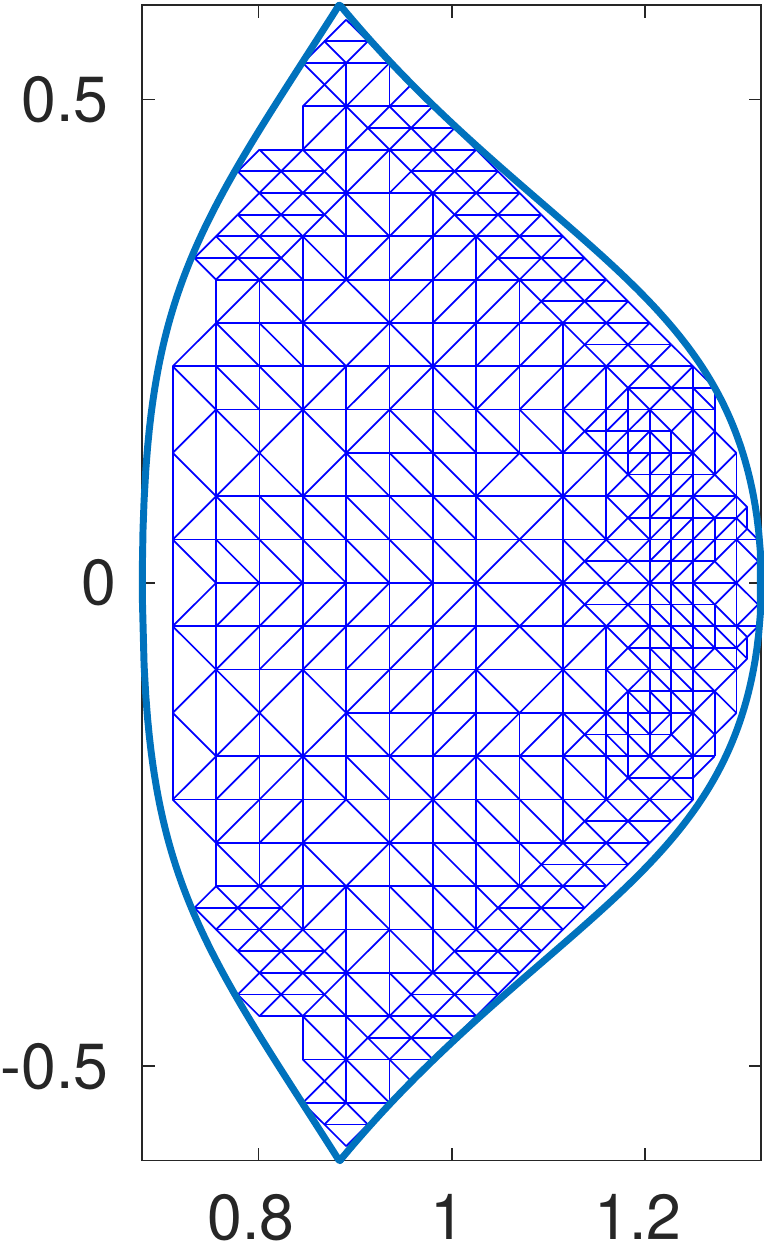} & 
\includegraphics[height=0.2\linewidth]
{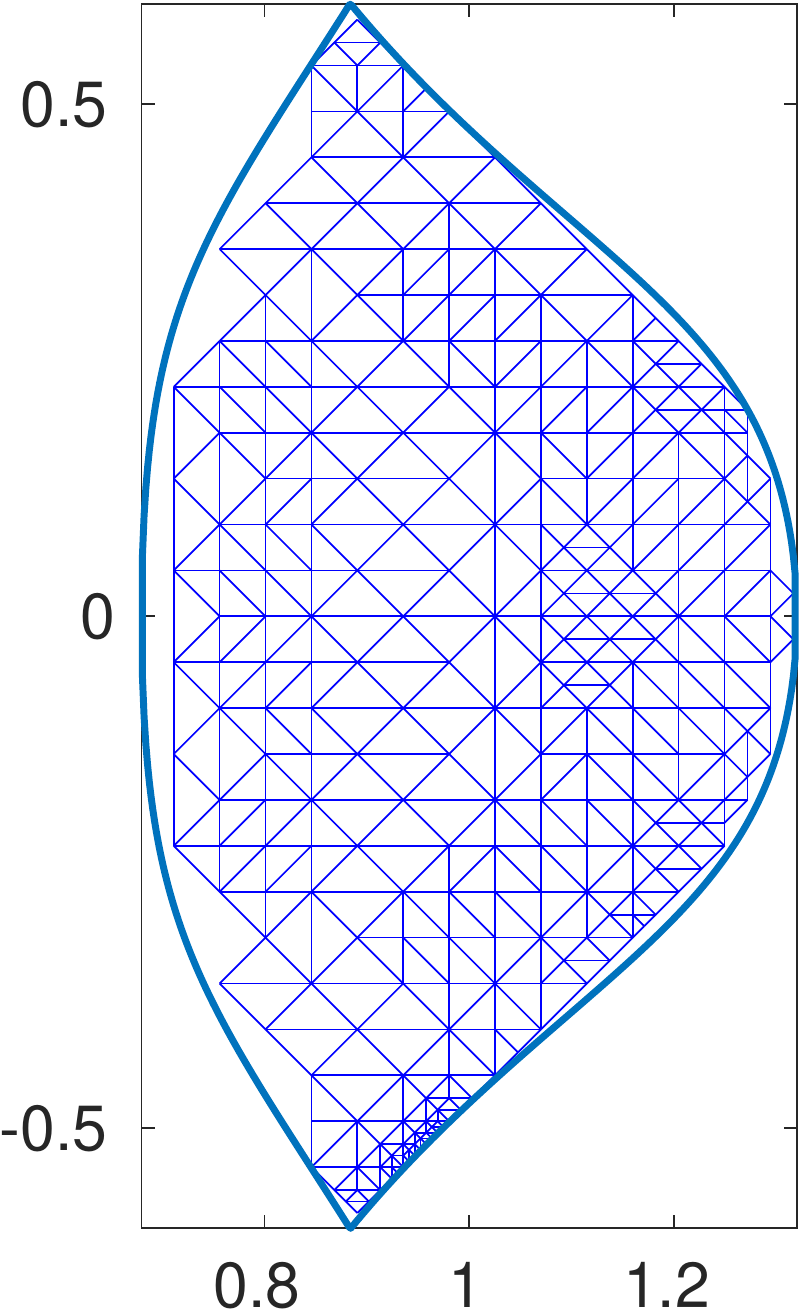}
\end{tabular}}
\caption{Left: Convergence history of the error estimator for the transport barrier after six levels of refinement with $\gamma=0.5$ for basis functions of degrees from $k=1$ to $k=4$. All the computations were carried out starting with the initial mesh displayed to the right of the convergence plot. The final computational grids after six levels of adaptive refinement for $k=1,\ldots,4$ are shown on the right.}\label{fig:TransportConvergence}
\end{figure}

Figure \ref{fig:TransportSimulations} shows the numerical solution for a transport barrier with parameter values $H=0.5$, $s = 40$, $\psi_0=0.3$, $a=4$, and $b=2$, corresponding to the steepest barrier depicted in Figure \ref{fig:TransportBarrier}. Plots of the pressure distribution and the source term in the confinement geometry are shown side by side. The geometry for this experiment corresponds to an up-down symmetric ITER-like configuration with two magnetic X-points \cite{CeFr:2010}. The solution displayed in the figure was computed adaptively using piecewise cubic polynomials. The initial mesh consisted of 129 elements, and the final grid consisted of 514 elements with a ratio $h_{max}=8.97\times 10^{-2}$ and $h_{min}=7.9\times 10^{-3}$. Both meshes can be seen in Figure \ref{fig:TransportConvergence}. 

Figure \ref{fig:TransportConvergence} shows the convergence of the global error estimator for basis functions of degrees varying from 1 to 4. All the experiments started on the initial mesh shown in Figure \ref{fig:TransportConvergence}, which subsequently underwent six levels of refinement with maximum marking using $\gamma=0.5$ (i.e. only those elements whose local contribution is at least 50\% of the maximum value of $\eta_k$ are refined at every level). Note that at the finest level of refinement all the examples in the figure have roughly the same number of degrees of freedom; however, as is expected, higher order discretizations result in coarser grids (i.e. consisting of fewer elements) and considerably smaller error. 

%
\subsection{A current hole}
Another challenging configuration of physical relevance is the family of equilibria with a so-called ``current hole". This name refers to equilibria for which there is an extended region in the plasma core where the toroidal current $J_\phi$ is nearly zero \cite{FujitaEtAl:2001,HaJaSt02,GoLe:2009,GoLeNe:2007,Fujita_2010}. The ``near absence" of the current (thus the name ``hole") corresponds to a core region with almost constant pressure. Such equilibria have been observed experimentally in several tokamaks, and are found to be remarkably robust \cite{Fujita_2010}. They are likely to naturally occur in future large scale tokamaks in fully non-inductive current drive operation. They have therefore gathered significant interest in recent years \cite{Fujita_2010} (and references therein).

\begin{figure}\centering\scalebox{.95}{
\begin{tabular}{ccccc}
\kern-2.5em {\small Source term} & \kern-1.5em{\small Source term and} $\psi^*_h$ {\small (cross sections)}  & \kern-2.5em $\psi^*_h$  & \kern-1em{\small Computational Mesh} &  \\
 \kern-1em \includegraphics[height=0.2\linewidth]{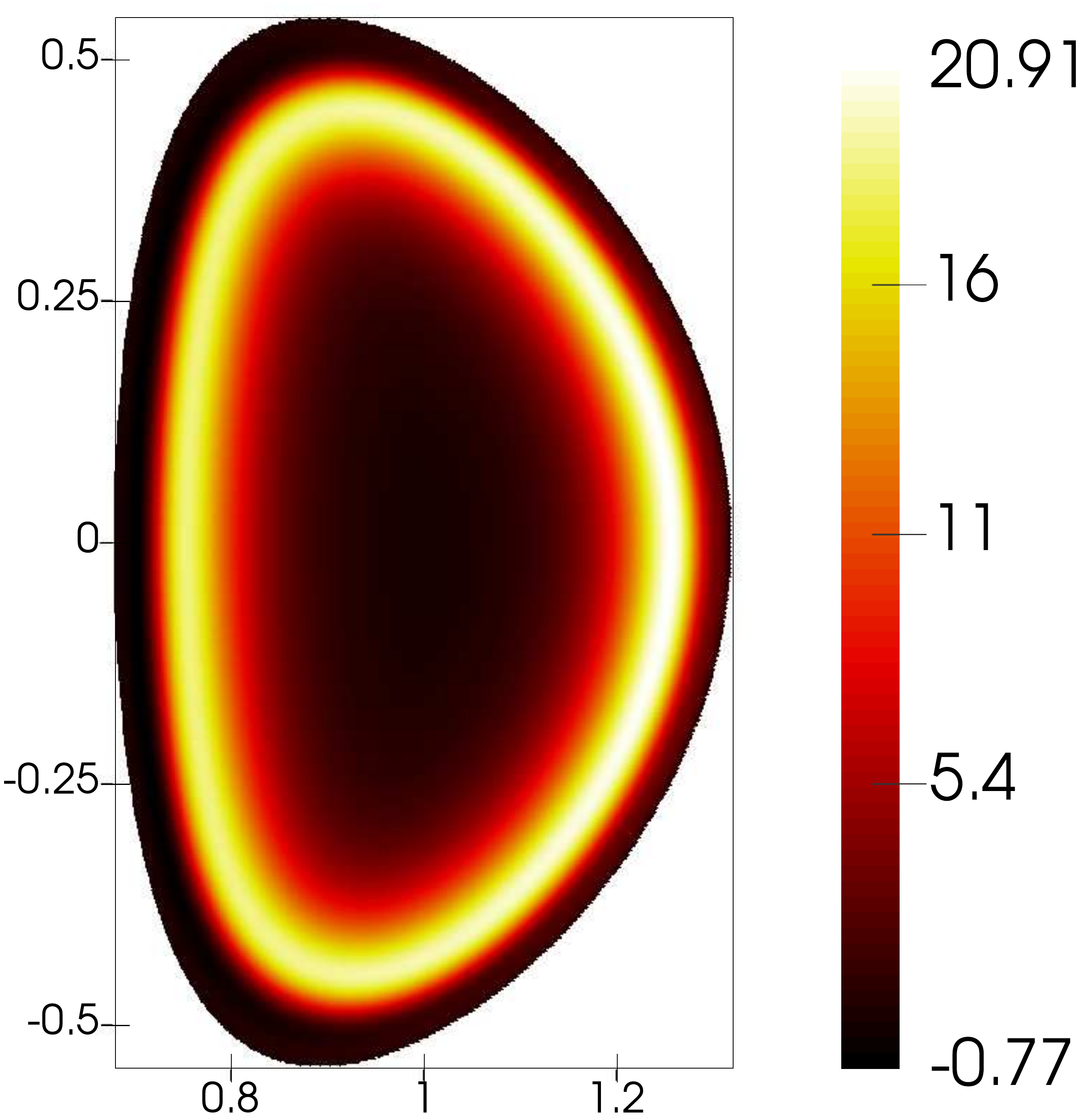} & \kern-2.5em \includegraphics[height=0.2\linewidth]{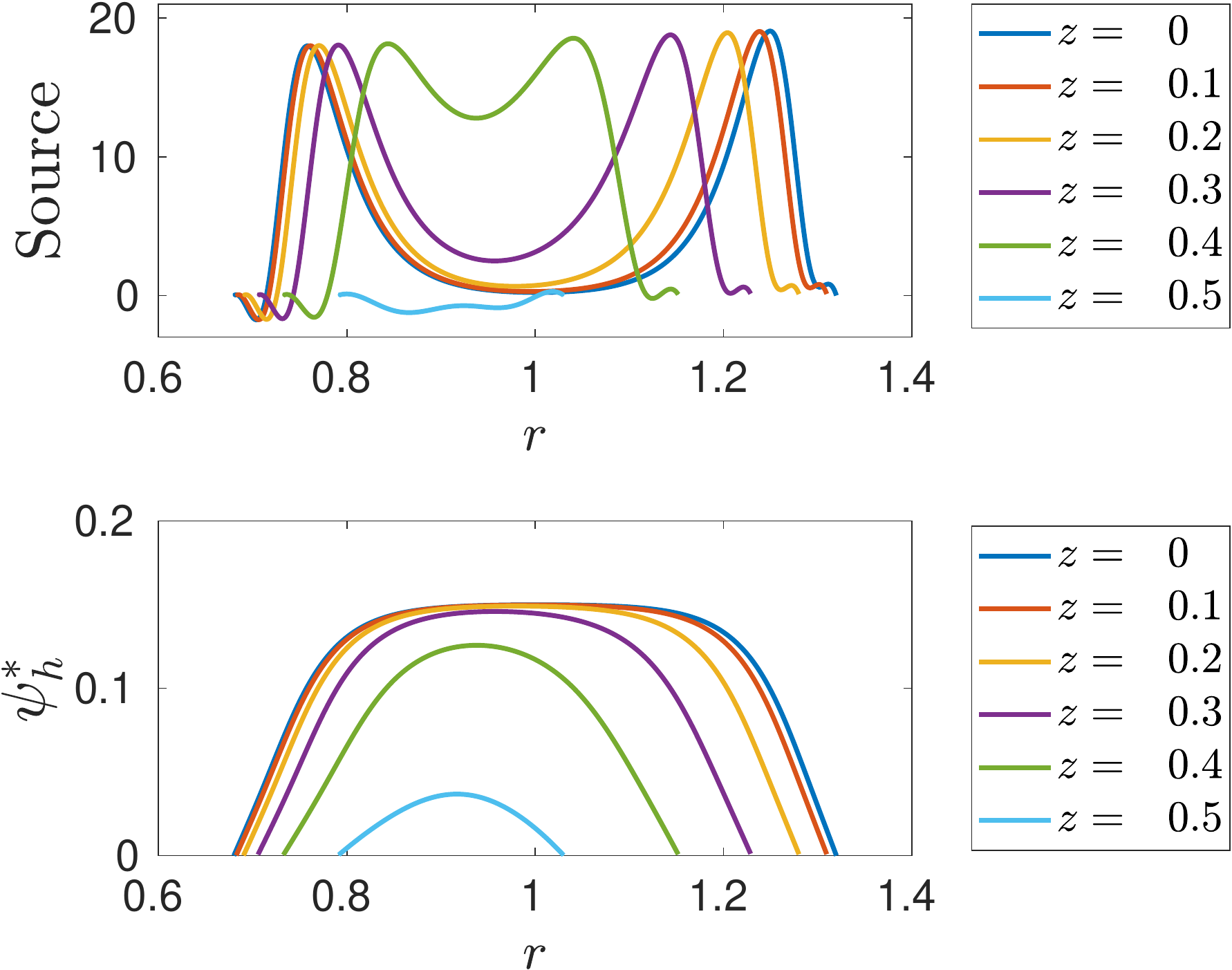} &  \kern-2em \includegraphics[height=0.2\linewidth]{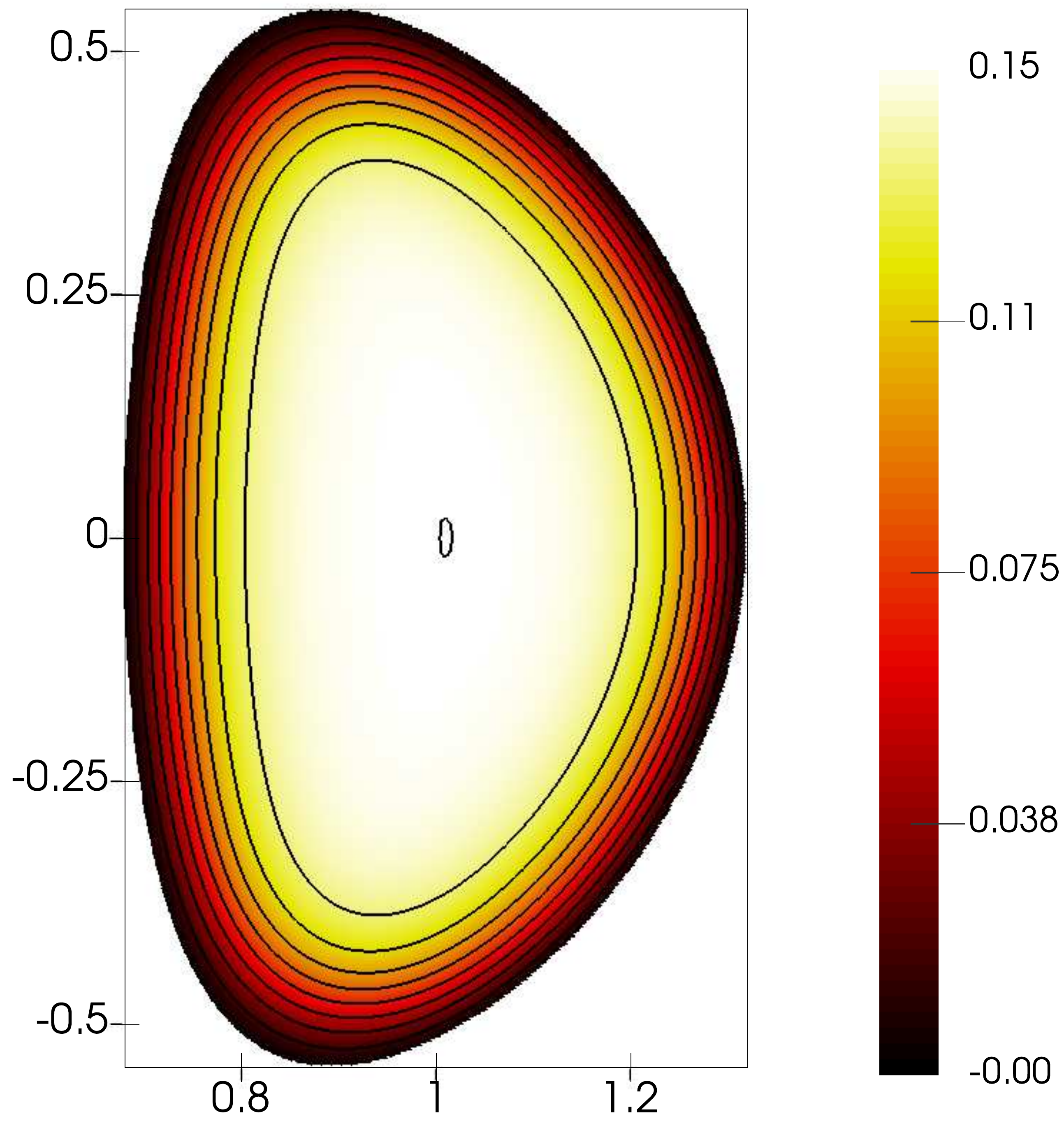}  & \kern-2em \includegraphics[height=0.2\linewidth]{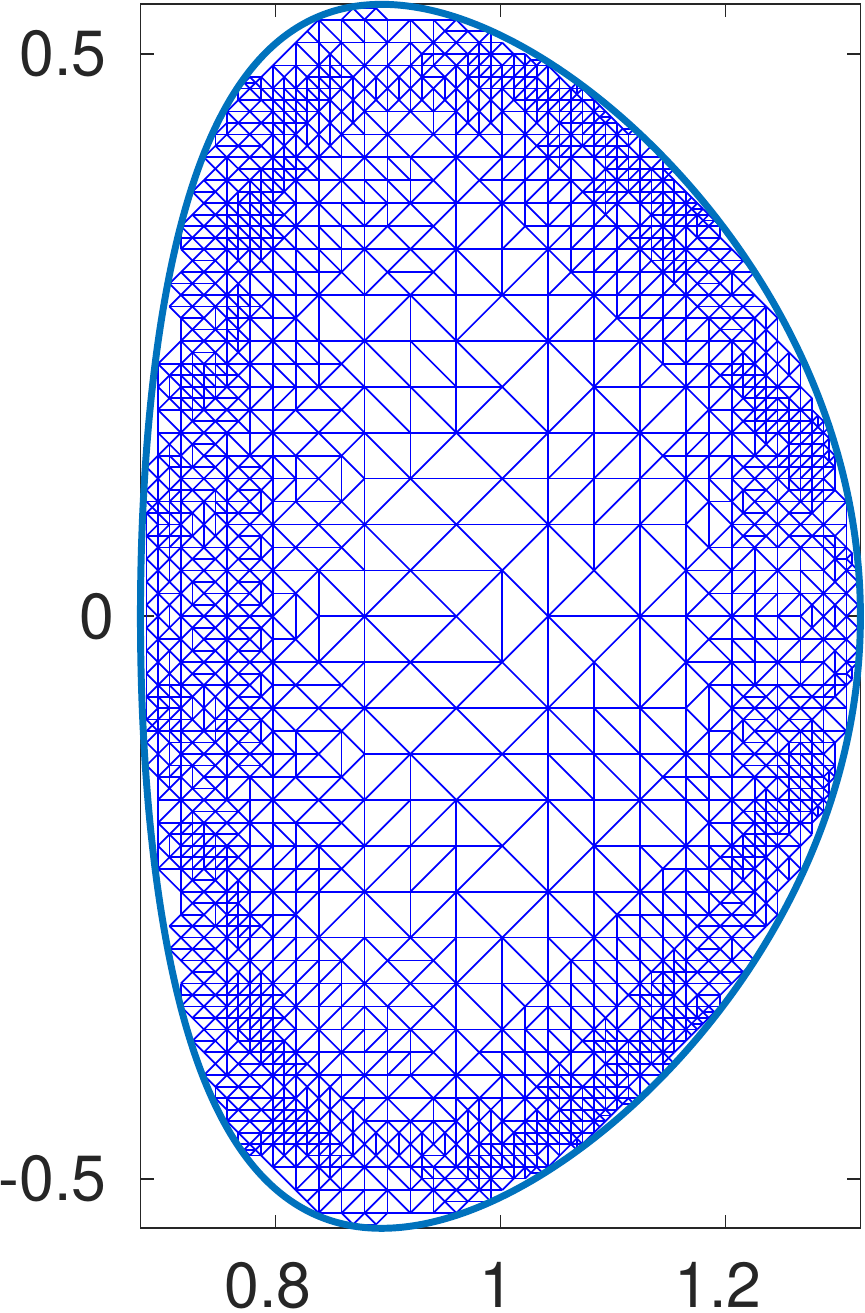}    & \kern-2em \includegraphics[height=0.2\linewidth]{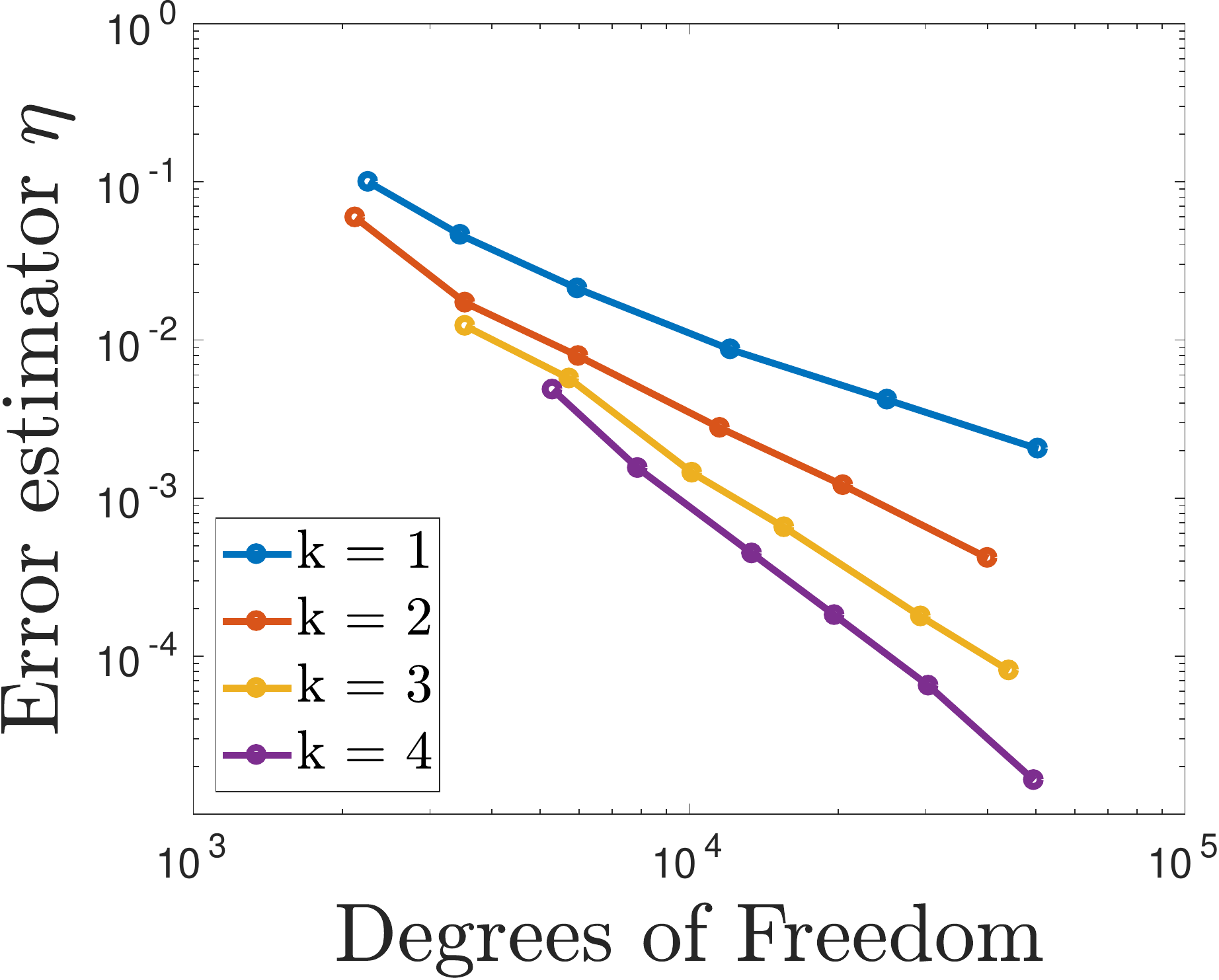} 
\end{tabular}}
\caption{Source term for an equilibrium with a current hole, given by  \eqref{eq:CurrentHoleSource} (first and top-second panels). This gives rise to mesa-like magnetic flux function (second-bottom and third panels). The refinement is automatically driven towards the boundary (fourth panel). The solution is up-down symmetric and only cross sections for the upper half are plotted. The evaluation of the source term was done using the post-processed approximation $\psi_h^*$.}\label{fig:CurrentHole}
\end{figure}

The following source term, adapted from that of the pressure pedestal, can give rise to such an equilibrium
\begin{equation}\label{eq:CurrentHoleSource}
F(r,\psi) = 2r^2\psi\left(c_2(1-e^{-(\psi/\sigma_1)^2}) + \frac{1}{\sigma_1^2}(c_1+c_2\psi^2)e^{-(\psi/\sigma_1)^2}\right) + c_3(1-e^{-(\psi/\sigma_2)^2})\cos{(c_4\psi)}.
\end{equation}
Choosing the values $c_1 = 0.4, c_2 = 0.1, c3=-18, c_4=10\pi$ as well as $\sigma^2_1 = 5\times 10^{-3}$ and $\sigma^2_2=3\times 10^{-3}$, the spatial distribution of the source term will be as shown in Figure \ref{fig:CurrentHole} for an up-down symmetric D-shaped geometry with ITER-like parameters (as described in \cite{PaCeFrGrOn:2013}). Recalling the relationship $\mu_{0}J_{\phi}=F(r,\psi)/r$, we see that the current drops to close to zero in the core of the confinement region and has sharp peaks near the boundary. As can be seen in the figure, there is a sharp contrast between the almost constant behavior of both the source and the solution in the central region and the large gradients at the edge. Consistently, the estimator focuses the computational effort on the edges and keeps a relatively coarse mesh in the core. The computation was ran over six levels of refinement with marking parameter $\gamma = 0.3$; for a polynomial basis of degree four the resulting final mesh shown in the figure consisted of 3310 elements where $h_{min}=5.1\times 10^{-3}$ and $h_{max}=8.16\times 10^{-2}$.

\begin{figure}\centering\scalebox{0.825}{
\begin{tabular}{cccccc}
$\partial_r\psi_h$ &  \!\!\! {\small Cross sections} &  \!\!\! $\partial_z\psi_h$  &   \!\!\!\!\!\! $\partial_{rr}\psi_h$ &  \!\!\! \!\!\! {\small Cross sections} &  \!\!\! \!\!\! $\partial_{zz}\psi_h$ \\
 \includegraphics[height=0.2\linewidth]{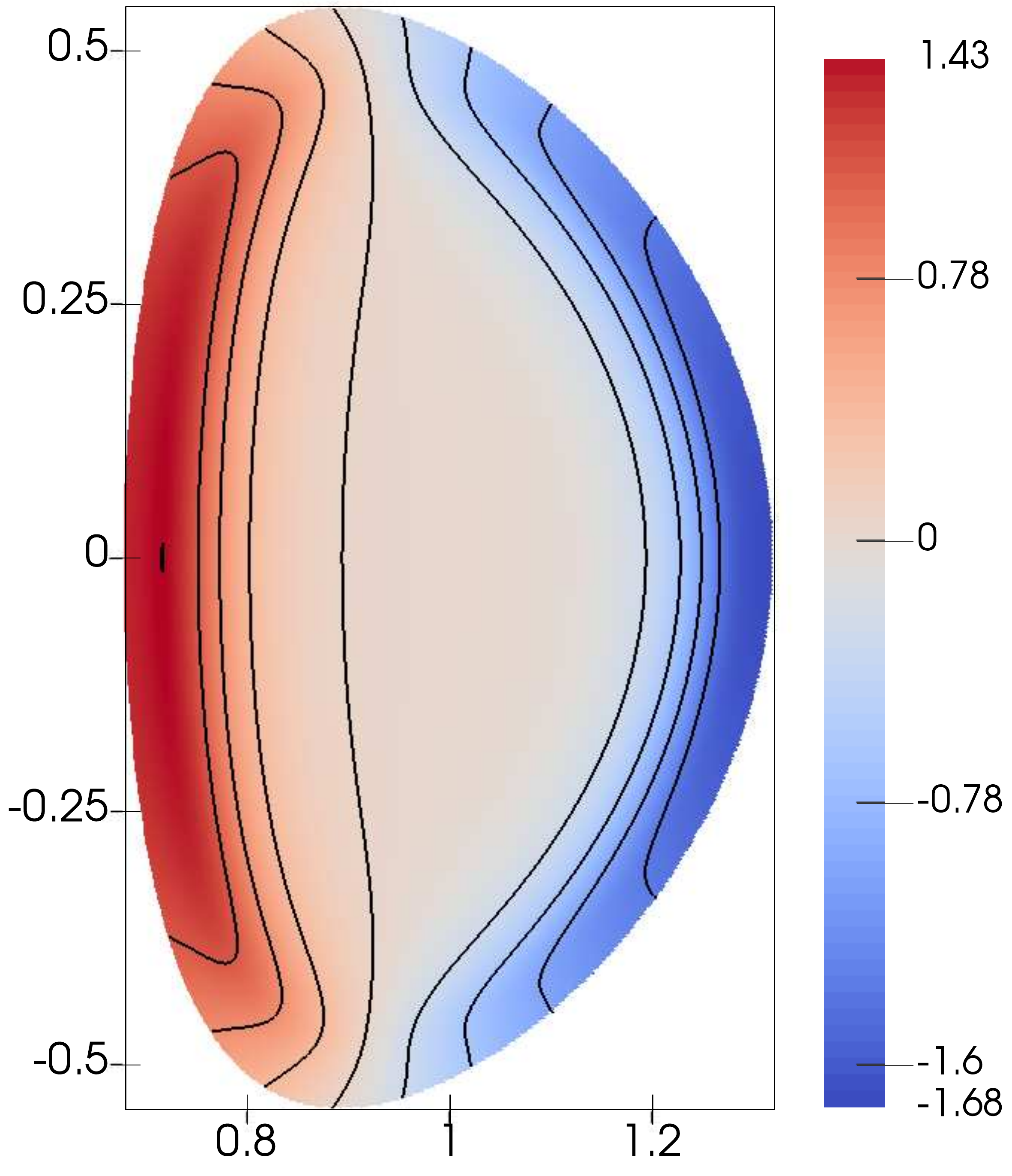} & \!\!\!\!\!\!\!\! \includegraphics[height=0.2\linewidth]{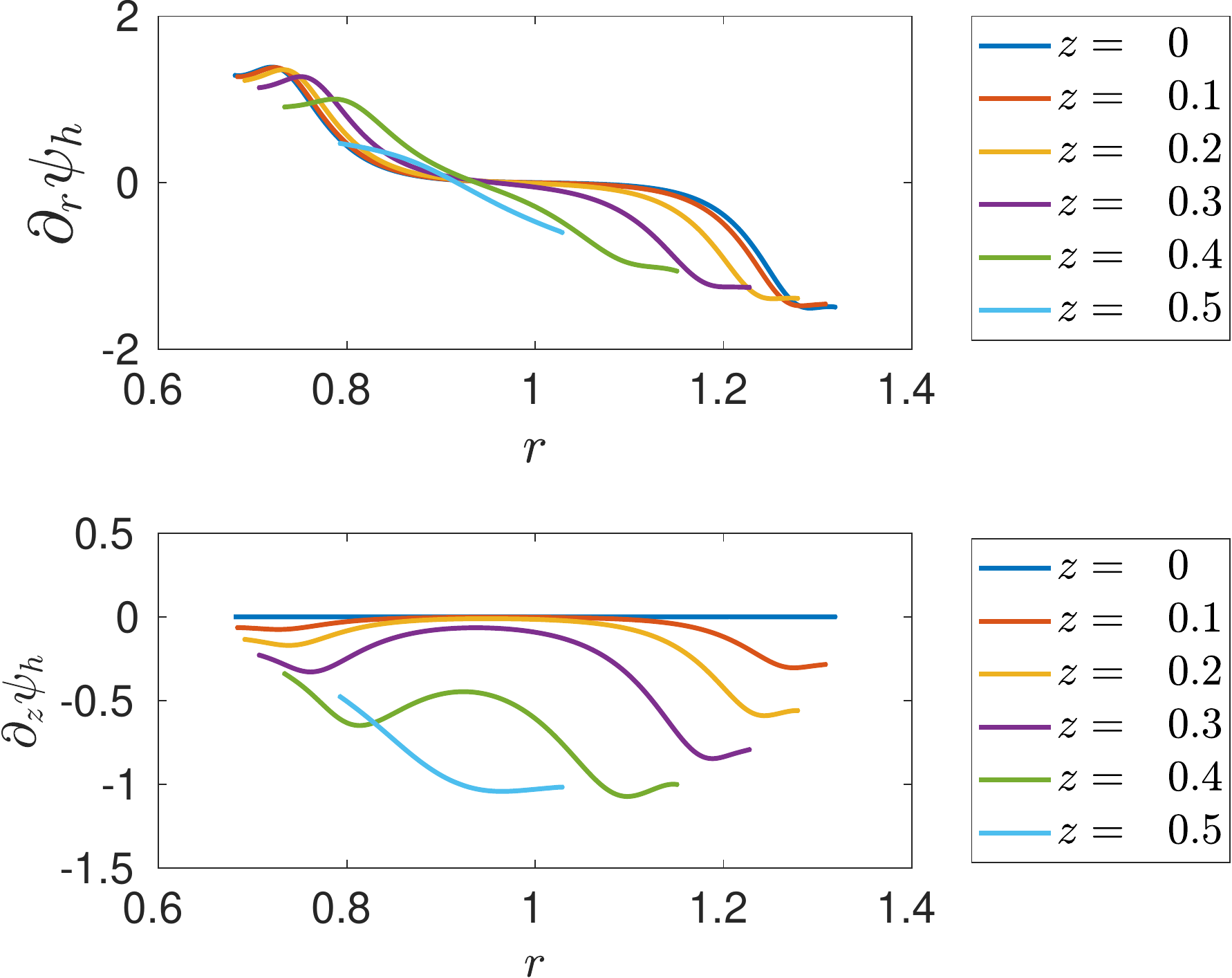} & \!\!\!\!\!\!\!\!  \includegraphics[height=0.2\linewidth]{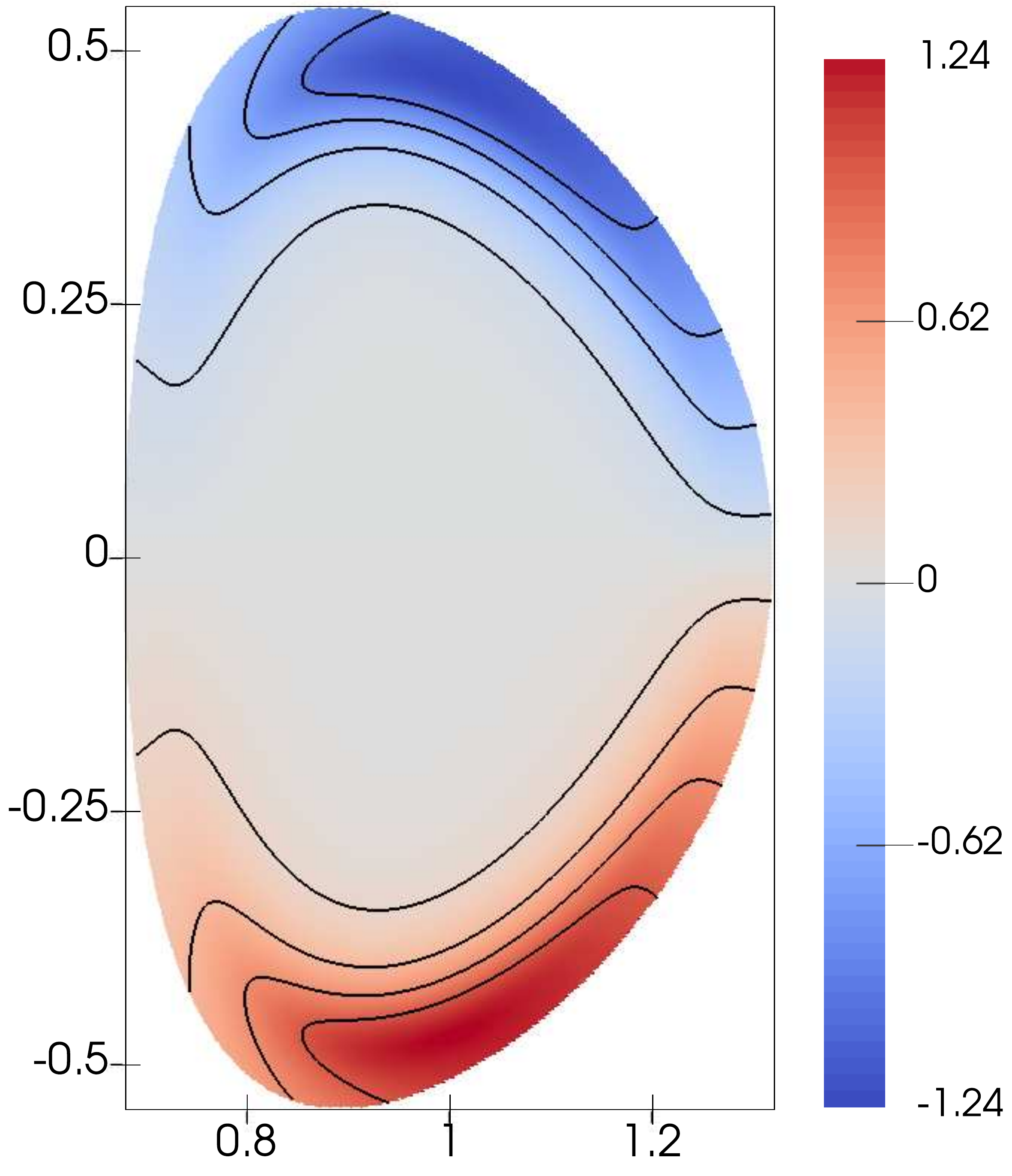}  & \!\!\!\!\!\!\!\! \includegraphics[height=0.2\linewidth]{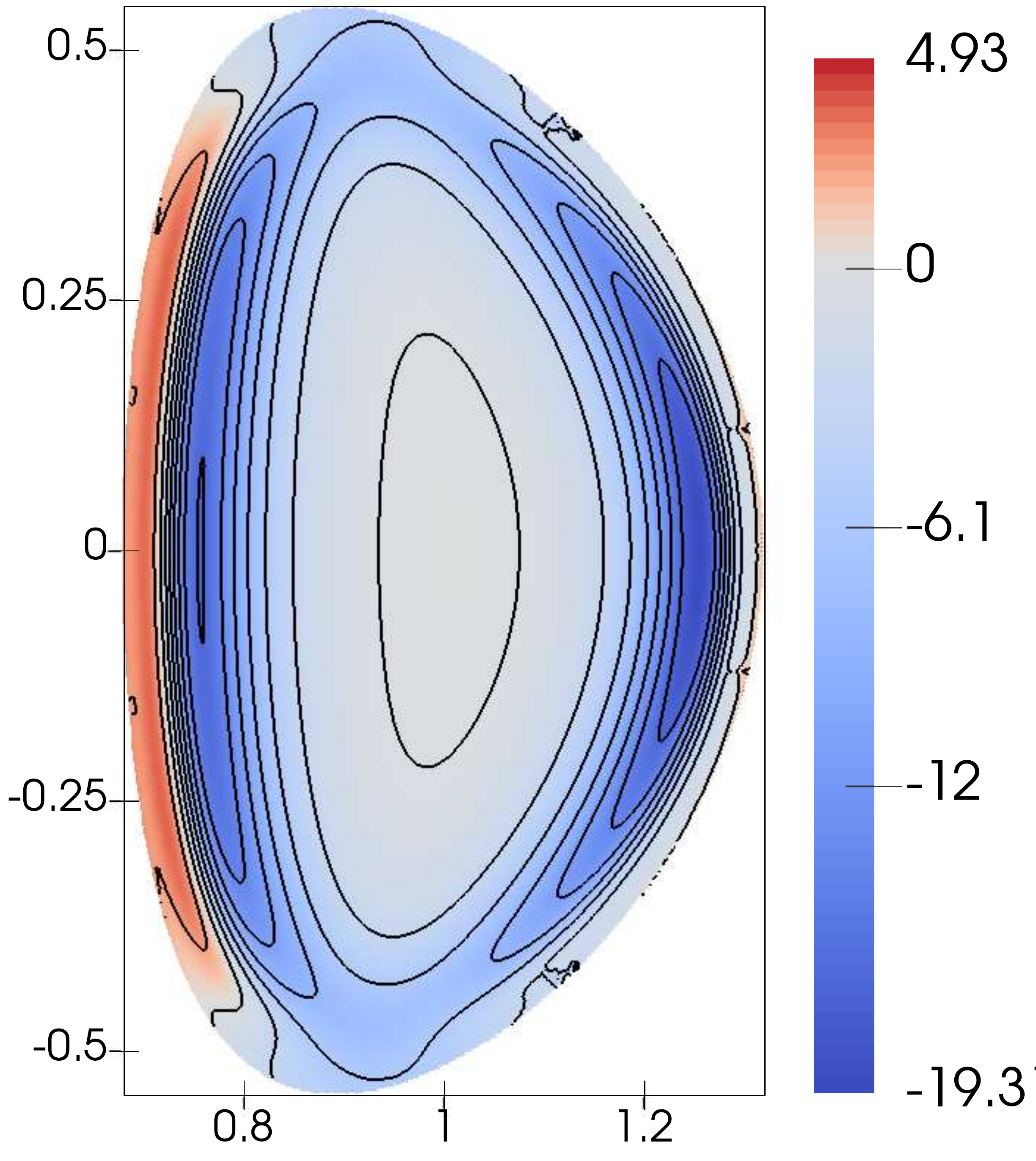} & \!\!\!\!\!\!\!\!\! \includegraphics[height=0.2\linewidth]{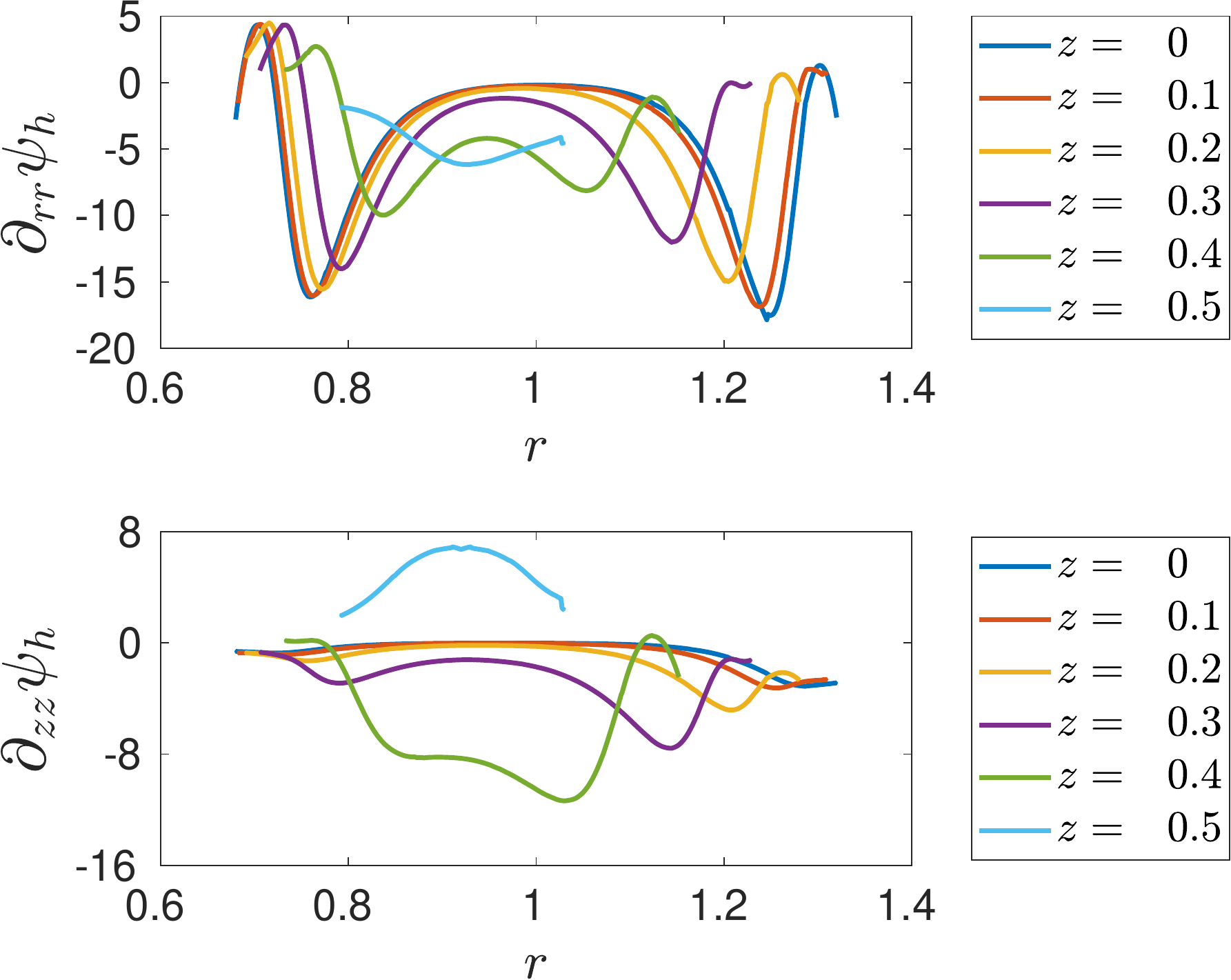} & \!\!\!\!\!\!\!\! \includegraphics[height=0.2\linewidth]{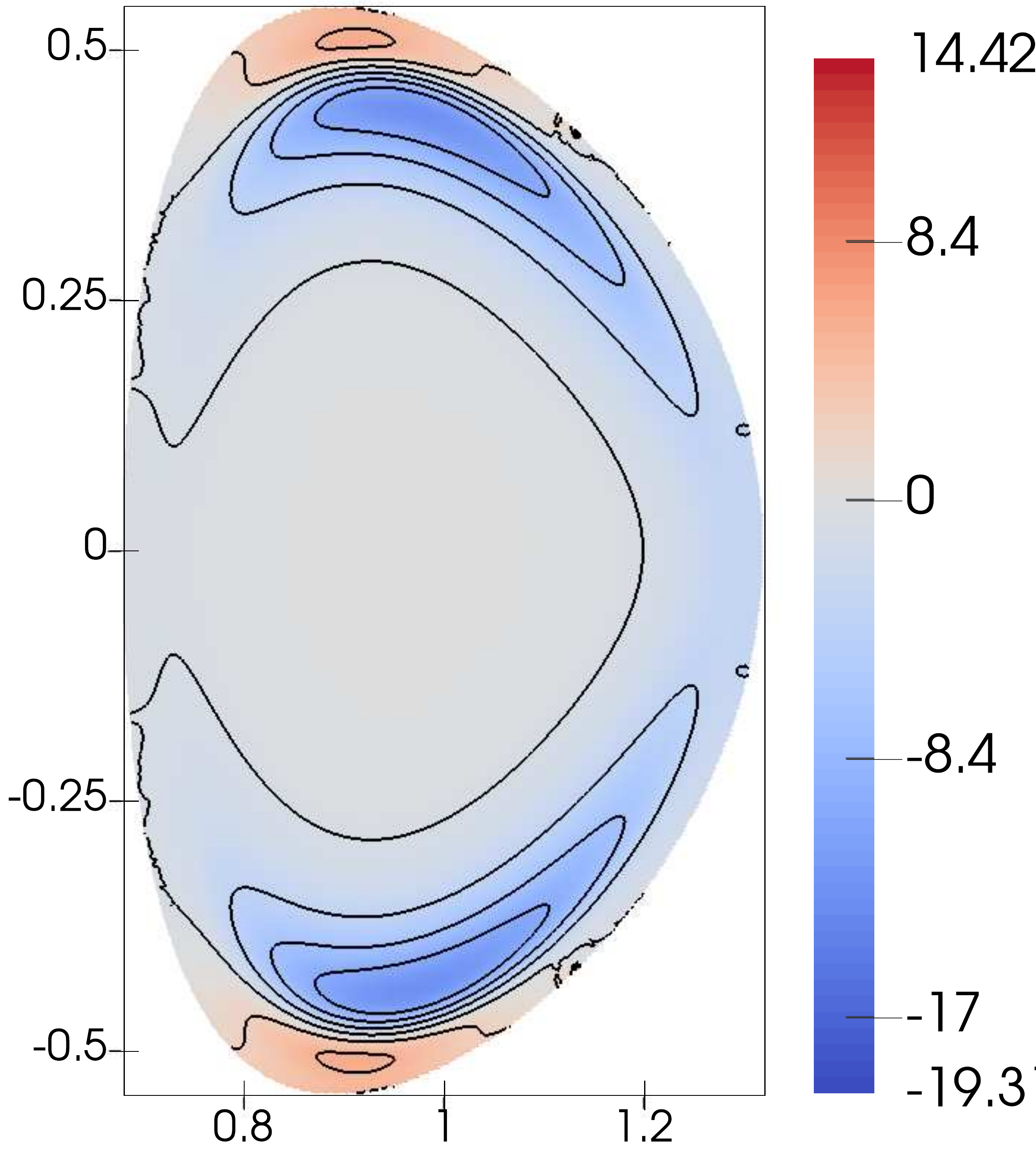} 
\end{tabular}}
\caption{Approximate first (left three panels) and second (rightmost three panels) partial derivatives  of $\psi_h$ for the equilibrium with the current hole, whose source term is given by given by  \eqref{eq:CurrentHoleSource}. Because of the up-down symmetry of the geometry, the cross sections are plotted only for values of $z$ ranging from 0 to 0.5.}\label{fig:CurrentHoleDerivatives}. 
\end{figure}

Figure \ref{fig:CurrentHoleDerivatives} shows the partial derivatives of $\psi_h$ obtained directly from the components of the flux $\boldsymbol q_h$ and their cross sections for different values of $z$. The second derivatives were computed by differentiating the local polynomial approximants of $\partial_r\psi_h$ and $\partial_z\psi_h$, which introduces additional error. However, the high polynomial order in combination with the focused mesh refinement imply that, even with the expected deterioration, the approximation remains within an acceptable range. 

%
\subsection{An internal layer}
\begin{figure}\centering\scalebox{.925}{
\begin{tabular}{ccccc}
\!\!{\small Source term} & {\small Source term (cross section)} & {\small Computational Mesh} & $\psi^*_h$ & $\psi^*_h$ {\small (cross section)} \\
\!\!\!\!\! \includegraphics[height=0.2\linewidth]{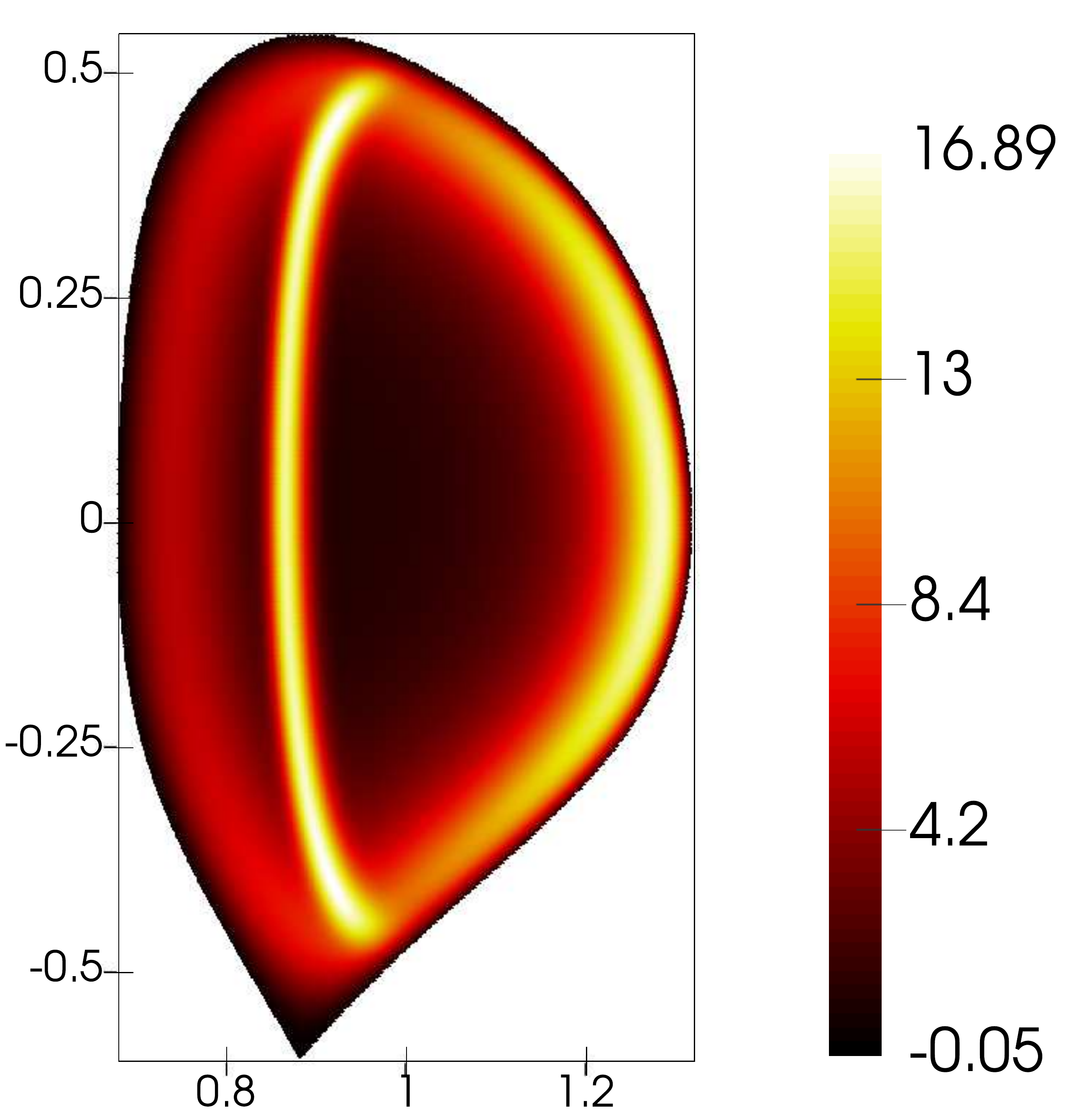} & \!\!\!\!\!\! \includegraphics[height=0.2\linewidth]{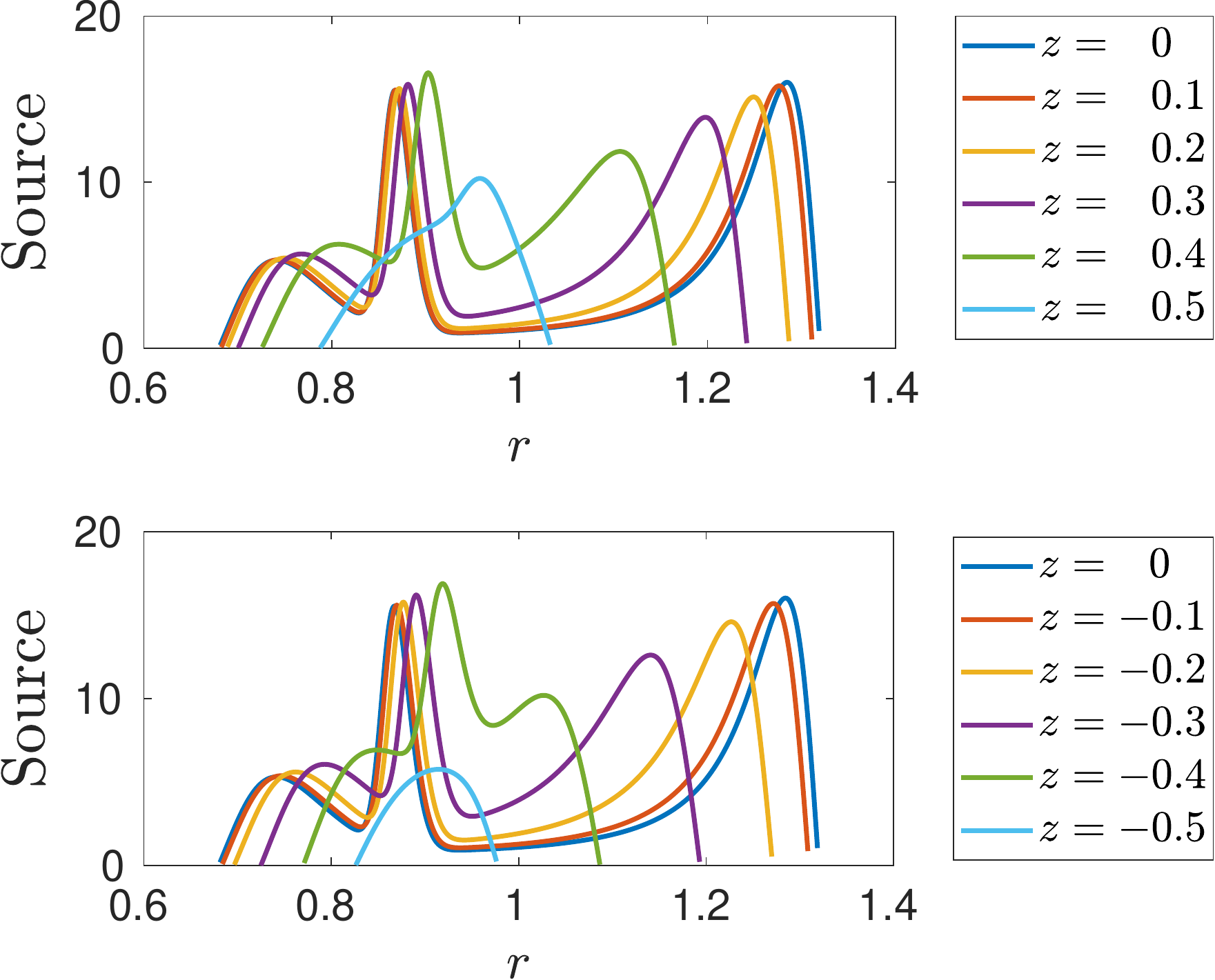} & \!\!\!\!\!\! \includegraphics[height=0.2\linewidth]{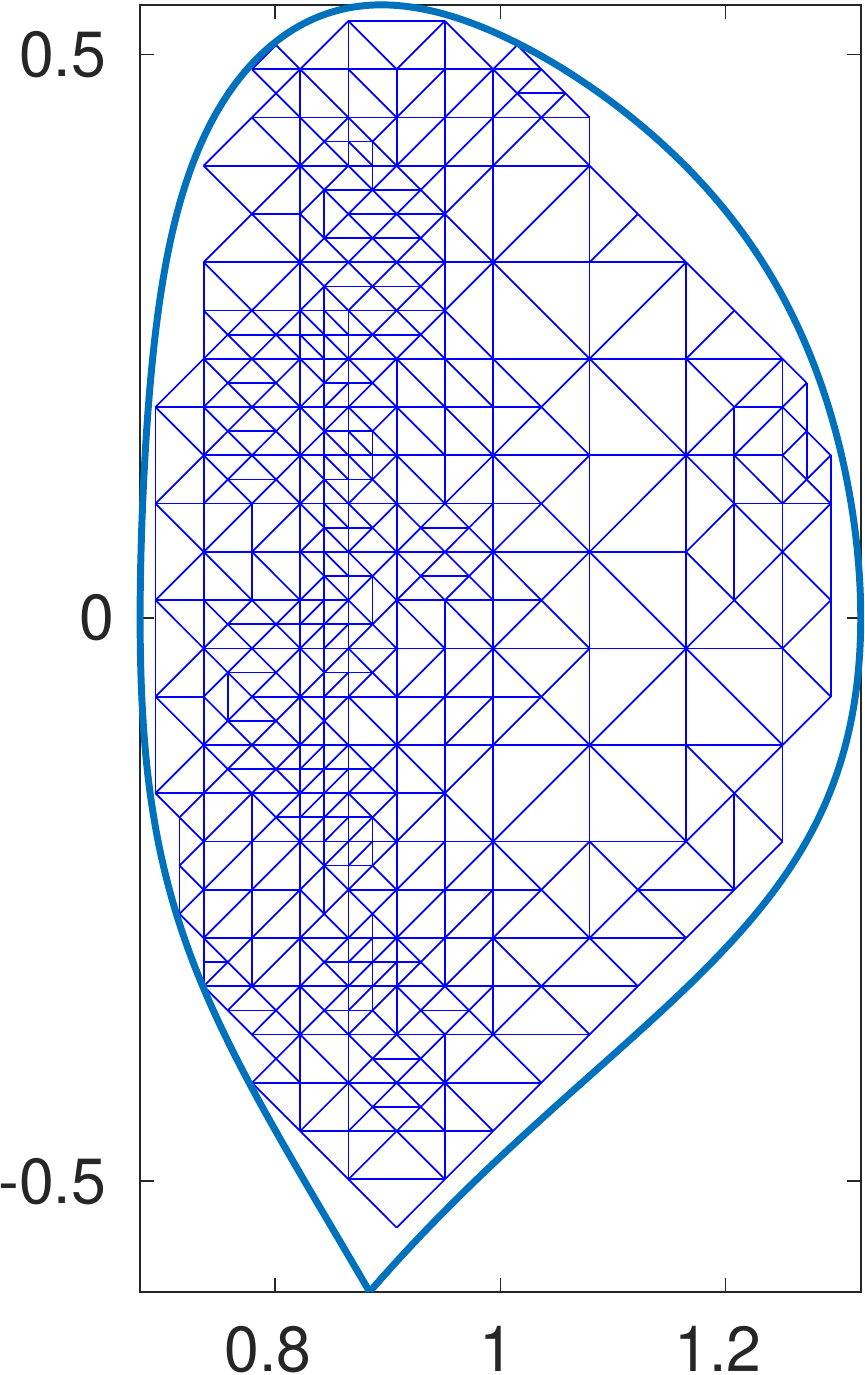} & \!\!\!\!\!\! \includegraphics[height=0.2\linewidth]{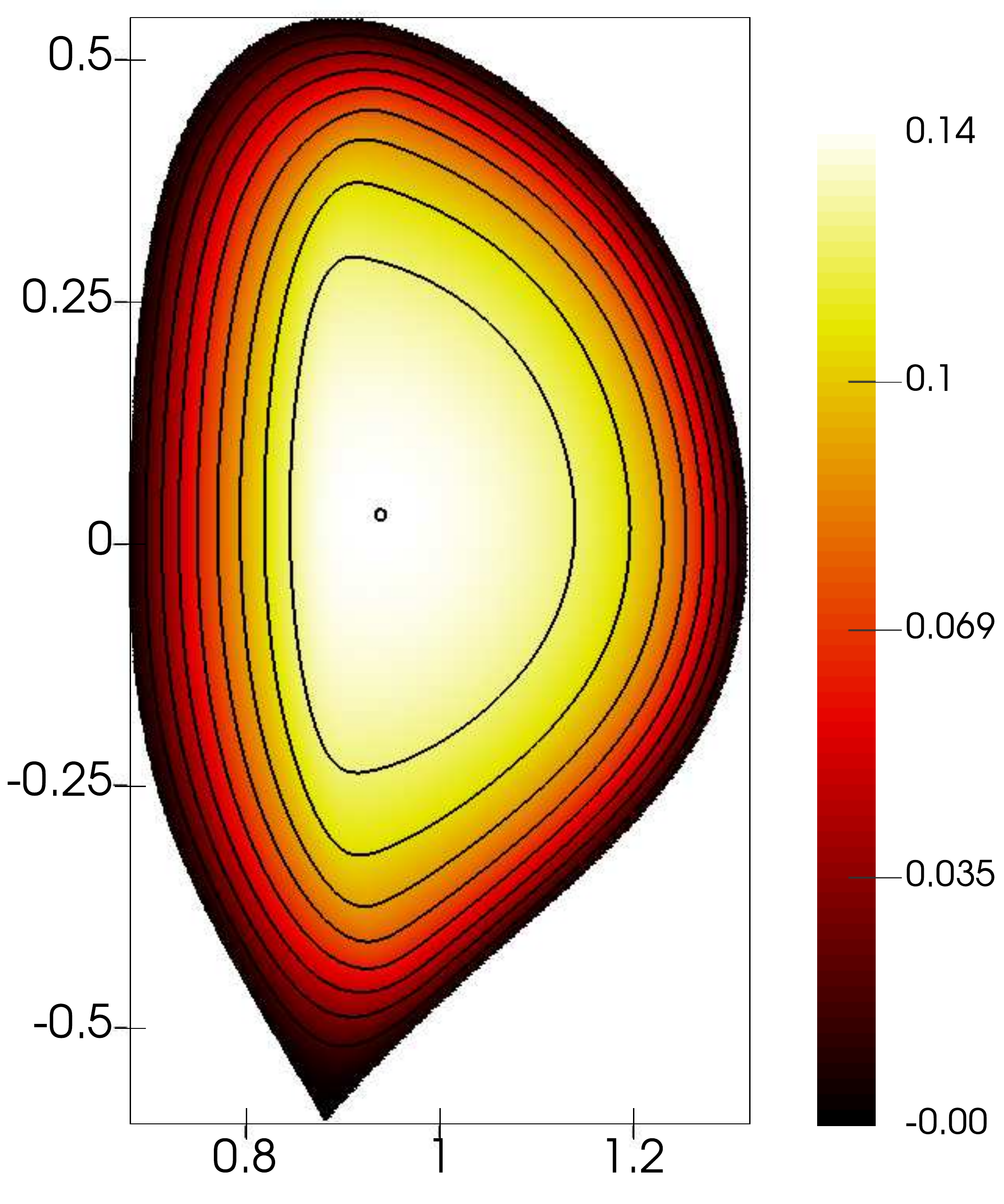} & \!\!\!\!\!\! \includegraphics[height=0.2\linewidth]{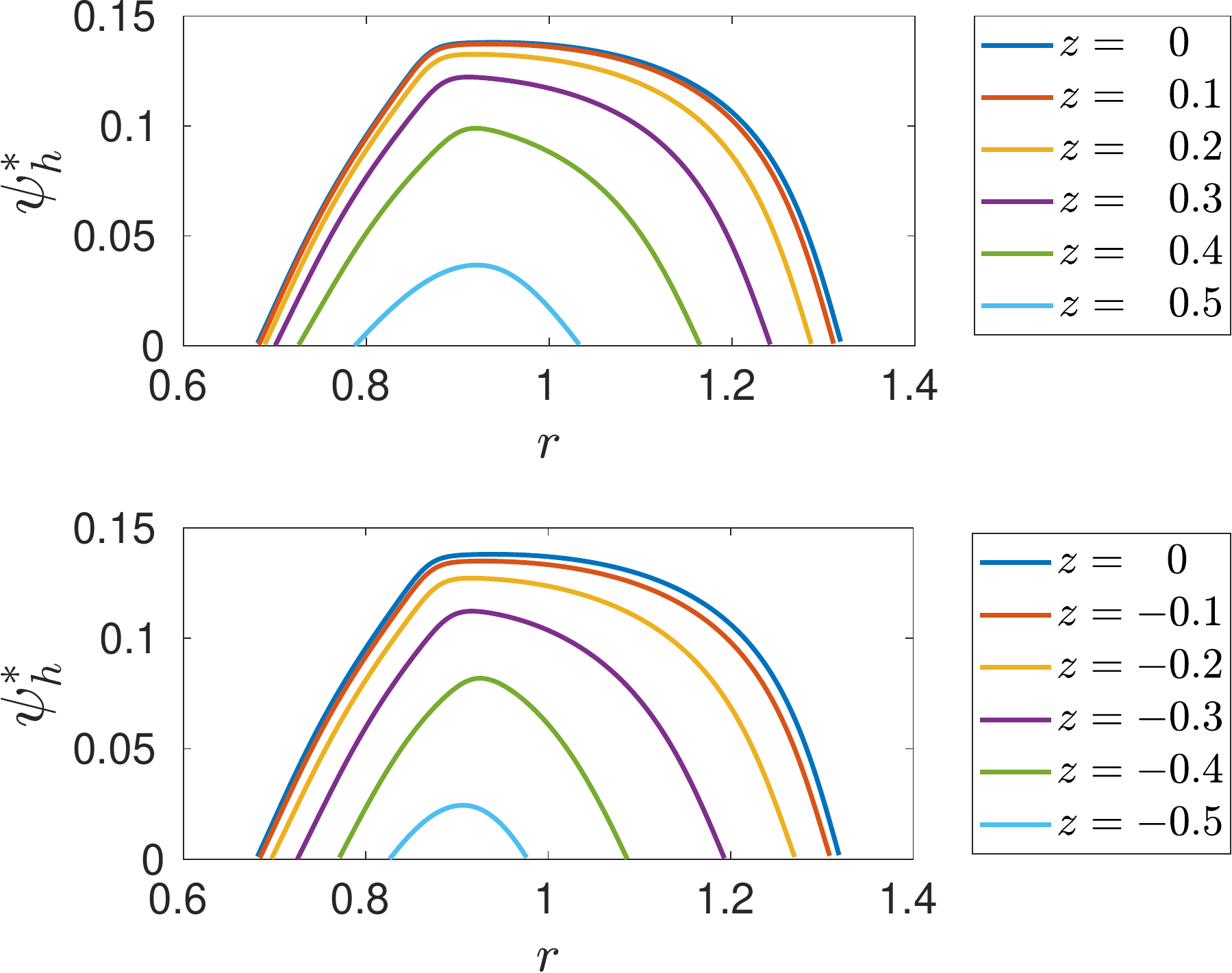}
\end{tabular}}
\caption{The combination of a pressure pedestal with highly localized internal structure of the form given by equation \eqref{eq:BarrierSource} for $\sigma^2_1 = 0.005$ and $\sigma^2_2 = 7.5\times 10^{-4}$ in an ITER-like geometry gives rise to a source term like the one displayed on the left. Cross sections of the source for $z$-values ranging from $-0.5$ to $0.5$ are shown in the center left panel. The computational mesh --shown in the center panel after six iterations-- is refined around the region of high curvature in the source. The numerical solution obtained using polynomials of degree $k=4$ and cross sections at different $z$-values are shown to the right. }\label{fig:BarrierIterX}
\end{figure}

Perhaps one of the most desirable features in an adaptive scheme is the ability to automatically detect localized features in the solution and to refine the computation locally in order to resolve them accurately. The pressure pedestal can be turned into a more challenging benchmark along these lines if the source term is modified to  
\begin{equation}\label{eq:BarrierSource}
F(r,\psi) = 2r^2\psi\left(c_2(1-e^{-(\psi/\sigma_1)^2}) + \frac{1}{\sigma_1^2}(c_1+c_2\psi^2)e^{-(\psi/\sigma_1)^2}\right) + c_3(1-e^{-(\psi/\sigma_1)^2})e^{-(1-r-\psi)^2/\sigma^2_2}.
\end{equation}
This source term is not physically relevant for magnetic confinement fusion applications, because it cannot be cast in the canonical form of the source in \eqref{eq:GradShafranov} due to the explicit appearance of the coordinate $r$ in the argument of the last exponential. Nevertheless, it represents a good benchmarking problem to test for the detection of internal layers. As can be seen in Figure \ref{fig:BarrierIterX} (left and center left) the source presents an internal layer that changes abruptly in addition to the large gradients at both edges of the confinement region. In the figure and the numerical experiment the constants were taken to be $c_1 = 0.8, c_2 = 0.2, c_3 = 15, \sigma^2_1 = 5\times 10^{-3},$ and $\sigma^2_2 = 7.5 \times 10^ {-4}$. 
The simulation parameters were as in the example with the pressure pedestal: the same ITER-like geometry with an x-point, the same starting grid, polynomial basis of degrees one to four and six levels of refinement with $\gamma=0.3$. As can be seen in the central panel of Figure \ref{fig:BarrierIterX}, the estimator successfully detects the development of internal features in the solution, and concentrates the refinement in that region. The final grid consists of only 601 elements with maximum mesh diameter $h_{max}=1.21\times 10^{-1}$ and minimum mesh diameter $h_{min}=2.14\times 10^{-2}$.

\begin{figure}\centering\scalebox{.9}{
\begin{tabular}{ccccc}
$\partial_{r}\psi_h$ & $\partial_{r}\psi_h$ {\small (cross section)} &  $\partial_{z}\psi_h$ & $\partial_{z}\psi_h$ {\small (cross section)} \\
\kern-1em\includegraphics[height=0.2\linewidth]{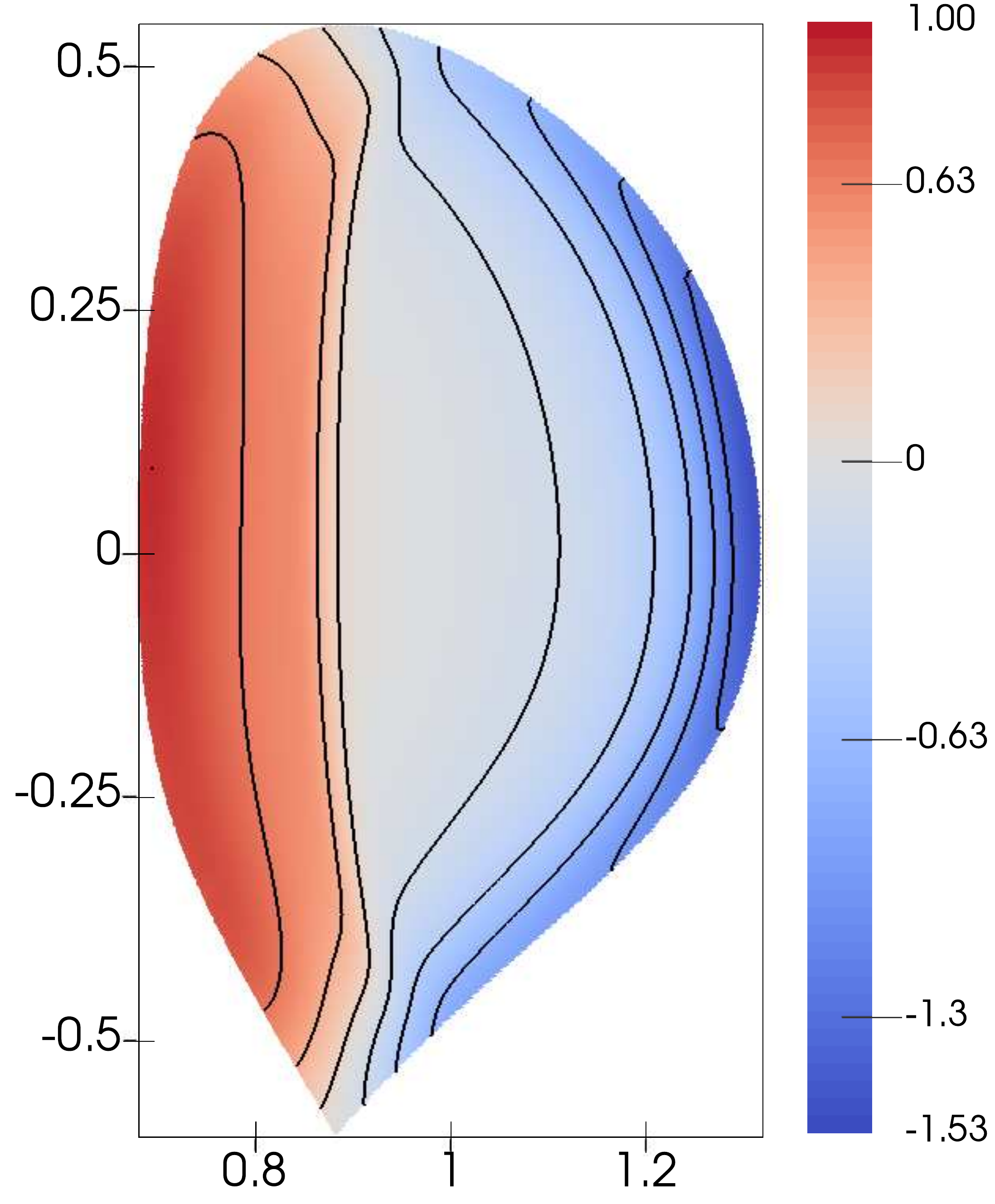} & \kern-1em \includegraphics[height=0.2\linewidth]{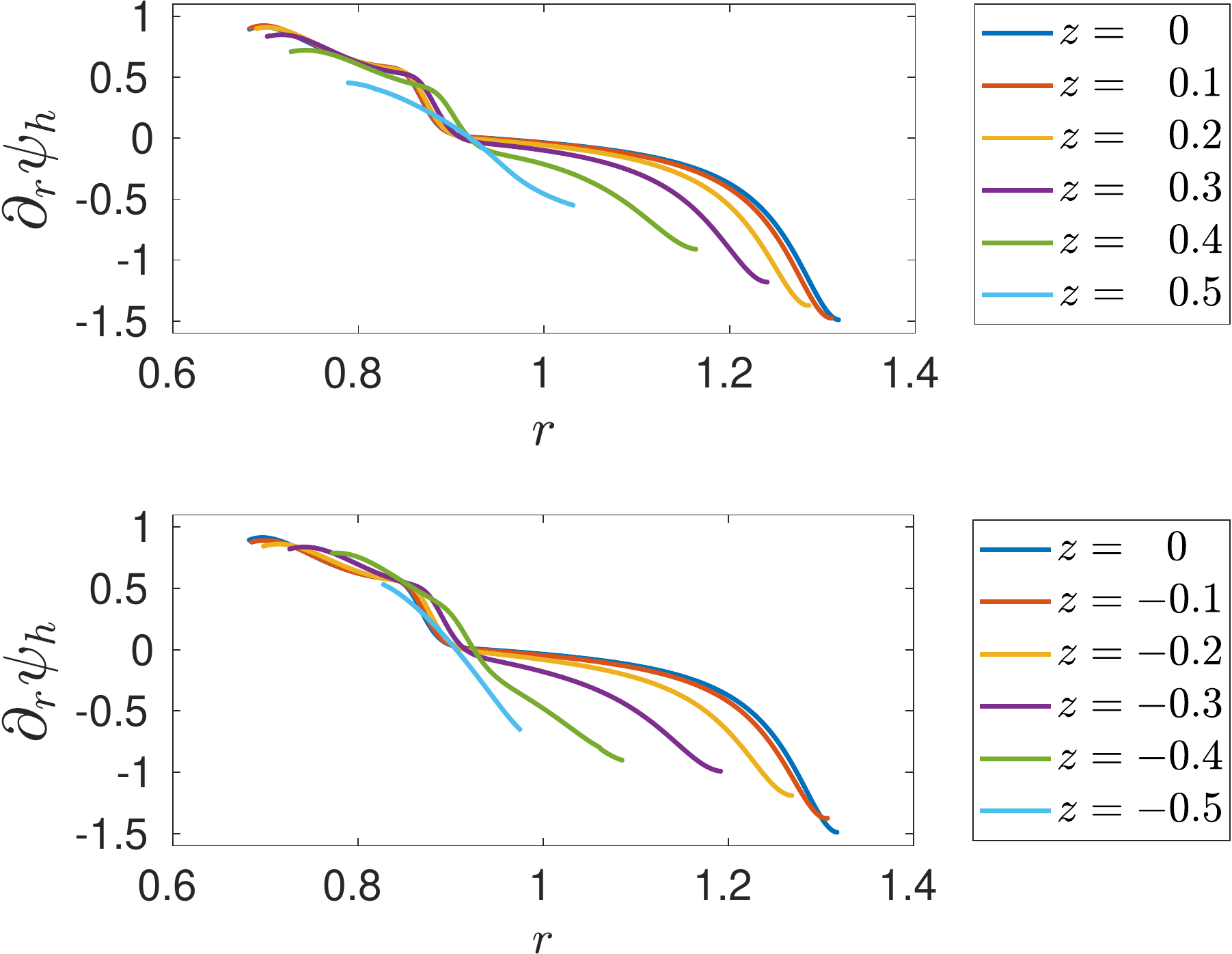} &  \kern-1em \includegraphics[height=0.2\linewidth]{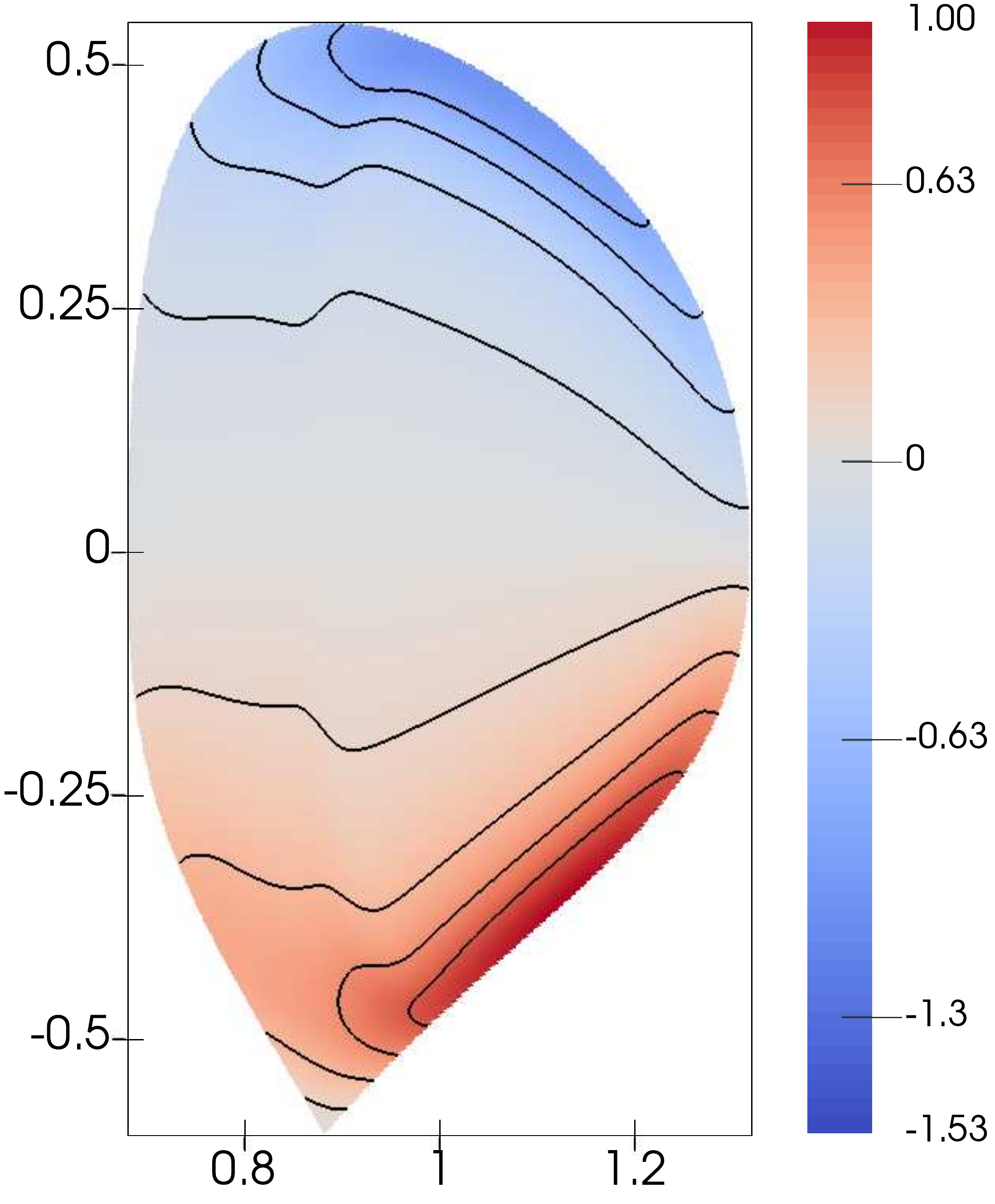} & \kern-1em \includegraphics[height=0.2\linewidth]{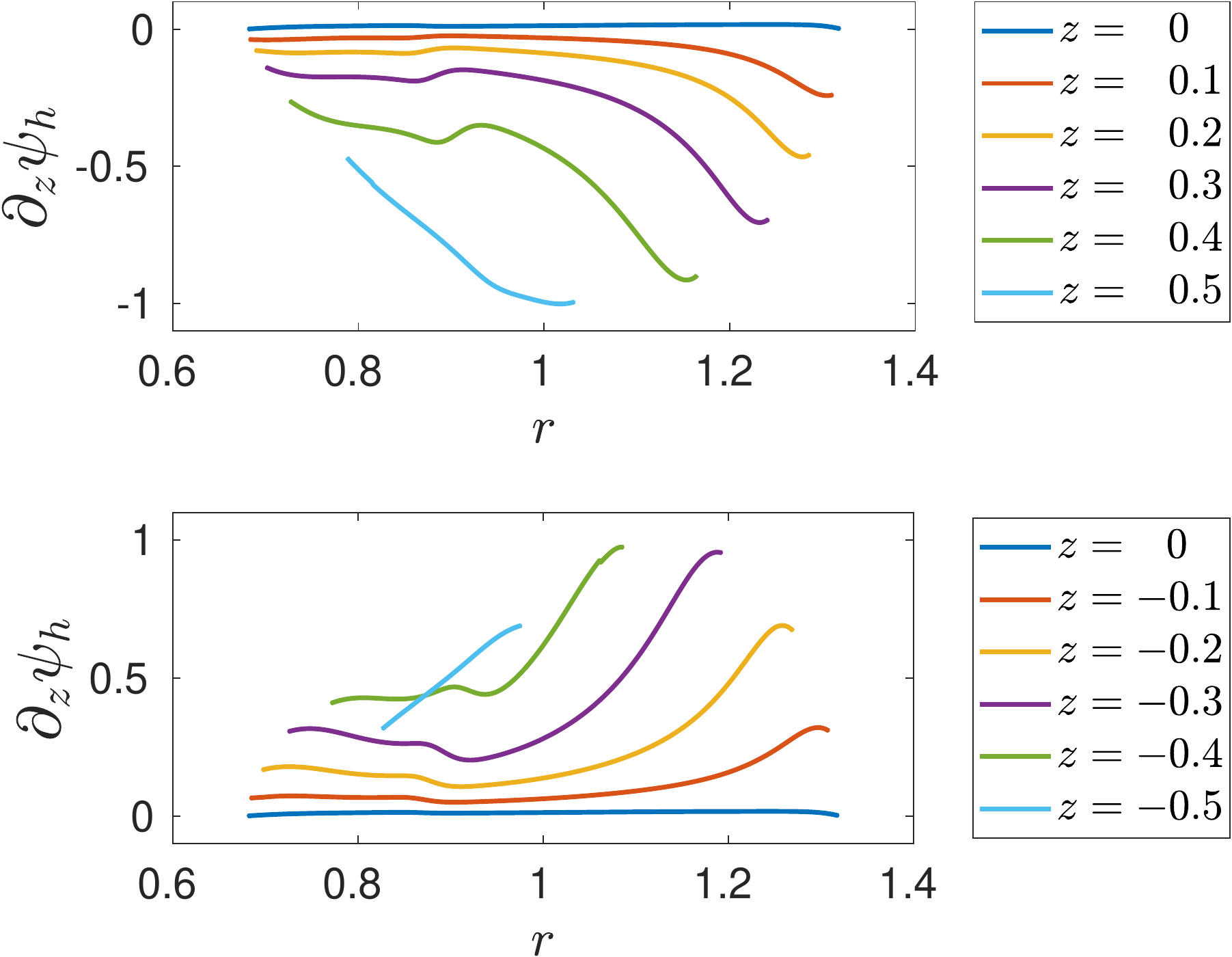} & \kern-1em \includegraphics[height=0.2\linewidth]{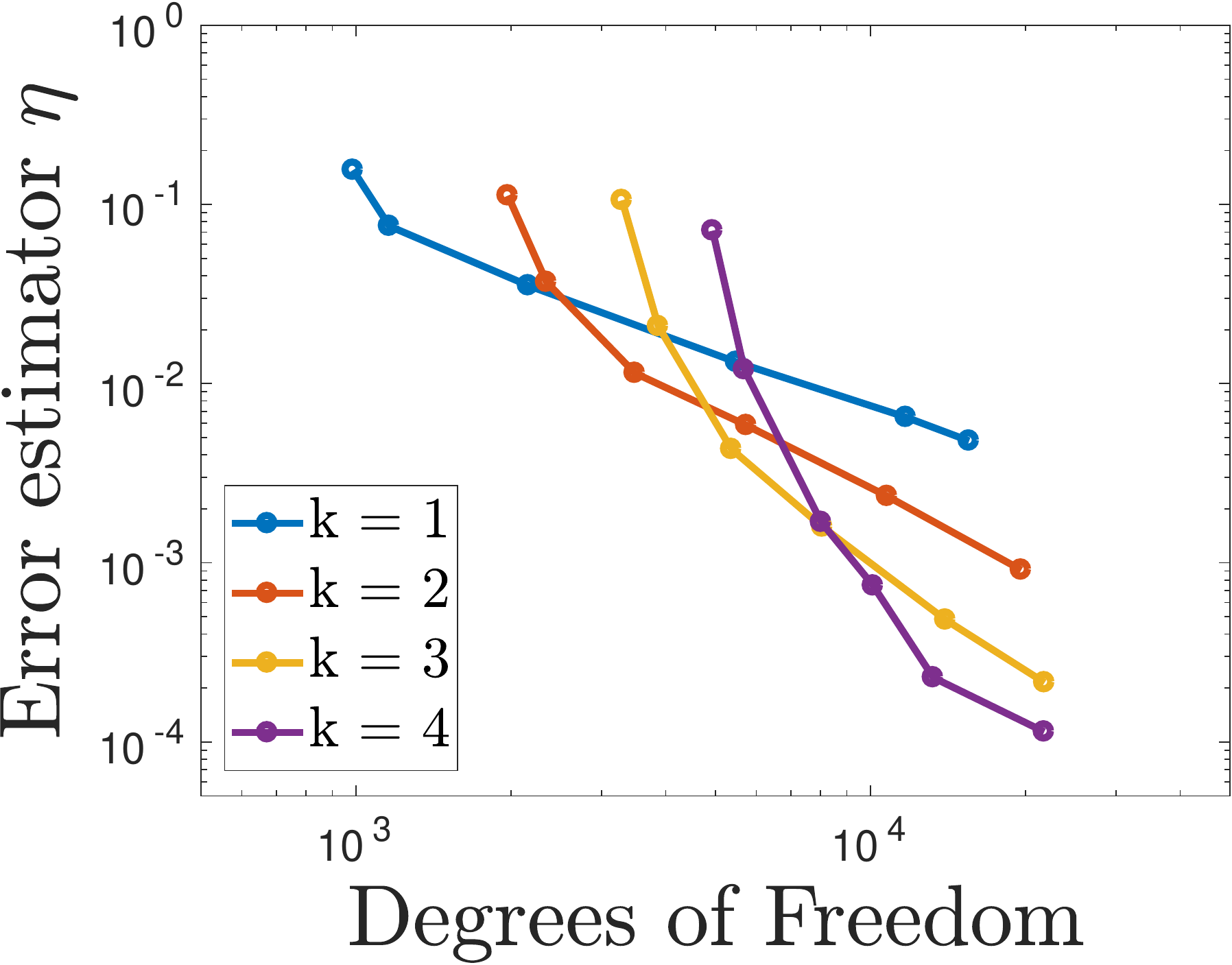}
\end{tabular}}
\caption{For the source term given by Equation \eqref{eq:BarrierSource}, the partial derivative of the solution in the $r$ direction (left) develops a step-like change (center left) due to the structure of the source term. Change in the $z$ direction (center and center right) is less dramatic. The convergence history for the error estimator is shown on the right for polynomial approximations of degrees one to five using maximum marking for $\gamma=0.3$.}\label{fig:BarrierIterXDerivatives}
\end{figure}

The post-processed numerical solution and cross sections at different heights are depicted on the right end of Figure \ref{fig:BarrierIterX}. The sharp change in the slope of the solution drives the interior refinement thus enabling the accurate approximation of the step-like behavior of the derivative in the horizontal direction (Figure \ref{fig:BarrierIterXDerivatives}). The approximate first derivatives are shown in Figure \ref{fig:BarrierIterXDerivatives} along with a plot of the convergence history of the global error estimator. 

%
\section{Conclusion} \label{sec:conclusion}
%
The solver for fixed-boundary Grad-Shafranov equilibria based on the hybridizable discontinuous Galerkin method we presented has several attractive features beyond the high order approximation properties for $\psi$ and its partial derivatives. The use of an HDG framework provides the code with a robust and reliable method that is naturally suited for parallelization and addresses the issue of unnecessarily large numbers of degrees of freedom---usually associated with discontinuous Galerkin discretizations---through hybridization.

The use of a polygonal subdomain to carry out the computations combined with the transfer algorithm to impose boundary conditions ``at a distance" allows for a simple, yet highly accurate, treatment of curved boundaries without having to resort to more complicated techniques like isoparametric mappings. It also enables a unified treatment of both smooth geometries and those with corners, corresponding to magnetic X-points. Moreover, in applications where the geometry of the confinement region needs to be updated, this technique provides the additional benefit of avoiding the need for constant updating of a fitted mesh. 

The addition of a local error estimator and adaptive mesh refinement allows for focused computational efforts.  As the numerical experiments show, this feature combined with the updating of the computational domain and the improved geometric approximation of the physical domain as the refinement progresses allows for efficient detection of relevant physical effects near the edge (as is the case for equilibria with a pressure pedestal) or the resolution of highly localized internal structures (as is the case for equilibia with an internal transport barrier). 
%
\section{Acknowledgements}
The computational implementation of the algorithm described in this paper benefited greatly from the detailed explanations and code templates for HDG and adaptive refinement provided respectively by Fu, Gatica and Sayas \cite{FuGaSa:2015}, and Funken, Praetorius and Wissgott \cite{FuPrWi:2011}. Finally, sampling of the confinement regions from the analytic expressions given in \cite{CeFr:2010} was done using {\tt chebfun} \cite{DrHaTr:2014}.

The authors are deeply grateful to Wrick Sengupta  (NYU) and Georg Stadler (NYU) for their valuable insights on the physical and mathematical aspects of the problem. They also thank Fran\c{c}ois Waelbroeck (UT-Austin) for suggesting the current hole problem as a benchmarking test. Antoine J. Cerfon. and Tonatiuh S\'anchez-Vizuet were partially funded by the US Department of Energy. Grant No. DE-FG02-86ER53233. Manuel E. Solano was partially funded by CONICYT--Chile through FONDECYT project No. 1160320 and by Project AFB170001 of the PIA Program: Concurso Apoyo a Centros Cientificos y Tecnologicos de Excelencia con Financiamiento
Basal.

\clearpage





\bibliographystyle{elsarticle-num}
\bibliography{References}

\begin{thebibliography}{10}
\expandafter\ifx\csname url\endcsname\relax
  \def\url#1{\texttt{#1}}\fi
\expandafter\ifx\csname urlprefix\endcsname\relax\def\urlprefix{URL }\fi
\expandafter\ifx\csname href\endcsname\relax
  \def\href#1#2{#2} \def\path#1{#1}\fi

\bibitem{GrRu:1958}
H.~Grad, H.~Rubin, Hydromagnetic equilibria and force{-}free fields, in: Proc.
  Second international conference on the peaceful uses of atomic energy,
  Geneva, Vol. 31,190, United Nations, New York, 1958.

\bibitem{Shafranov:1958}
V.~D. Shafranov, On magnetohydrodynamical equilibrium configurations, Soviet
  Physics JETP 6 (1958) 545--554.

\bibitem{LuSc:1957}
R.~L\"{u}st, A.~Schl\"{u}ter, Axialsymmetrische magnetohydrodynamische
  {G}leichgewichtskonfigurationen, Z. Naturf 12a (1957) 850--854.

\bibitem{Brambilla1999}
M.~Brambilla,
  \href{https://doi.org/10.1088%2F0741-3335%2F41%2F1%2F002}{Numerical
  simulation of ion cyclotron waves in tokamak plasmas}, Plasma Physics and
  Controlled Fusion 41~(1) (1999) 1--34.
\newblock \href {http://dx.doi.org/10.1088/0741-3335/41/1/002}
  {\path{doi:10.1088/0741-3335/41/1/002}}.
\newline\urlprefix\url{https://doi.org/10.1088%2F0741-3335%2F41%2F1%2F002}

\bibitem{Fable2013}
E.~Fable, C.~Angioni, A.~Ivanov, K.~Lackner, O.~Maj, S.~Yu, Medvedev,
  G.~Pautasso, G.~Pereverzev,
  \href{https://doi.org/10.1088%2F0029-5515%2F53%2F3%2F033002}{A stable scheme
  for computation of coupled transport and equilibrium equations in tokamaks},
  Nuclear Fusion 53~(3) (2013) 033002.
\newblock \href {http://dx.doi.org/10.1088/0029-5515/53/3/033002}
  {\path{doi:10.1088/0029-5515/53/3/033002}}.
\newline\urlprefix\url{https://doi.org/10.1088%2F0029-5515%2F53%2F3%2F033002}

\bibitem{HoSo:2014}
E.~Howell, C.~Sovinec,
  \href{http://www.sciencedirect.com/science/article/pii/S001046551400040X}{Solving
  the {G}rad{-}{S}hafranov equation with spectral elements}, Computer Physics
  Communications 185~(5) (2014) 1415 -- 1421.
\newblock \href {http://dx.doi.org/https://doi.org/10.1016/j.cpc.2014.02.008}
  {\path{doi:https://doi.org/10.1016/j.cpc.2014.02.008}}.
\newline\urlprefix\url{http://www.sciencedirect.com/science/article/pii/S001046551400040X}

\bibitem{Kerner1998}
W.~Kerner, J.~Goedbloed, G.~Huysmans, S.~Poedts, E.~Schwarz,
  \href{http://www.sciencedirect.com/science/article/pii/S0021999198959101}{Castor:
  Normal-mode analysis of resistive mhd plasmas}, Journal of Computational
  Physics 142~(2) (1998) 271 -- 303.
\newblock \href {http://dx.doi.org/https://doi.org/10.1006/jcph.1998.5910}
  {\path{doi:https://doi.org/10.1006/jcph.1998.5910}}.
\newline\urlprefix\url{http://www.sciencedirect.com/science/article/pii/S0021999198959101}

\bibitem{Lapillonne2009}
X.~Lapillonne, S.~Brunner, T.~Dannert, S.~Jolliet, A.~Marinoni, L.~Villard,
  T.~G\"{o}rler, F.~Jenko, F.~Merz,
  \href{https://doi.org/10.1063/1.3096710}{Clarifications to the limitations of
  the s-{$\alpha$} equilibrium model for gyrokinetic computations of
  turbulence}, Physics of Plasmas 16~(3) (2009) 032308.
\newblock \href {http://arxiv.org/abs/https://doi.org/10.1063/1.3096710}
  {\path{arXiv:https://doi.org/10.1063/1.3096710}}, \href
  {http://dx.doi.org/10.1063/1.3096710} {\path{doi:10.1063/1.3096710}}.
\newline\urlprefix\url{https://doi.org/10.1063/1.3096710}

\bibitem{Lee2017}
J.~Lee, J.~Freidberg, A.~Cerfon, M.~Greenwald,
  \href{https://doi.org/10.1088%2F1741-4326%2Faa6877}{An analytic scaling
  relation for the maximum tokamak elongation against
  n{\hspace{0.167em}}{\hspace{0.167em}}={\hspace{0.167em}}{\hspace{0.167em}}0
  {MHD} resistive wall modes}, Nuclear Fusion 57~(6) (2017) 066051.
\newblock \href {http://dx.doi.org/10.1088/1741-4326/aa6877}
  {\path{doi:10.1088/1741-4326/aa6877}}.
\newline\urlprefix\url{https://doi.org/10.1088%2F1741-4326%2Faa6877}

\bibitem{Lee2019}
J.~Lee, D.~Smithe, E.~F. Jaeger, R.~W. Harvey, P.~T. Bonoli,
  \href{https://doi.org/10.1063/1.5066288}{Similarity of the coupled equations
  for rf waves in a tokamak}, Physics of Plasmas 26~(1) (2019) 012505.
\newblock \href {http://arxiv.org/abs/https://doi.org/10.1063/1.5066288}
  {\path{arXiv:https://doi.org/10.1063/1.5066288}}, \href
  {http://dx.doi.org/10.1063/1.5066288} {\path{doi:10.1063/1.5066288}}.
\newline\urlprefix\url{https://doi.org/10.1063/1.5066288}

\bibitem{TaTo:1991}
T.~Takeda, S.~Tokuda,
  \href{http://www.sciencedirect.com/science/article/pii/002199919190074U}{Computation
  of {MHD} equilibrium of tokamak plasma}, Journal of Computational Physics
  93~(1) (1991) 1 -- 107.
\newblock \href
  {http://dx.doi.org/https://doi.org/10.1016/0021-9991(91)90074-U}
  {\path{doi:https://doi.org/10.1016/0021-9991(91)90074-U}}.
\newline\urlprefix\url{http://www.sciencedirect.com/science/article/pii/002199919190074U}

\bibitem{LuBoRo:1992}
H.~L\"{u}tjens, A.~Bondeson, A.~Roy,
  \href{http://www.sciencedirect.com/science/article/pii/001046559290167W}{Axisymmetric
  {MHD} equilibrium solver with bicubic {H}ermite elements}, Computer Physics
  Communications 69~(2) (1992) 287 -- 298.
\newblock \href
  {http://dx.doi.org/https://doi.org/10.1016/0010-4655(92)90167-W}
  {\path{doi:https://doi.org/10.1016/0010-4655(92)90167-W}}.
\newline\urlprefix\url{http://www.sciencedirect.com/science/article/pii/001046559290167W}

\bibitem{Lutjens1996}
H.~L\"{u}tjens, A.~Bondeson, O.~Sauter,
  \href{http://www.sciencedirect.com/science/article/pii/001046559600046X}{The
  {CHEASE} code for toroidal {MHD} equilibria}, Computer Physics Communications
  97~(3) (1996) 219 -- 260.
\newblock \href
  {http://dx.doi.org/https://doi.org/10.1016/0010-4655(96)00046-X}
  {\path{doi:https://doi.org/10.1016/0010-4655(96)00046-X}}.
\newline\urlprefix\url{http://www.sciencedirect.com/science/article/pii/001046559600046X}

\bibitem{PaKoFe:2016}
A.~Palha, B.~Koren, F.~Felici,
  \href{http://www.sciencedirect.com/science/article/pii/S0021999116300341}{A
  mimetic spectral element solver for the {G}rad{-}{S}hafranov equation},
  Journal of Computational Physics 316~(Supplement C) (2016) 63 -- 93.
\newblock \href {http://dx.doi.org/https://doi.org/10.1016/j.jcp.2016.04.002}
  {\path{doi:https://doi.org/10.1016/j.jcp.2016.04.002}}.
\newline\urlprefix\url{http://www.sciencedirect.com/science/article/pii/S0021999116300341}

\bibitem{SaSo:2018}
T.~S\'anchez-Vizuet, M.~E. Solano,
  \href{http://www.sciencedirect.com/science/article/pii/S0010465518303278}{A
  {H}ybridizable {D}iscontinuous {G}alerkin solver for the {G}rad-{S}hafranov
  equation}, Computer Physics Communications 235 (2019) 120 -- 132.
\newblock \href {http://dx.doi.org/https://doi.org/10.1016/j.cpc.2018.09.013}
  {\path{doi:https://doi.org/10.1016/j.cpc.2018.09.013}}.
\newline\urlprefix\url{http://www.sciencedirect.com/science/article/pii/S0010465518303278}

\bibitem{PaCeFrGrOn:2013}
A.~Pataki, A.~J. Cerfon, J.~P. Freidberg, L.~Greengard, M.~O’Neil,
  \href{http://www.sciencedirect.com/science/article/pii/S0021999113001721}{A
  fast, high{-}order solver for the {G}rad{-}{S}hafranov equation}, Journal of
  Computational Physics 243~(Supplement C) (2013) 28 -- 45.
\newblock \href {http://dx.doi.org/https://doi.org/10.1016/j.jcp.2013.02.045}
  {\path{doi:https://doi.org/10.1016/j.jcp.2013.02.045}}.
\newline\urlprefix\url{http://www.sciencedirect.com/science/article/pii/S0021999113001721}

\bibitem{LeCe:2015}
J.~Lee, A.~Cerfon,
  \href{http://www.sciencedirect.com/science/article/pii/S0010465515000351}{{ECOM}:
  A fast and accurate solver for toroidal axisymmetric {MHD} equilibria},
  Computer Physics Communications 190 (2015) 72 -- 88.
\newblock \href {http://dx.doi.org/https://doi.org/10.1016/j.cpc.2015.01.015}
  {\path{doi:https://doi.org/10.1016/j.cpc.2015.01.015}}.
\newline\urlprefix\url{http://www.sciencedirect.com/science/article/pii/S0010465515000351}

\bibitem{CoGoLa:2009}
B.~Cockburn, J.~Gopalakrishnan, R.~Lazarov,
  \href{http://dx.doi.org/10.1137/070706616}{Unified hybridization of
  discontinuous {G}alerkin, mixed, and continuous {G}alerkin methods for second
  order elliptic problems}, SIAM J. Numer. Anal. 47~(2) (2009) 1319--1365.
\newblock \href {http://dx.doi.org/10.1137/070706616}
  {\path{doi:10.1137/070706616}}.
\newline\urlprefix\url{http://dx.doi.org/10.1137/070706616}

\bibitem{SaSaSo:2018}
N.~S\'anchez, T.~S\'anchez{-}Vizuet, M.~E. Solano, A priori and a posteriori
  error analysis of an {HDG} method for semi-linear elliptic problems in curved
  domains, (In preparation).

\bibitem{CoSo:2012}
B.~Cockburn, M.~Solano, \href{http://dx.doi.org/10.1137/100805200}{Solving
  {D}irichlet boundary-value problems on curved domains by extensions from
  subdomains}, SIAM J. Sci. Comput. 34~(1) (2012) A497--A519.
\newblock \href {http://dx.doi.org/10.1137/100805200}
  {\path{doi:10.1137/100805200}}.
\newline\urlprefix\url{http://dx.doi.org/10.1137/100805200}

\bibitem{Guyan:1965}
R.~J. Guyan, \href{https://doi.org/10.2514/3.2874}{Reduction of stiffness and
  mass matrices}, {AIAA} Journal 3~(2) (1965) 380--380.
\newblock \href {http://dx.doi.org/10.2514/3.2874} {\path{doi:10.2514/3.2874}}.
\newline\urlprefix\url{https://doi.org/10.2514/3.2874}

\bibitem{Fraeijs:1965}
B.~X. Fraeijs~de Veubeke,
  \href{https://ci.nii.ac.jp/naid/10003737730/en/}{Displacement and equilibrium
  models in the finite element method}, in: O.~C. Zienkiewicz, G.~S. Holister
  (Eds.), Stress Analysis, Wiley, New York, 1965, pp. 275--284.
\newline\urlprefix\url{https://ci.nii.ac.jp/naid/10003737730/en/}

\bibitem{Cockburn:2016}
B.~Cockburn, \href{https://doi.org/10.1007/978-3-319-41640-3_5}{Static
  condensation, hybridization, and the devising of the {HDG} methods}, in:
  Lecture Notes in Computational Science and Engineering, Springer
  International Publishing, 2016, pp. 129--177.
\newblock \href {http://dx.doi.org/10.1007/978-3-319-41640-3_5}
  {\path{doi:10.1007/978-3-319-41640-3_5}}.
\newline\urlprefix\url{https://doi.org/10.1007/978-3-319-41640-3_5}

\bibitem{Cockburn:2010}
B.~Cockburn, The hybridizable discontinuous {G}alerkin methods, in: Proceedings
  of the {I}nternational {C}ongress of {M}athematicians. {V}olume {IV},
  Hindustan Book Agency, New Delhi, 2010, pp. 2749--2775.

\bibitem{Anderson:1965}
D.~G. Anderson, \href{http://dx.doi.org/10.1145/321296.321305}{Iterative
  procedures for nonlinear integral equations}, J. Assoc. Comput. Mach. 12
  (1965) 547--560.
\newblock \href {http://dx.doi.org/10.1145/321296.321305}
  {\path{doi:10.1145/321296.321305}}.
\newline\urlprefix\url{http://dx.doi.org/10.1145/321296.321305}

\bibitem{ToKe:2015}
A.~Toth, C.~T. Kelley, \href{http://dx.doi.org/10.1137/130919398}{Convergence
  analysis for {A}nderson acceleration}, SIAM J. Numer. Anal. 53~(2) (2015)
  805--819.
\newblock \href {http://dx.doi.org/10.1137/130919398}
  {\path{doi:10.1137/130919398}}.
\newline\urlprefix\url{http://dx.doi.org/10.1137/130919398}

\bibitem{WaNi:2011}
H.~F. Walker, P.~Ni, \href{http://dx.doi.org/10.1137/10078356X}{Anderson
  acceleration for fixed-point iterations}, SIAM J. Numer. Anal. 49~(4) (2011)
  1715--1735.
\newblock \href {http://dx.doi.org/10.1137/10078356X}
  {\path{doi:10.1137/10078356X}}.
\newline\urlprefix\url{http://dx.doi.org/10.1137/10078356X}

\bibitem{Stenberg:1991}
R.~Stenberg, \href{https://doi.org/10.1051/m2an/1991250101511}{Postprocessing
  schemes for some mixed finite elements}, ESAIM: M2AN 25~(1) (1991) 151--167.
\newblock \href {http://dx.doi.org/10.1051/m2an/1991250101511}
  {\path{doi:10.1051/m2an/1991250101511}}.
\newline\urlprefix\url{https://doi.org/10.1051/m2an/1991250101511}

\bibitem{CoGoSa:2010}
B.~Cockburn, J.~Gopalakrishnan, F.-J. Sayas,
  \href{http://dx.doi.org/10.1090/S0025-5718-10-02334-3}{A projection-based
  error analysis of {HDG} methods}, Math. Comp. 79~(271) (2010) 1351--1367.
\newblock \href {http://dx.doi.org/10.1090/S0025-5718-10-02334-3}
  {\path{doi:10.1090/S0025-5718-10-02334-3}}.
\newline\urlprefix\url{http://dx.doi.org/10.1090/S0025-5718-10-02334-3}

\bibitem{CoZh:2012}
B.~Cockburn, W.~Zhang, \href{https://doi.org/10.1007/s10915-011-9522-2}{A
  posteriori error estimates for {HDG} methods}, Journal of Scientific
  Computing 51~(3) (2012) 582--607.
\newblock \href {http://dx.doi.org/10.1007/s10915-011-9522-2}
  {\path{doi:10.1007/s10915-011-9522-2}}.
\newline\urlprefix\url{https://doi.org/10.1007/s10915-011-9522-2}

\bibitem{CoZh:2013}
B.~Cockburn, W.~Zhang, A posteriori error analysis for hybridizable
  discontinuous {G}alerkin methods for second order elliptic problems, SIAM
  Journal on Numerical Analysis 51~(1) (2013) 676--693.
\newblock \href {http://dx.doi.org/10.1137/120866269}
  {\path{doi:10.1137/120866269}}.

\bibitem{CoNoZh:2016}
B.~Cockburn, R.~H. Nochetto, W.~Zhang,
  \href{https://doi.org/10.1090/mcom/3014}{Contraction property of adaptive
  hybridizable discontinuous {G}alerkin methods}, Math. Comp. 85~(299) (2016)
  1113--1141.
\newblock \href {http://dx.doi.org/10.1090/mcom/3014}
  {\path{doi:10.1090/mcom/3014}}.
\newline\urlprefix\url{https://doi.org/10.1090/mcom/3014}

\bibitem{Dorfler:1996}
W.~D\"orfler, \href{https://doi.org/10.1137/0733054}{A {C}onvergent {A}daptive
  {A}lgorithm for {P}oisson's {E}quation}, SIAM Journal on Numerical Analysis
  33~(3) (1996) 1106--1124.
\newblock \href {http://dx.doi.org/10.1137/0733054}
  {\path{doi:10.1137/0733054}}.
\newline\urlprefix\url{https://doi.org/10.1137/0733054}

\bibitem{Dorfler:1995}
W.~D\"orfler, \href{https://doi.org/10.1007/BF02238484}{A robust adaptive
  strategy for the nonlinear {P}oisson equation}, Computing 55~(4) (1995)
  289--304.
\newblock \href {http://dx.doi.org/10.1007/BF02238484}
  {\path{doi:10.1007/BF02238484}}.
\newline\urlprefix\url{https://doi.org/10.1007/BF02238484}

\bibitem{BaRh:1978}
I.~Babu\v{s}ka, W.~C. Rheinboldt, \href{https://doi.org/10.1137/0715049}{Error
  estimates for adaptive finite element computations}, SIAM J. Numer. Anal.
  15~(4) (1978) 736--754.
\newblock \href {http://dx.doi.org/10.1137/0715049}
  {\path{doi:10.1137/0715049}}.
\newline\urlprefix\url{https://doi.org/10.1137/0715049}

\bibitem{MoSiVe:2008}
P.~Morin, K.~G. Siebert, A.~Veeser,
  \href{https://doi.org/10.1142/S0218202508002838}{A basic convergence result
  for conforming adaptive finite elements}, Math. Models Methods Appl. Sci.
  18~(5) (2008) 707--737.
\newblock \href {http://dx.doi.org/10.1142/S0218202508002838}
  {\path{doi:10.1142/S0218202508002838}}.
\newline\urlprefix\url{https://doi.org/10.1142/S0218202508002838}

\bibitem{Rivara:1984a}
M.-C. Rivara, \href{https://doi.org/10.1137/0721042}{Mesh refinement processes
  based on the generalized bisection of simplices}, SIAM J. Numer. Anal. 21~(3)
  (1984) 604--613.
\newblock \href {http://dx.doi.org/10.1137/0721042}
  {\path{doi:10.1137/0721042}}.
\newline\urlprefix\url{https://doi.org/10.1137/0721042}

\bibitem{Rivara:1984b}
M.-C. Rivara, \href{https://doi.org/10.1002/nme.1620200412}{Algorithms for
  refining triangular grids suitable for adaptive and multigrid techniques},
  Internat. J. Numer. Methods Engrg. 20~(4) (1984) 745--756.
\newblock \href {http://dx.doi.org/10.1002/nme.1620200412}
  {\path{doi:10.1002/nme.1620200412}}.
\newline\urlprefix\url{https://doi.org/10.1002/nme.1620200412}

\bibitem{BaShWe:1983}
R.~E. Bank, A.~H. Sherman, A.~Weiser, Refinement algorithms and data structures
  for regular local mesh refinement, in: Scientific computing ({M}ontreal,
  {Q}ue., 1982), IMACS Trans. Sci. Comput., I, IMACS, New Brunswick, NJ, 1983,
  pp. 3--17.

\bibitem{Solov'ev:1968}
L.~S. Solov'ev, The theory of hydromagnetic stability of toroidal plasma
  configurations, Soviet Physics JETP 26 (1968) 400--407.

\bibitem{CeFr:2010}
A.~J. Cerfon, J.~P. Freidberg,
  \href{http://dx.doi.org/10.1063/1.3328818}{{``O}ne size fits all{"} analytic
  solutions to the {G}rad{-}{S}hafranov equation}, Physics of Plasmas 17~(3)
  (2010) 032502.
\newblock \href {http://dx.doi.org/10.1063/1.3328818}
  {\path{doi:10.1063/1.3328818}}.
\newline\urlprefix\url{http://dx.doi.org/10.1063/1.3328818}

\bibitem{Wolf:2003}
R.~C. Wolf, \href{http://stacks.iop.org/0741-3335/45/i=1/a=201}{Internal
  transport barriers in tokamak plasmas}, Plasma Physics and Controlled Fusion
  45~(1) (2003) R1.
\newline\urlprefix\url{http://stacks.iop.org/0741-3335/45/i=1/a=201}

\bibitem{Cedres:2015}
H.~Heumann, J.~Blum, C.~Boulbe, B.~Faugeras, G.~Selig, J.-M. An\'{e}, S.~Br\'{e}mond,
  V.~Grandgirard, P.~Hertout, E.~Nardon, et~al.,
  \href{https://www.cambridge.org/core/journals/journal-of-plasma-physics/article/quasistatic-freeboundary-equilibrium-of-toroidal-plasma-with-cedres-computational-methods-and-applications/4358A3CED81708EBBF4BD64FC8D290A1}{Quasi-static
  free-boundary equilibrium of toroidal plasma with {CEDRES}: {C}omputational
  methods and applications}, Journal of Plasma Physics 81~(3) (2015) 905810301.
\newblock \href {http://dx.doi.org/10.1017/S0022377814001251}
  {\path{doi:10.1017/S0022377814001251}}.
\newline\urlprefix\url{https://www.cambridge.org/core/journals/journal-of-plasma-physics/article/quasistatic-freeboundary-equilibrium-of-toroidal-plasma-with-cedres-computational-methods-and-applications/4358A3CED81708EBBF4BD64FC8D290A1}

\bibitem{FujitaEtAl:2001}
T.~Fujita, T.~Oikawa, T.~Suzuki, S.~Ide, Y.~Sakamoto, Y.~Koide, T.~Hatae,
  O.~Naito, A.~Isayama, N.~Hayashi, H.~Shirai,
  \href{https://link.aps.org/doi/10.1103/PhysRevLett.87.245001}{Plasma
  equilibrium and confinement in a tokamak with nearly zero central current
  density in {JT-60U}}, Phys. Rev. Lett. 87 (2001) 245001.
\newblock \href {http://dx.doi.org/10.1103/PhysRevLett.87.245001}
  {\path{doi:10.1103/PhysRevLett.87.245001}}.
\newline\urlprefix\url{https://link.aps.org/doi/10.1103/PhysRevLett.87.245001}

\bibitem{HaJaSt02}
G.~W. Hammett, S.~C. Jardin, B.~C. Stratton,
  \href{https://doi.org/10.1063/1.1608935}{Non-existence of normal tokamak
  equilibria with negative central current}, Physics of Plasmas 10~(10) (2003)
  4048--4052.
\newblock \href {http://arxiv.org/abs/https://doi.org/10.1063/1.1608935}
  {\path{arXiv:https://doi.org/10.1063/1.1608935}}, \href
  {http://dx.doi.org/10.1063/1.1608935} {\path{doi:10.1063/1.1608935}}.
\newline\urlprefix\url{https://doi.org/10.1063/1.1608935}

\bibitem{GoLe:2009}
P.-A. Gourdain, J.-N. Leboeuf, \href{https://doi.org/10.1063/1.3247073}{Hollow
  current profile scenarios for advanced tokamak reactor operations}, Physics
  of Plasmas 16~(11) (2009) 112506.
\newblock \href {http://arxiv.org/abs/https://doi.org/10.1063/1.3247073}
  {\path{arXiv:https://doi.org/10.1063/1.3247073}}, \href
  {http://dx.doi.org/10.1063/1.3247073} {\path{doi:10.1063/1.3247073}}.
\newline\urlprefix\url{https://doi.org/10.1063/1.3247073}

\bibitem{GoLeNe:2007}
P.-A. Gourdain, J.-N. Leboeuf, R.~Y. Neches,
  \href{https://doi.org/10.1063/1.2807024}{Stability of highly shifted
  equilibria in a large aspect ratio low-field tokamak}, Physics of Plasmas
  14~(11) (2007) 112513.
\newblock \href {http://arxiv.org/abs/https://doi.org/10.1063/1.2807024}
  {\path{arXiv:https://doi.org/10.1063/1.2807024}}, \href
  {http://dx.doi.org/10.1063/1.2807024} {\path{doi:10.1063/1.2807024}}.
\newline\urlprefix\url{https://doi.org/10.1063/1.2807024}

\bibitem{Fujita_2010}
T.~Fujita,
  \href{https://doi.org/10.1088%2F0029-5515%2F50%2F11%2F113001}{Tokamak
  equilibria with nearly zero central current: the current hole}, Nuclear
  Fusion 50~(11) (2010) 113001.
\newblock \href {http://dx.doi.org/10.1088/0029-5515/50/11/113001}
  {\path{doi:10.1088/0029-5515/50/11/113001}}.
\newline\urlprefix\url{https://doi.org/10.1088%2F0029-5515%2F50%2F11%2F113001}

\bibitem{FuGaSa:2015}
Z.~Fu, L.~F. Gatica, F.-J. Sayas,
  \href{http://dx.doi.org/10.1145/2658992}{Algorithm 949: {MATLAB} tools for
  {HDG} in three dimensions}, ACM Trans. Math. Software 41~(3) (2015) Art. 20,
  21.
\newblock \href {http://dx.doi.org/10.1145/2658992}
  {\path{doi:10.1145/2658992}}.
\newline\urlprefix\url{http://dx.doi.org/10.1145/2658992}

\bibitem{FuPrWi:2011}
S.~Funken, D.~Praetorius, P.~Wissgott,
  \href{https://doi.org/10.2478/cmam-2011-0026}{Efficient implementation of
  adaptive {P}1-{FEM} in {M}atlab}, Comput. Methods Appl. Math. 11~(4) (2011)
  460--490.
\newblock \href {http://dx.doi.org/10.2478/cmam-2011-0026}
  {\path{doi:10.2478/cmam-2011-0026}}.
\newline\urlprefix\url{https://doi.org/10.2478/cmam-2011-0026}

\bibitem{DrHaTr:2014}
T.~A. Driscoll, N.~Hale, L.~N. Trefethen,
  \href{http://www.chebfun.org/docs/guide/}{Chebfun Guide}, Pafnuty
  Publications, 2014.
\newline\urlprefix\url{http://www.chebfun.org/docs/guide/}

\end{thebibliography}
\end{document}